\documentclass[journal]{vgtc}                     % final (journal style)
\onlineid{0}

\vgtccategory{Research}

\vgtcpapertype{please specify}

\newcommand{\papertitle}{Designing Semantically-Resonant Abstract Patterns\\ for Data Visualization}
\newcommand{\sr}{se\-man\-ti\-cal\-ly-re\-so\-nant\xspace}

\title{\papertitle}

\author{%
  \authororcid{Zihan Lu}{0009-0003-1610-7435},
  \authororcid{Tingying He}{0000-0002-0500-7995},
  \authororcid{Jiayi Hong}{0000-0002-1332-5045},
  \authororcid{Lijie Yao}{0000-0002-4208-5140},
  \authororcid{Tobias Isenberg}{0000-0001-7953-8644}
}

\authorfooter{
  \item \textls[-5]{Z.\ Lu (\includegraphics[height=6pt]{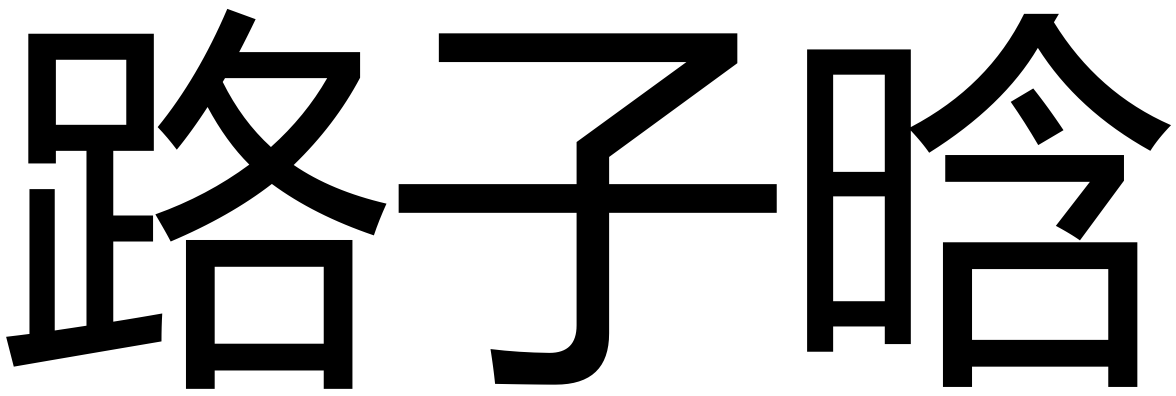}) and T.\ Isenberg are with Univ.\ Paris-Saclay, CNRS, Inria, LISN, France. Emails: zyhen.loo@gmail.com, given\_name.family\-name@inria.fr,}
  \item T.\ He (\includegraphics[height=6pt]{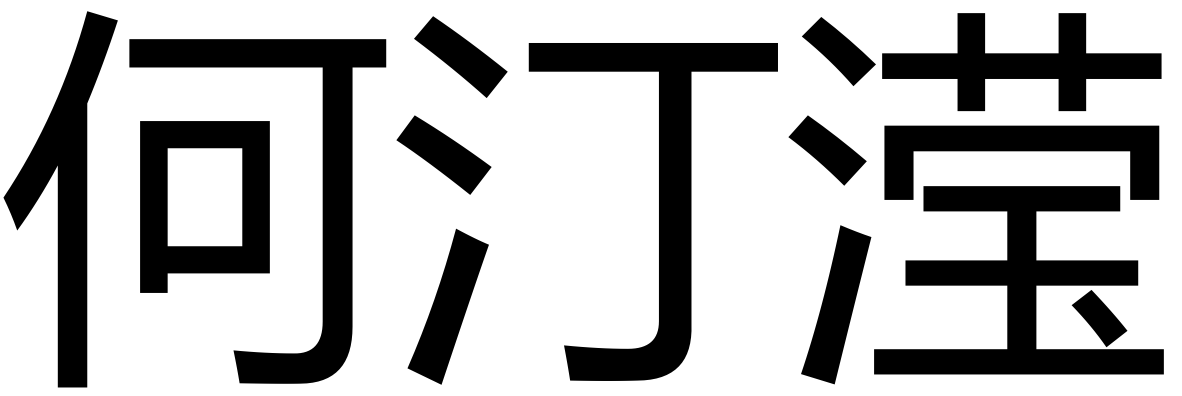}) is with Université Paris-Saclay, CNRS, Inria, LISN, France and the University of Utah, USA. E-mail: hetingying.hty@gmail.com, 
  \item J.\ Hong (\includegraphics[height=6pt]{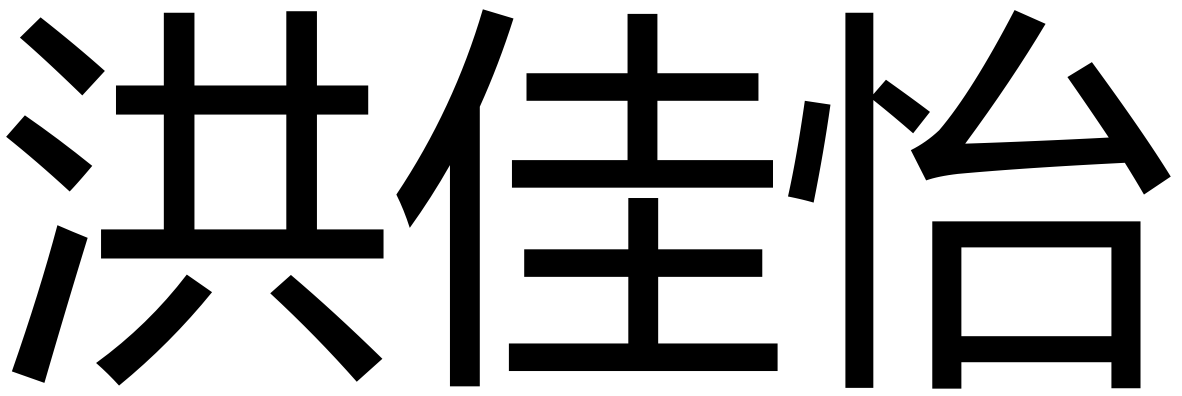}) is with the Arizona State University, USA. Email: jiayihong.vis@gmail.com,
  \item L.\ Yao (\includegraphics[height=6pt]{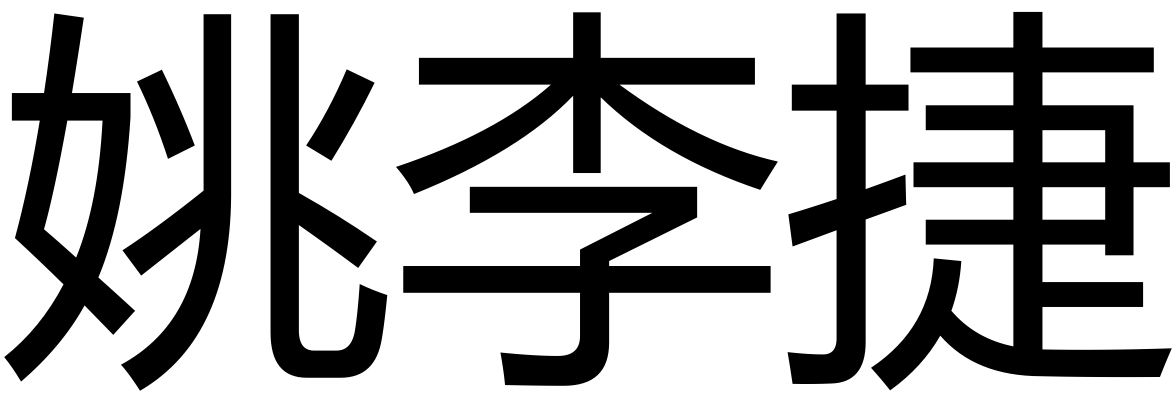}) is with Xi'an Jiaotong-Liverpool University, China.  Email: yaolijie0219@gmail.com\\
  \item Z.\ Lu and T.\ He share first authorship. T.\ He is corresponding author.
  
}

\abstract{%
  We present a structured design methodology for creating semantically-resonant abstract patterns, making the pattern design process accessible to the general public. Semantically-resonant patterns are those that intuitively evoke the concept they represent within a specific set (e.g., in a vegetable concept set, small dots for olives and large dots for tomatoes), analogous to the concept of semantically-resonant colors (e.g., using olive green for olives and red for tomatoes). Previous research has shown that semantically-resonant colors can improve chart reading speed, and designers have made attempts to integrate semantic cues into abstract pattern designs. However, a systematic framework for developing such patterns was lacking. To bridge this gap, we conducted a series of workshops with design experts, resulting in a design methodology that summarizes the methodology for designing semantically-resonant abstract patterns. We evaluated our design methodology through another series of workshops with non-design participants. The results indicate that our proposed design methodology effectively supports the general public in designing semantically-resonant abstract patterns for both abstract and concrete concepts.

}


\graphicspath{{figs/}{figures/}{pictures/}{images/}{./}} % where to search for the images

\usepackage{tabu}                      % only used for the table example
\usepackage{booktabs}                  % only used for the table example
\usepackage{lipsum}                    % used to generate placeholder text
\usepackage{mwe}                       % used to generate placeholder figures

\usepackage{mathptmx}                  % use matching math font

\usepackage{cuted} % appendix
\usepackage{ccicons} %copyright icons

\usepackage{makecell}

\newcommand{\inlinevis}[3]{\raisebox{#1}[0pt][0pt]{\includegraphics[height=#2]{#3}}}

\newcommand{\eg}{e.\,g.}
\newcommand{\ie}{i.\,e.}

\newcommand{\hty}[1]{\textcolor{SeaGreen}{#1}} % Tingying He
\newcommand{\ti}[1]{\textcolor{RoyalBlue}{#1}}

\renewcommand{\hty}[1]{\textcolor{black}{#1}} % Tingying He
\renewcommand{\ti}[1]{\textcolor{black}{#1}}

\newcommand{\designexperts}{13\xspace}
\newcommand{\patternnumber}{273\xspace}
\newcommand{\nonexperts}{12\xspace}

\usepackage{flushend}

\begin{document}

%%%%%%%%%%%%%%%%%%%%%%%%%%%%%%%%%%%%%%%%%%%%%%%%%%%%%%%%%%%%%%%%
%%%%%%%%%%%%%%%%%%%%%% START OF THE PAPER %%%%%%%%%%%%%%%%%%%%%%
%%%%%%%%%%%%%%%%%%%%%%%%%%%%%%%%%%%%%%%%%%%%%%%%%%%%%%%%%%%%%%%%

% \firstsection{Introduction}

\maketitle

%% \section{Introduction} %for journal use above \firstsection{..} instead

\section{Introduction}
\label{sec:intro}

The concept of ``\sr'' design refers to choices when creating visual data mappings that evoke specific concepts or associations in a viewer, as originally defined by Lin et al. \cite{lin:2013:selecting} in the context of color. Their study showed that \sr colors (\eg, representing bananas in yellow rather than blue) enhance chart reading speed, especially when strong con\-cept-co\-lor associations exist. Follow-up work \cite{schloss:2018:color} also showed that color-concept associations have the potential to help viewers interpret visualizations more intuitively than with non-resonant mappings. In contexts where color is unavailable or insufficient, black-and-white patterns offer an important alternative visual variable option \cite{He:2024:Encoding}. These patterns have shown benefits for visualization, from historical applications to modern-life practices. Leveraging pattern-concept associations may similarly improve chart reading and interpretation as color-concept associations.

Prior research by He et al. \cite{He:2024:DCB} explored the influence of using patterns in visualization by asking designers to create black-and-white patterns for visualizations, focusing on two types: abstract geometric patterns (using simple lines or dots) \inlinevis{-1pt}{1em}{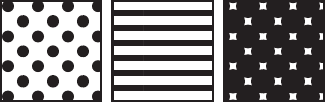} and iconic patterns (such as a banana icon for bananas) \inlinevis{-1pt}{1em}{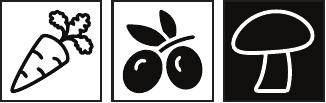}. Although iconic patterns are inherently \sr, they did not improve chart reading speed as expected \cite{He:2024:DCB}. In contrast, abstract geometric patterns showed promise, particularly in enhancing reading speed in pie charts. An intriguing observation in He et al.'s study \cite{He:2024:DCB} showed participants attempted to embed semantic meaning into patterns while they created abstract geometric patterns. One designer mainly used the strategy of creating associations between patterns and the concepts they represent (\autoref{fig:BG1} in \autoref{appendix:semantic-association-in-previous-study} shows such design, and we discuss related strategies in \autoref{appendix:semantic-association-in-previous-study}), \eg, used small and large dots to represent olives and tomatoes, respectively, reflecting their relative physical sizes. Such design received the highest aesthetic pleasure rating among 14 expert-designed bar charts with abstract patterns. 
\hty{In their past work, He et al. \cite{He:2024:DCB} focused broadly on pattern design---without discussing in-depth how to systematically embed semantic meaning into abstract geometric patterns. Their findings, however, suggest that such abstract patterns have the potential to carry meaningful associations that enhance both the readability and aesthetics of the visuals that use the patterns. Building on this insight, our work specifically investigates how to effectively embed semantic meaning into abstract geometric patterns.}

We first conducted a series of design workshops with \designexperts design experts to design \sr abstract patterns for three concept sets that range from the concrete to the abstract: vegetables, music genres, and emotions. This workshop yielded \patternnumber pattern designs and associated design strategies. Based on our qualitative coding of the feedback we received from the design experts, we developed a structured \hty{design methodology for how to design} \sr abstract patterns. Our \hty{design methodology categorizes the approaches into} two key steps: first, to identify the \hty{underlying} content to be visualized and, second, to encode the content as suitable patterns. Based on the codes we analyzed, we 
identified \hty{possible approaches} that are applicable at each stage of this process. Subsequently, we conducted another series of workshops with \nonexperts non-expert participants to assess whether our \hty{design methodology can} effectively guide them in creating \sr patterns. This second group of workshops demonstrated the effectiveness of our \hty{design methodology}, as most participants consider it useful, easy to use, and helpful for both abstract and concrete concepts. The results of the evaluation workshops also show that our \hty{design methodology} comprehensively covers potential thinking directions for designing \sr patterns. 

\hty{In summary, our contributions include (1) a structured design methodology, which summarizes approaches for designing abstract \sr patterns that work for both abstract and concrete concepts, derived from ex\-pert-led design workshops, along with a curated set of pattern designs. We also contribute (2) an evaluation of our design methodology with non-ex\-perts, which demonstrates its effectiveness in supporting \sr pattern design without requiring specialized design knowledge. Overall, our methodology advances the understanding of the visual variable \emph{pattern}, expands the toolkit available to visualization designers, and lowers barriers for non-experts to engage in visualization design.}
% \marginpar{\tiny\todo{if we can somehow shorten (i.e., cut by rephrasing) this last line here, then section 2 would start on the first page}}}

\section{Background and related work}
\label{sec:related-work}
We first discuss the difference between abstract and concrete concepts, followed by associations between visual variables and concepts.

\subsection{Concrete and abstract concepts}
According to research in psychology, concrete concepts are tangible and directly experienced through sensory perception (e.g., ``cat'' or ``chair''), while abstract concepts are intangible and not directly perceived with our sensors (e.g., ``love'' or ``freedom'') \cite{Paivio:1968:Concreteness, Wiemer-Hastings:2005:Content, Connell:2012:Strength, brysbaert:2014:concreteness, nedjadrasul:2017:abstract}.

People understand these two types of concepts in different ways. According to embodied cognition theory \cite{shapiro:2014:RoutledgeEmbodiedCognition}, people comprehend concepts through perceptual experiences, which can include sensorimotor (\eg, vision or hearing) and emotional experiences. Concrete concepts are often strongly associated with sensorimotor experiences, while abstract concepts are understood more through emotional and metaphorical connections \cite{Adams:2010:Embodied, Vigliocco:2009:Toward, Skipper:2014:semantic, Wang:2024:Unlocking}. Based on embodied cognition, conceptual metaphor theory \cite{lakoff:1980:metaphors} also provides a framework for our meaning-making process and a theoretical foundation for understanding the visualization process \cite{Parsons:2018:Conceptual, Preim:2024:Survey, Pokojná:2025:Language}. It states that we understand the concepts in one concept domain by associating them with another concept in another concept domain. Therefore, to understand an abstract concept, we can link this abstract concept to a more concrete one \cite{sloutsky:2019:categories, gentner:2019:metaphoric, dijkstra:2014:embodied}.

%Conceptual metaphor theory is rooted in the idea of embodied cognition, where human thought is grounded in sensory and motor experiences. 

To visually represent these two types of concepts, there might also be differences. 
% Visual semiotics provides a foundational theory for visualizations \cite{MacEachren:2012:Visual}. 
In semiotics, Peirce categorized signs into three types based on their directness level to the concept they convey: icon, which physically resembles the concept; index, which has some direct connection to the concept; and symbol, which relies on the learned association. Based on semiotics theory, visual semiotics provides a framework for understanding how visual representations convey information, and it focuses on articulating which visual variables we can vary to visually convey information \cite{MacEachren:2012:Visual}---which is also foundation of visual mapping. For concrete concepts, visual representations can directly resemble their physical appearance, whereas abstract concepts require alternative strategies for visual representation. 
Ultimately, this difference between abstract and concrete concepts motivates us to explore \sr patterns for both types of concepts \hty{for generality}.\footnote{Please notice that, while we cover \textbf{concepts \emph{that range from abstract to concrete}}, we only investigate their encoding using \textbf{\emph{abstract} patterns}.}
% \marginpar{\tiny\todo{I added this footnote to at least once point the reader onto the fact that the concepts range from abstract to concrete, but the patterns are only abstract. Ok? Or remove the footnote?}}

\subsection{Linking visual variables to concepts in visualization}
Data often have semantic aspects---typically represented by concepts such as category names in categorical visualizations. Visualization designers manipulate the visual features (\eg, color hue, size, shape, etc.) of graphical elements in charts to encode data, including these concepts. We commonly \hty{refer to} these visual features as visual variables or visual channels \cite{Munzner:2015:VisualizationAnalysisAndDesign}. For viewers to decode data from these visual representations, they must understand the associations between the visual variables and the concepts they represent.

The visual features used as visual variables may have either intuitive or non-intuitive links to the concepts. In their studies of co\-lor-con\-cept association, Schloss et al. \cite{schloss:2020:semantic} define \emph{``the degree to which observers can infer a unique mapping between visual features and concepts, based on the visual features and concepts alone (\ie, without legends or labels)''} as semantic discriminability, and further quantify it with a metric called semantic distance, which depends on \emph{``the relative association strengths between each color and each concept in the context of an encoding system.''} Based on Schloss et al.'s \cite{schloss:2020:semantic} framework, we define a \textbf{\sr visual variable} as one that exhibits high semantic discriminability and that ideally can be easily interpreted (or at least its meaning easily recalled) without a legend or labels.
% \marginpar{\tiny\todo{TI: I added the part in brackets, ok? I think that it is not only the direct interpretation w/o a legend, but also the easier recall without the legend that leads to the faster performance; maybe this could be a part to be added to our discussion at the end of this paper?}}

Among visual variables, color is the most extensively studied w.r.t.\ its associations with concepts. 
It has been suggested \cite{Schoenlein:2022:Color} that the co\-lor-con\-cept association is learned from past experience and also influenced by both environmental and cognitive factors. For concrete concepts such as vegetables, the color associated to the concepts can be the concepts' physical look. For abstract concepts (\eg, plastic, paper, metal), people still tend to select similar colors for them \cite{schloss:2020:semantic, schloss:2018:color}.
Important for visualization, researchers found that the color-concept association can affect chart reading and interpretation \cite{lin:2013:selecting}.

For visual variables beyond color, associations between the variable of shape and concepts have also been explored, primarily in studies of map symbols and icon-based visual representations. For example, MacEachren \cite{MacEachren:2004:HowMapsWork} describes that we can convey meaning with map symbols from concrete to abstract by gradually reducing the details of a concrete shape, extracting the most common features at each step until the shape is abstracted into a simple geometric one.
Pictographs are typical visual data representations using icon-based language, which have many benefits \cite{zhang:2020:dataquilt}, such as ISOTYPE visualization\cite{neurath:2010:hieroglyphics}. Haroz et al. \cite{haroz:2015:isotype} found that they improve viewers' working memory and engagement to visualizations. Burns et al. \cite{burns:2021:designing} also found that pictographs help people envision the content of visualization. 

Orientation-concept association is less studied but also plays a role, especially according to conceptual metaphor theory \cite{lakoff:1980:metaphors}. For example, Parsons \cite{Parsons:2018:Conceptual} discussed that the conceptual metaphor plays an important role in visualization and pointed out that there are schemas we can use to make complex data more intuitive to understand. Parsons mentioned \emph{``more is up''} as an example common in everyday life. Pokojná et al. \cite{Pokojná:2025:Language} discussed scientific infographics (\ie, visualizations of scientific concept [usually] targeted at the general public) in the context of conceptual metaphor theory and identified four conceptual metaphor types. Among these four types, one is ``orientational,'' \ie, it makes a link between two concepts based on spatial orientation and/or composition. Wang et al. \cite{Wang:2024:Unlocking} empirically studied color, shape, and orientation individually in bar charts and found that for all these visual variables, conceptually relevant visual features improve chart reading speed.

In contrast to some of those examples, in our work, we focus on black-and-white abstract patterns \hty{because the creation of semantic associations using them has not been studied so far, to the best of our knowledge.} We exclude color and icons and also do not study any visual variables individually. Patterns form a composite visual variable that have many attributes we can vary \cite{He:2024:Encoding}, including aspects such as shape and orientation. We aim to explore how designers can combine these parameters to embed semantic meaning within abstract patterns.

\section{Design workshop}
\label{sec:workshop}

As we were not aware of past work on methods for creating \sr patterns, we organized a design workshop with visualization experts to understand their approaches to generating these patterns. We pre-registered our work on OSF (\href{https://osf.io/h62yb}{\texttt{osf\discretionary{}{.}{.}io\discretionary{/}{}{/}h62yb}}) and received IRB approval (Inria COERLE, \textnumero\,2023-01).

\begin{figure*}
    \centering
    \includegraphics[width=\textwidth]{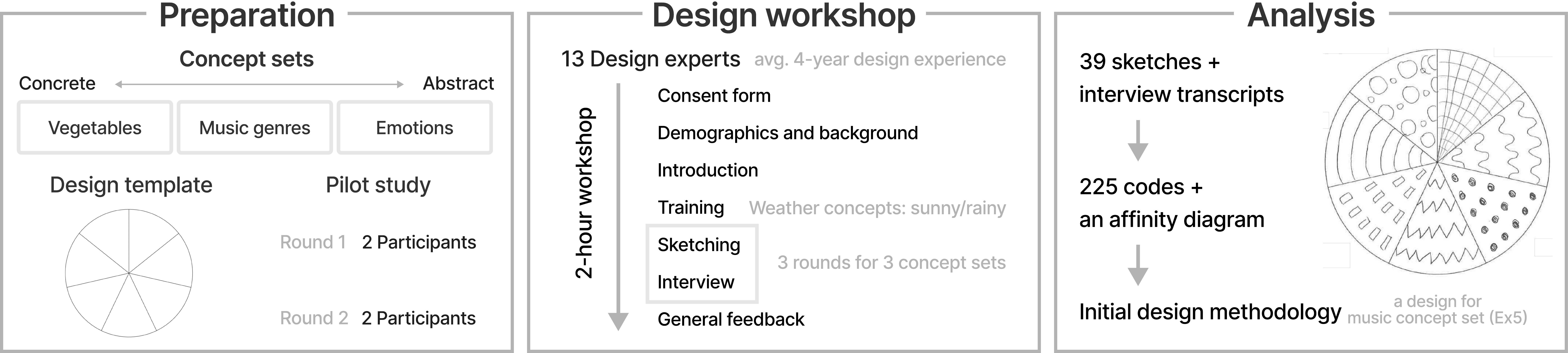}
    \caption{Procedure of the design workshop with each design expert, as described in detail in \autoref{sec:workshop}. Left: Preparation before the workshop, including the three concept sets, a design template with a blank pie chart, and two rounds of pilot studies. Middle: Steps of the formal workshop conducted individually with 13 design experts. Right: Data collected and the corresponding analysis results.}
    \label{fig:workshop-procedure}
\end{figure*}
% To collect semantically-resonant patterns, we organized a design workshop. Before starting the formal experiment, we conducted two preparatory steps, including selecting concept sets and conducting pilot studies. Following these preparations, we conducted the formal experiment and analyzed data.

\subsection{Preparation of the concept sets}
\label{sec:preparation-concept-sets}
We first prepared the concept sets for which we wanted the design experts to create \sr patterns. We followed Lin et al.'s \cite{lin:2013:selecting} approach, which investigated \sr colors, and selected \hty{concept sets. Each concept set has a specific theme and contains seven concepts per set.}
We planned to cover topics that range from concrete to abstract ones. Yet to keep the workshop duration to a reasonable time, we decided to limit our investigation to three concept sets: one concrete, one intermediate, and one abstract. For this purpose, we relied on Brysbaert et al.'s \cite{brysbaert:2014:concreteness} concreteness scale, in which the authors provide concreteness scores in a range from 1 to 5 for approximately 40,000 English word lemmas. In this scale, 1 represents highly abstract concepts, while 5 represents highly concrete ones. 

For our most concrete concept set, we directly adopted Lin et al.'s \cite{lin:2013:selecting} \emph{vegetable} set due to its exceptionally high mean concreteness rating of 4.92. It consists of seven commonly recognized vegetables: corn, carrot, eggplant, mushroom, olive, celery, and tomato, all of which are listed in Brysbaert et al.'s scale \cite{brysbaert:2014:concreteness}. We wanted design experts to focus on developing and expressing their design ideas without being burdened by the meaning of the concepts. Therefore, for the other two concept sets, we selected topics and concepts familiar to many people's daily lives and experiences to minimize misunderstanding or confusion. To ensure broad recognition and reliable evaluation of concreteness, we limited our selection to those for which all selected concepts were listed in Brysbaert et al.'s table. In addition, we used Google Image Search to verify their practical applicability in visualizations by ensuring they appeared together in at least one categorical chart \hty{(for examples see \autoref{sec:real-world-references-concept-selection})}. 
Ultimately, we settled on \emph{music genres} (including pop music, country music, blues, rock music, dance music, hip hop, and folk music) and \emph{emotions} (including happy, loving, angry, afraid, bored,surprised, and disgust) as listed in \autoref{tab:concept-sets-design-workshop-scores}.

% The three concept sets thus have average concreteness scores of 4.92, 3.56, and 2.46, respectively. Although the \emph{emotions} set, with a score of 2.46, is not the most abstract one we identified in our process, we chose it over possible alternatives (\eg, \emph{quality metrics} with an average score 2.07, including the terms efficiency, usability, integrity, flexibility, reliability, correctness, and portability) due to its higher expected familiarity to participants and thus to avoid excessive difficulty in designing \sr patterns.

The three concept sets have mean concreteness scores of 4.92, 3.56, and 2.46, respectively. Although the \emph{emotions} set, with a score of 2.46, is not the most abstract concept set we identified, we selected it for its greater familiarity and manageable complexity in designing patterns. Potential alternatives, such as the \emph{quality metrics} set (including terms like efficiency, usability, integrity, flexibility, reliability, correctness, and portability), might be more abstract, with a lower mean score of 2.07. We concluded, however, that the \emph{quality metrics} set's high level of abstraction might be too abstract and too difficult for designing \sr patterns. By choosing the \emph{emotions} set, we aimed to provide participants with a less daunting design process, to ensure they can deliver meaningful designs.

\begin{table}
\centering
\caption{Concreteness scores according to Brysbaert et al.'s \cite{brysbaert:2014:concreteness} concreteness scale for vegetables, music genres, and emotions, and the means for each concept group. The scale ranges from 1 to 5, where 1 represents the most abstract and 5 represents the most concrete.}
\resizebox{\columnwidth}{!}{%
\begin{tabu}{lclclc}
\toprule
\textbf{\emph{vegetable}} & \textbf{score} & \textbf{\emph{music genre}} & \textbf{score} & \textbf{\emph{emotion}} & \textbf{score} \\
\midrule
corn      & 4.96 & pop music     & 3.89 & happy     & 2.56 \\ 
carrot    & 5.00 & country music & 3.86 & loving    & 1.73 \\ 
eggplant  & 4.97 & blues         & 2.31 & angry     & 2.53 \\ 
mushroom  & 4.83 & rock music    & 4.00 & afraid    & 2.70 \\ 
olive     & 4.90 & dance music   & 3.88 & bored     & 2.13 \\ 
celery    & 4.80 & hip hop       & 3.33 & surprised & 2.50 \\ 
tomato    & 5.00 & folk music    & 3.68 & disgust   & 3.07 \\
\midrule
\textbf{mean}      & \textbf{4.92} & \textbf{mean}       & \textbf{3.56} & \textbf{mean}   & \textbf{2.46} \\
\bottomrule
\end{tabu}%
}
\label{tab:concept-sets-design-workshop-scores}
\end{table}

\subsection{Design template: Chart type}

We provided participants with prepared design templates. Each template included an identical-sized blank pie chart representing the seven categories corresponding to the seven concepts in each concept set. Designing patterns within a chart, rather than on blank paper, helped participants engage with the context of data representation. We chose pie charts based on He et al.'s study \cite{He:2024:DCB}, which showed that abstract geometric patterns in pie charts improved people's chart reading speed compared to non-pattern or iconic pattern fills. In addition, \hty{the pie chart is a widely used visualization type} and also provides ample space for participants to showcase their pattern designs.

Since our study focused on designing patterns to represent concepts (the category names), we did not prioritize specific values for each pie slice. To simplify the task, we divided the pie chart into seven equal segments. Although filling space in real-world applications can affect pattern design---for example, when pie slices become too thin to accommodate recognizable patterns---we aimed to start from the simplest case and minimize barriers as we conduct the first study in \sr pattern design.

% \marginpar{\tiny\ti{[TI: just as a remark: a reviewer may argue that this is all fine and good, what happens when the pies in a chart become too thin to be able to recognize the pattern, should you not have accounted for such a case?]}}%\marginpar{\tiny\hjy{I do not quite understand this sentence? ``pie piece'' sounds a little bit weird in this sentence.}}
% \marginpar{\tiny\ti{[TI: Say something about the actually chosen values for the pie chart? Or does this come later? If later, why not also discuss the pie charts later?]}}

\subsection{Pilot study}
We conducted two rounds of pilot studies, each with two participants, to evaluate the workshop procedure and estimate its duration. Based on the feedback from these pilots, we made several improvements: (1) we introduced a training session to better prepare participants; (2) we removed the introduction to the design methods that we had initially used as it unintentionally constrained participants' \hty{design considerations to only} the provided methods; and (3) we simplified the design process by eliminating an initial sketching of the pattern without context. Separating this process from the transfer of the design to the charts proved to be unnecessary and time-con\-su\-ming, so in the formal study we asked participants to directly design on the chart outlines we provided, with the option to draft in blank areas of the template if they so preferred.

\subsection{Participants}
We invited 
% \ti{unpaid}
% \marginpar{\tiny\ti{[I added this, correct?]}\hty{Not exactly ``unpaid.'' Zihan offered each of them a Marabou chocolate bar valued around 6 euros. I added at the end of this paragraph}} 
13 \hty{design experts} to participate in our workshop via direct e-mail. The group comprised 8 females and 5 males; 4 aged 18--24, 7 aged 25--34, 1 aged 35--44, and 1 aged 55--64. Participants' highest degrees were: 6 Bachelor's, 4 Master's, and 3 doctoral. Their prior experience in visualization or design averaged 4 years. All participants demonstrated at least foundational familiarity with the three concept sets, but most had limited knowledge of the concept of ``semantic association.'' In accordance with payment policy of the first authors' institution at the time (Inria), we were not permitted to compensate non-crowdsourced participants with cash. Instead, we offered each of them a chocolate bar to show our appreciation.
% (Never encountered: 6; Heard but Don't Know the Meaning: 4; Know the Meaning but Haven't Applied: 2; Used in Research/Design: 1). 

% \subsection{Workshop set up}

\subsection{Procedure}
\label{sec:workshop1-procedure}
We conducted 11 experiments in the lab in person and 2 experiments via a remote connection. For both on-site and remote experiments, the experimental procedure was the same. 
We provided on-site participants with pens and design templates printed on A4 paper. The two remote participants were unable to print the design template. Therefore, we allowed one participant (Ex5) to use a digital version of the template on an iPad and the other (Ex6) to hand-draw the blank chart on A4 paper. We instructed both participants to ensure that the charts matched the dimensions of the printed template.

At the start of the workshop we asked participants to complete a consent form and a questionnaire\footnote{We share all study materials in our OSF repository (\href{https://osf.io/9h5nd}{\texttt{osf\discretionary{}{.}{.}io\discretionary{/}{}{/}9h5nd}}).} to collect demographic and background information. Next, we explained the concept of se\-man\-ti\-cal\-ly-re\-so\-nant patterns and asked participants to design patterns that allow viewers to match patterns with corresponding data without the need for a legend. In addition to ensuring a strong semantic association, we encouraged them to prioritize readability and aesthetics, as these two criteria are key objectives in pattern design, according to design experts in previous work \cite{He:2024:DCB}. 
To inspire participants, we provided examples that had been designed by one of the authors for a ball game concept set.
%\marginpar{\tiny\todo{Please include this example set in the study materials on OSF.} \hty{It is in the slides page 8 https://osf.io/j5nxu}} 
\hty{We also introduced the visual variables that can be manipulated in a pattern \cite{He:2024:Encoding} to the participants, for their reference.} We emphasized that the variation in a shape should be limited to abstract geometric forms, with no iconic shapes. Next, we gave participants 5 minutes to practice---asking them to design two se\-man\-ti\-cal\-ly-re\-so\-nant patterns, one for ``sunny'' and another for ``rainy,'' to familiarize themselves with the design objectives and process. We clarified that the results of this training session would not be included in the data analysis.

Following the training, participants moved on to the main task: sketching se\-man\-ti\-cal\-ly-re\-so\-nant patterns for the three concept sets, with one concept set per session. Each session began with a sketching phase, followed by an interview and a break. We fully coun\-ter-ba\-lanced the order of the three concept sets across participants in 3\,\texttimes\,2\,\texttimes\,1 $=$ 6 sequences and used each sequence for 2 participants for the first 12 participants; For the last participant (Ex13), one of the six sequences was randomly assigned. For each session, we provided participants with a prepared design template, which included a blank pie chart divided into seven equal slices. We asked participants to sketch their patterns for each concept on one of the slices. During the interviews, we inquired about the design strategies they used and their thoughts on their designs. After completing all three sessions, we gathered general feedback from the participants. The entire workshop lasted approximately two hours. We recorded all interviews using an iPhone's built-in recorder application. By default, we used English as the language of communication during the workshop. However, in cases that both the experimenter and the participant shared the same non-English native language, we allowed them to use their native language (\ie, Mandarin Chinese; 8\texttimes) to ensure they can fully express their thoughts.

\subsection{Data analysis} 
\label{sec:workshop1-data-analysis}
From the workshops, we collected 39 pie chart designs featuring 273 semantically resonant patterns created by our participants (we show all of them in \hyperref[fig:ex001-ve]{Figures~}\ref{fig:ex001-ve}--\ref{fig:ex013-emo} in \autoref{appendix:all-designs-workshop1}), along with the interview recordings in which participants discussed their design strategies. We scanned all sketches and transcribed the audio using TurboScribe, followed by manual corrections to ensure transcription accuracy. For the eight Mandarin Chinese recordings, we used ChatGPT 4.0 \cite{chatgpt4.0} to translate them into English, also followed by a manual proof-reading and correction pass to ensure a correct translation. \hty{We provide all the original transcriptions and corrected transcriptions in our OSF repository \href{https://osf.io/9h5nd/}{\texttt{osf\discretionary{}{.}{.}io\discretionary{/}{}{/}9h5nd\discretionary{/}{}{/}}}.}

Using this English corpus, \hty{two authors qualitatively coded the experts' design strategies using Atlas.ti 8.4.4 \cite{ATLAS.ti}}. Both coders referred to the corresponding sketches during the coding process. In an iterative process of discussion and revision, we finalized 225 codes and developed an affinity diagram (\autoref{fig:design-space-v1} in \autoref{appendix:iteration-design-space}) that outlines an initial \hty{design methodology}, which, in turn, captures a rich set of strategies for creating \sr patterns as we discuss next.

% (\href{https://atlasti.com}{\texttt{atlas\discretionary{}{.}{.}com}}) 

% From the design workshop, we collected 39 pie charts filled with patterns, discovering 273 semantic associations between concepts and patterns. After , we transcribed the audio using TurboScribe, followed by manual corrections to ensure accuracy. Five of the recordings were originally in Chinese (as both the interviewer and interviewee were native Chinese speakers). Once we obtained the initial transcriptions, we used ChatGPT 4.0 to translate them into English. After transcribing the interviews, we coded the design strategies mentioned by participants for each pattern, as well as their comments on the design requirements, using Atlas.ti 8.4.4. This process generated 301 initial codings. Then, two researchers summarized and integrated these initial codings based on the steps participants followed in designing the patterns and classified them according to different design strategies. Through discussions between the two coders, the coding scheme was continuously revised and refined, simplifying the codings to 225 final entries. This process resulted in the final version of the affinity diagram, which was then analyzed (see \autoref{sec:affinity} in Appendix).

\section{\hty{Design methodology}}
\label{sec:design-space}
% (\todo{XX \texttimes})
% (P\textsubscript{ex}10)

\begin{figure}
    \centering
    \includegraphics[width=\columnwidth]{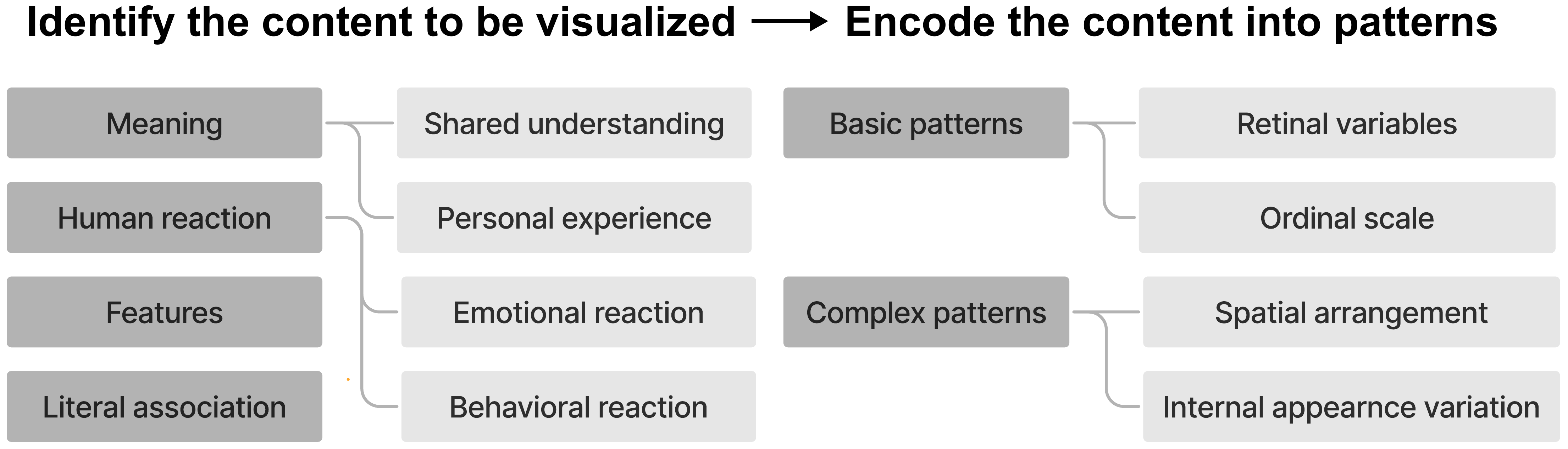}
    \caption{Overview of our \hty{design methodology}. The \hty{design methodology categorize the methodology into} two steps, with each step categorizing potential approaches for designing \sr patterns.}
    \label{fig:design-space}
\end{figure}

In our \hty{design methodology} we now systematically structure the potential \hty{approaches} for creating \sr patterns. Starting with our initial coding of design strategies used by design experts in the workshop, we iteratively refined the structure of this \hty{design methodology} over two iterations: (1) after completing the data analysis of the design workshop and during the evaluation study design and (2) after analyzing data from the evaluation study. Below, we present the \emph{\textbf{final version}} of our two-step \hty{design methodology} (\autoref{fig:design-space}), \hty{and we present the evolution process of our design methodology in \autoref{appendix:iteration-design-space}.}
% while we show the initial iterations of the structure including the coding that led to our final \hty{design methodology} in \autoref{appendix:iteration-design-space}.
% \marginpar{\tiny\ti{[I did not see \autoref{fig:design-space} to be cited here, but it should, so I added it. Please check that this is ok like this. In addition, I have the feeling that the headers of the subsubsections in 4.1 do not entirely match the terminology used in \autoref{fig:design-space}, so please make sure that they match.]}}

\subsection{Step 1: Identifying the content to be visualized}

In our observations we saw that virtually all visualization experts began to create \sr patterns for a target concept by first identifying the relevant elements or associations for encoding in a pattern. Such elements or associations are typically a concept or idea linked to the target concept and, are usually more concrete than the original target concept, especially in the case of abstract concepts. We categorize the identification processes according to three different bases and describe \hty{the corresponding approaches under each category}.

\subsubsection{Based on the target concept's meaning}
% \marginpar{\tiny
% \ti{[I feel that here and also in the subsections below you make claims for psychology and perception were a reference for the claims would be good, even if this is well-established knowledge in that field.]} \hty{I don't understand what ``you make claims for psychology and perception were a reference for the claims'' means. Could you please explain more or give me an example?}} 
Using this \hty{approach}, a designer creates particular associations by linking the target concept with a concrete concept based on a human's general understanding, which aligns well with metaphor theory \cite{lakoff:1980:metaphors}. Designers can achieve this association through \textbf{shared understanding}---according to universal or culturally shared knowledge. For example, when talking about \emph{love}, humans typically associate it with a \emph{heart}. Similarly, within specific cultural contexts, \emph{dance music} can be linked to \emph{flamingos}. 
Such an association can also be derived from \textbf{personal experience}---an individual's unique experiences, memories, or imagination. These associations are highly subjective and vary between design experts. For example, in our design workshop with experts (\autoref{sec:workshop}), one design expert associated \emph{love} to \emph{cursive script on decorative paper}.

\subsubsection{Based on human reaction to the target concept}
In this \hty{approach}, a designer links the target concept to human reactions, which can be emotional or behavioral. As embodied cognition theory suggests, our cognition is rooted in bodily experiences and our interaction with the world \cite{Adams:2010:Embodied, Vigliocco:2009:Toward, Skipper:2014:semantic, Wang:2024:Unlocking}. \textbf{Emotional reaction} refers to those emotional impressions or personal feelings that are evoked by the concept. For example, 3 of our \designexperts design experts thought \emph{bored} is \emph{dull}, and 2 said \emph{loving} feels \emph{changeable}.
\textbf{Behavioral reaction} refers to psychological or physiological responses to the target concept. Instances include that 9 designers link \emph{happy} with a \emph{smily face}---a common body reaction when being happy.

\subsubsection{Based on the target concept's features}
When using this \hty{approach}, a designer isolates specific inherent or intrinsic attributes of the target concept that can directly be mapped to a visual variable of a visualization (\eg, music's rhythm, vegetables' shape). For example, 11 of the \designexperts design experts associated vegetables with their external appearance, such as using the typical \emph{semicircular shape} of a mushroom cap to represent the \emph{mushroom} concept.

\subsubsection{Based on the target concept's literal meaning}
In a last possible \hty{approach}, designers can create the association based on a direct semantic mapping and link the target concept to the literal meaning of the target concept's name. This association can happen when the name of the target concept refers to a more concrete and familiar concept. For instance, in our workshop, one design expert (Ex9) associated \emph{rock music} with \emph{rocks}.

\subsection{Step 2: Encoding the content into patterns}

After the visualization experts had identified the relevant elements or associations, they went on to encode them into an abstract pattern representation. This second step uses a number of visual variables available to patterns to represent the semantics of the refined concept. The possible visual variables arise from a view of patterns as a composite visual variables of a group of pattern primitives \cite{He:2024:Encoding} that can be characterized by three sets of attributes: (1) spatial arrangement relationships among pattern primitives, (2) appearance relationships among pattern primitives, and (3) individual appearance of pattern primitives.

\begin{figure}
    \centering
    \includegraphics[width=\columnwidth]{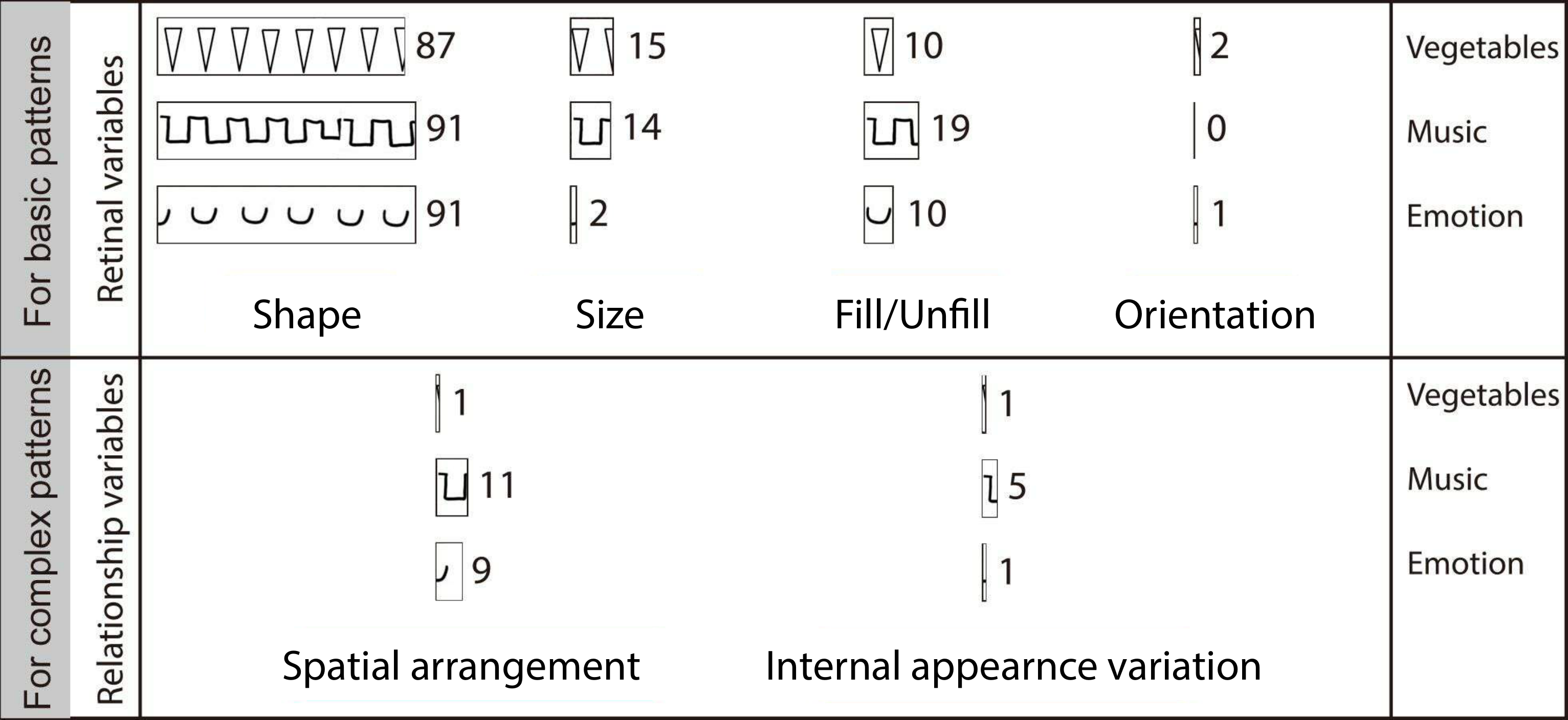}
    \caption{Frequencies of use of each visual variable in our design workshop across the three concept sets. Each use of a visual variable for a concept was counted once. We use part of Ex5's patterns for the three concept sets \hty{to fill the bars in the corresponding rows}; we show the whole designs in \hyperref[fig:ex005-ve]{Figures~}\ref{fig:ex005-ve}, \ref{fig:ex005-mus}, and \ref{fig:ex005-emo} in \autoref{appendix:all-designs-workshop1}.}
    \label{fig:visual-variables-count}
\end{figure}

In our workshop, the design experts used visual variables across all three dimensions, and we counted how many times they were used (which we report in \autoref{fig:visual-variables-count}). We categorize the patterns that were generated into two types that we describe next: basic patterns, which vary only appearance attributes (\ie, (3) in He's \cite{He:2024:Encoding} framework) repetitively, and complex patterns, which vary spatial or appearance relationships (\ie, (1) and (2) in He's framework).

\subsubsection{Basic patterns}
% \begin{figure}[t]
%     \centering
% 		\includegraphics[width=0.7\linewidth,trim={12 19 23 19},clip]{figures//Vegetable_dataset/ex005-ve.pdf}
%     \caption{An example of basic pattern design, from Ex5 for vegetable concept set, collected in our design workshop.}
%     \label{fig:ex005-ve-example-basic}
% \end{figure}

For basic patterns, we can vary the following \textbf{retinal variables} of pattern primitives repetitively: shape, size, orientation, or fill/unfill (see \autoref{fig:expert-design-examples}(left) for an example). Design experts most commonly varied the shape of pattern primitives to distinguish concepts 
% (87\texttimes, 91\texttimes, and 91\texttimes\ for \emph{vegetables}, \emph{music genres}, and \emph{emotions}, respectively), 
\hty{(87\texttimes\ for \emph{vegetables}, 91\texttimes\ for \emph{music genres}, and 91\texttimes\ for \emph{emotions},}
followed by size (15\texttimes, 14\texttimes, 2\texttimes) and fill/unfill (10\texttimes, 19\texttimes, 10\texttimes). Orientation was the least used visual variable (2\texttimes, 0\texttimes, 1\texttimes).

A specific \hty{approach} is \textbf{building an ordinal scale}: using an ordinal visual variable to visualize a common feature shared by all concepts within the concept set. 
For example, 1 design expert extracted the negativity of emotions in the emotion concept set and referenced the Bouba/Kiki effect \cite{cwiek:2022:bouba} to visualize it. They stated that ``the number of spikes in geometric shapes conveys more negative emotions.'' In their design, the most positive emotion, \emph{love}, had no spikes, while the most negative emotion, \emph{anger}, had eight spikes (see \autoref{fig:ex013-emo} in \autoref{appendix:all-designs-workshop1}).

\subsubsection{Complex patterns}
\begin{figure}[t]
    \centering
    \begin{subfigure}[t]{0.49\linewidth}
        \centering
        \includegraphics[width=\linewidth,trim={12 19 23 19},clip]{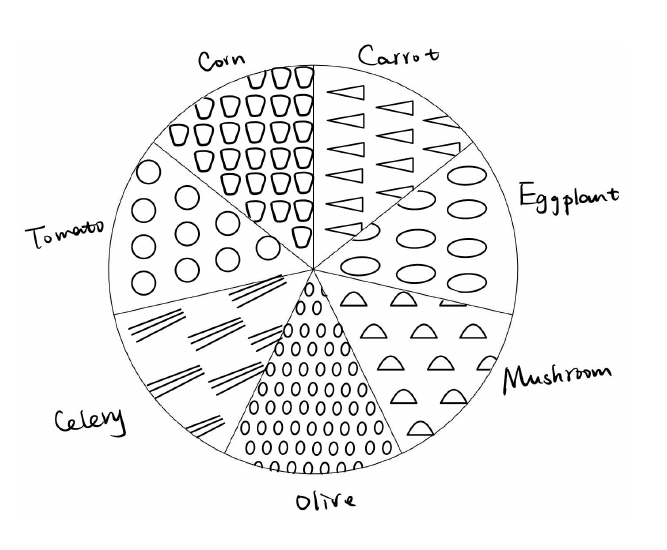}
        % \caption{An example of basic pattern design, from Ex5 for the vegetable concept set, collected in our design workshop.}
        % \label{fig:ex005-ve-example-basic}
    \end{subfigure}
    \hfill
    \begin{subfigure}[t]{0.49\linewidth}
        \centering
        \includegraphics[width=\linewidth,trim={4 11 12 22},clip]{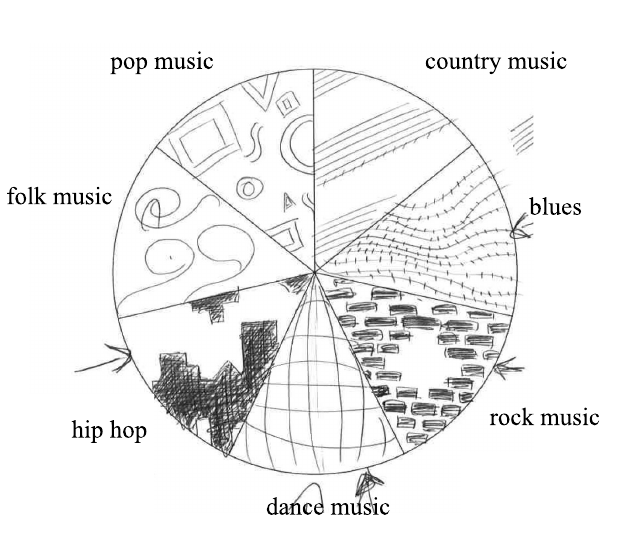}
        % \caption{An example of complex pattern design, from Ex7 for the music concept set, collected in our design workshop.}
        % \label{fig:ex007-mus-example-complex}
    \end{subfigure}
    \caption{Examples of pattern design from our design workshop: (left) Basic patterns for the vegetable concept set by Ex5, and (right) Complex patterns for the music concept set by Ex7. We show larger versions of these figures in \autoref{fig:ex005-ve} and \autoref{fig:ex007-mus}, respectively, in \autoref{appendix:all-designs-workshop1}.}
    \label{fig:expert-design-examples}
\end{figure}

Complex patterns are those patterns that involve varying spatial arrangements and appearance relationships among pattern primitives. Design experts, for instance, occasionally used randomness in \textbf{spatial arrangements} to convey uncertainty (1\texttimes, 11\texttimes, 9\texttimes).
Also, some design experts modified \textbf{appearance relationships} to embed richer semantics, resulting in more complicated and visually diverse patterns (1\texttimes, 5\texttimes, 1\texttimes). 
\autoref{fig:expert-design-examples}(right) illustrates an example of complex patterns. In this design, the pattern primitives are less repetitive, which presents challenges for computational generation compared to basic patterns (\eg, \autoref{fig:expert-design-examples}(left)). Despite their abstract nature, however, these complex patterns are typically more figurative than basic patterns, offering potential for encoding nuanced semantics (\eg, in the blues music slice, the participant drew a farm fence to represent the historical background of blues music).

%Although some of these complex patterns present challenges for computational generation, they offer potential for encoding nuanced semantics.
% \marginpar{\tiny\ti{[in what way? elaborate! at least give an example of what you mean]}}

% \begin{figure}[t]
%     \centering
%     \includegraphics[width=0.7\linewidth,trim={4 11 12 22},clip]{figures//Music_dataset/ex007-mus.pdf}
%     \caption{An example of complex pattern design, from Ex7 for the music concept set, collected in our design workshop.}
%     \label{fig:ex007-mus-example-complex}
% \end{figure}

\subsection{Summary}

While seemingly relatively simple, our observation-driven \hty{design methodology} offers a systematic framework that will allow future designers of \sr patterns for the visualization of conceptual data to conceptualize the needed process. The two-step process, which we derived from both the initial design expert workshop and the evaluation on which we report next, provides a structured approach and the specific cognitive \hty{approach}es that we have outlined can serve as inspiration for the design process.

\section{\hty{Design methodology} evaluation}
\label{sec:evaluation}

To evaluate the effectiveness and practical utility of our proposed \hty{design methodology}, we followed Shi et al.'s \cite{Shi:2021:Communicating} methodology and conducted another workshop with people who had limited to no design knowledge and skills. 
We aimed to evaluate whether the \hty{design methodology} could be extended to broader concept sets and provide guidance to the general public who have limited design or visualization expertise in creating effective and semantically resonant patterns. 
\hty{The goal of this evaluation is to assess how our design methodology supports participants in the design process, and intentionally not to evaluate the quality of their specific designs. Ultimately, our aim is to help the general public engage in effective and innovative data communication. Accordingly, our analysis primarily focuses on how the methodology aids participants throughout the design process. This support may show in various ways, such as helping participants initiate their designs more easily or generate a higher number of design ideas, even if the final sketches are not yet highly refined---particularly given that our participants had no prior experience with sketching. Therefore, we purposefully did not evaluate the quality of the participants' final designs.}

\hty{We note that in this evaluation stage we did not use the final version of the \hty{design methodology} as we just described it in \autoref{sec:design-space}. Instead, we assessed a previous version that already included all the final components, as we detail it in \autoref{appendix:iteration-design-space}. The differences between the evaluated version and the final version are purely structural and terminological. In fact, this very evaluation led to several refinements in the overall terminology and structure of the design methodology, ultimately resulting in the final (presented) version we discussed in \autoref{sec:design-space}.}

For our experiment we received ethics approval from our institution's ethics review board (XJTLU, \textnumero\,ER-LRR-11000180720241018134919) and pre-registered the experiments on OSF (\href{https://osf.io/ystm3}{\texttt{osf\discretionary{}{.}{.}io\discretionary{/}{}{/}ystm3}}).

\subsection{Concept selection}
For our evaluation workshop we selected two new concept sets, different from those we had used in developing the \hty{design methodology}, to test its versatility: a concrete concept set containing names of \emph{ball games} and an abstract concept set containing described terms of \emph{personality traits}. Like before, each concept set comprises seven terms. For the \emph{ball games} concept set we used the terms badminton, basketball, billiards, ice hockey, rugby, table tennis, and volleyball. For the \emph{personality traits} we selected honest, ambitious, selfish, creative, stubborn, hardworking, and lazy.

A crucial factor in selecting concepts for our \hty{design methodology} is ensuring that all selected terms and concepts are familiar to participants and unambiguous, because we did not want to evaluate our participants' understanding of the concepts themselves. We thus followed a process of systematic filtering of concept candidates to ensure a balanced and representative selection, as we detail next.

\subsubsection{Criteria for selecting ball game concepts}
We began by collecting a list of ball games from the Olympic Games\footnote{\url{https://olympics.com/en/sports/}, accessed October 2024} and included additional widely recognized games, such as billiards. From this initial collection, we filtered out games that are less commonly recognized or widely viewed, \eg, handball. We further excluded games with regional ambiguities or varying interpretations. For instance, ``football'' can refer to different sports depending on a given speaker's region, which could lead to confusion---despite the clear definition on the Olympic Games website. Finally, to ensure that the concept set is representative in the sense of capturing a diverse group of things, we selected games with characteristics that capture a wide range of ball game features (\autoref{tab:concept-sets-ball-games}). 

\begin{table}
\centering
\caption{Set of characteristics we considered to pick the final concept set for ball games for our evaluation.}
\resizebox{\columnwidth}{!}{%
\begin{tabu}{ll}
\toprule
\textbf{feature} & \textbf{examples} \\
\midrule
\makecell[l]{\textbf{shape}\\shapes of balls}     & \makecell[l]{\textbf{irregular}: badminton\\ 
                                                \textbf{spherical}: basketball, billiards, ice hockey,\\ table tennis, volleyball\\
                                                \textbf{oval}: rugby}\\ 
\midrule
\makecell[l]{\textbf{size}\\sizes of balls}       & \makecell[l]{\textbf{big}: basketball, rugby, volleyball\\
                                                    \textbf{medium}: badminton\\
                                                    \textbf{small}: billiards, ice hockey, table tennis}\\
\midrule
\makecell[l]{\textbf{equipment}\\equipment to hit the ball}    
                                            & \makecell[l]{\textbf{hand}: basketball, rugby, volleyball\\
                                                    \textbf{racket}: badminton, table tennis\\
                                                    \textbf{stick}: billiards, ice hockey}\\
\midrule
\makecell[l]{\textbf{venue}\\location or Environment\\ of playing official games}
        & \makecell[l]{\textbf{indoors/on floor}: badminton, basketball\\
                        \textbf{indoors/on table}: billiards, table tennis\\
                        \textbf{outdoors}: rugby, volleyball\\
                        \textbf{special}: ice hockey}\\
\midrule
\makecell[l]{\textbf{headcount}\\team size of players on\\ each side}
        & \makecell[l]{\textbf{one/two}: badminton, billiards, table tennis\\
                        \textbf{multiple}: basketball, ice hockey, rugby,\\ volleyball}\\
\bottomrule
\end{tabu}%
}
\label{tab:concept-sets-ball-games}
\end{table}

% Absence of Ambiguity: Sports with ambiguous or region-specific terminology were excluded (e.g., we excluded both "soccer" and "football" due to their differing regional usage).
% Relevance to the Olympics and Common Usage: The selected sports are either commonly featured in the Olympic Games or widely recognized through common usage (e.g., billiards).
% Diverse Characteristics:
% Shape: Whether the ball is spherical, oval, or otherwise.
% Use of Equipment: Whether players use tools like rackets or sticks.
% Distinctive Format: Such as singles vs. doubles play in table tennis.
% Size: The physical dimensions of the ball.
% Venue: Whether the game is typically played indoors (e.g., on tables) or outdoors (e.g., on fields).
% Number of Players: Whether the game involves individual or team-based play.
% The selected ball game concepts include: badminton, basketball, billiards, ice hockey, rugby, table tennis and volleyball.

\subsubsection{Criteria for selecting personality concepts}
For the personality concept set, we first collected a list of commonly used personality-related terms, based on a Google search for words that describe personality traits and on some we proposed ourselves.
% vocabulary list \footnote{https://www.vocabulary.com/lists/179970}
To maintain consistency in word types, we only kept adjectives. We focused exclusively on adjectives that were clearly recognizable as personality descriptors, excluding ambiguous terms such as ``funny.'' From this refined list, four authors independently selected representative concepts. We again strove to create a balanced set by including terms with varied emotional tones (positive, negative, and neutral). After compiling these initial selections, we reviewed and discussed each concept as a group, and voted to determine the most representative terms for the final set. 

\subsubsection{Concreteness ratings of selected concepts}
As in the concept selection process of our previous workshop (\autoref{sec:preparation-concept-sets}), we based our selections on the concreteness ratings provided by Brysbaert et al.~\cite{brysbaert:2014:concreteness} and ensured that all chosen concepts were listed in their table. \autoref{tab:concept-sets-evaluation-workshop-scores} shows the concreteness ratings for each concept in our two concept sets. The mean rating for the \emph{ball games} (concrete concept set) is 4.72, which is clearly high on a 1-to-5 scale, while the \emph{personality traits} (abstract concept set) has a mean rating of 2.09, which is distinctly low. These values confirm the respective concreteness and abstractness of our two concept sets.

\begin{table}
\centering
\caption{Concreteness scores according to Brysbaert et al.'s \cite{brysbaert:2014:concreteness} concreteness scale for ball games and personalities, and the means for each concept group. The scale ranges from 1 to 5, where 1 represents the most abstract and 5 represents the most concrete.}
% \resizebox{\columnwidth}{!}{%
\begin{tabu}{lclc}
\toprule
\textbf{\emph{ball game}} & \textbf{score} & \textbf{\emph{personality}} & \textbf{score} \\
\midrule
badminton                   & 4.7                                & ambitious                     & 1.81                               \\
basketball                  & 4.97                               & creative                      & 1.93                               \\
billiards                   & 4.61                               & hard working                  & 2.48                               \\
ice hockey                  & 4.64                               & honest                        & 1.66                               \\
rugby                       & 4.33                               & lazy                          & 2.67                               \\
table tennis                & 4.83                               & selfish                       & 1.92                               \\
volleyball                  & 4.93                               & stubborn                      & 2.18                               \\ 
\midrule
\textbf{mean}      & \textbf{4.72} & \textbf{mean}       & \textbf{2.09}  \\
\bottomrule\end{tabu}
% }
\label{tab:concept-sets-evaluation-workshop-scores}
\end{table}

\subsection{Pilot Studies}
\label{sec:evaluation-pilot-studies}
We conducted two rounds of pilot studies to determine a suitable time limit for the participants in their drawing sessions and to refine the experimental procedure. 
In the first round, four of this paper's authors completed the drawing tasks to provide an initial time estimate. For the ball game concept set, the authors took 5:10 minutes, 4:20, 4:30, and 6:11, respectively. For the personality concept set, they took 3:07, 4:35, 5:32, and 6:04. The longest designing time any of us needed was 6:11 and, given our own design expertise, we estimated that non-expert participants would require roughly twice this time. We thus set a 12-minute limit for each design session. All authors' drawings from this round of the pilot study are provided in \autoref{appendix:all-designs-workshop2-pilot-authors}.
In the second round, we conducted a pilot with four non-expert participants to confirm the timing and further refine the procedure. We show all drawings collected from participants in this round in \autoref{appendix:all-designs-workshop2-pilot-participants}. In addition, all subjective rating data and interview transcriptions from this round are available in our OSF repository (\href{https://osf.io/9h5nd/}{\texttt{osf\discretionary{}{.}{.}io\discretionary{/}{}{/}9h5nd\discretionary{/}{}{/}}}). We did not include the data from the pilot studies in our formal data analysis.

\subsection{Procedure}
%Procedure see: https://docs.google.com/presentation/d/1aVWz9hgBe-8f16_N2zjj8oC9UwVetapEFHWbxhG8_Ao/edit#slide=id.g30d1c938faa_0_43
We conducted all the studies in person, using the Qualtrics platform \cite{qualtrics} to distribute and collect questionnaires. Before starting the experiment, we asked participants to read and sign the informed consent form, agreeing to take part in the study. We then collected background and demographic information, including gender identification, age range, education level, skill level in visual design and sketching, frequency in using design tools and hand sketching, and familiarity with the concept of semantic association. 

After filling out the questionnaire, we introduced participants to the background of the project and our proposed \hty{design methodology} with our prepared slides.\footnote{We used the structure of \autoref{fig:design-space-v2}, which includes all approaches in the final structure we presented in \autoref{sec:design-space}.} 
\hty{Participants then completed two rounds of sketching sessions, each corresponding to a different set of concepts: \emph{ball games} and \emph{personality traits}. The order of these two sets was fully counterbalanced across participants. Within each session, participants first indicated their familiarity with each concept in the assigned set before beginning the sketching task. We provided participants with pens and design templates printed on A4 paper. During the sketching process, participants could refer to a slide summarizing our design methodology as a cheat sheet (\autoref{fig:design-space-v2} in \autoref{appendix:iteration-design-space}) at any time. Based on the time determined in our pilot study in \autoref{sec:evaluation-pilot-studies}, each sketching session was limited to 12 minutes.}

Following the sketching of each concept set, we ask the participants to rate how difficult it was to design each concept on a 7-point Likert scale. 
\hty{We then conducted a post-study interview to gather feedback on our design methodology and their design strategies. First, participants rated the \emph{usefulness}, \emph{ease of use}, and \emph{helpfulness} of our design methodology on a 7-point Likert scale, followed by explanations for their ratings. Next, we asked them to describe their use of the approaches in our design methodology, their own design strategies, and whether they changed their approach while completing the tasks. We also asked for general feedback after both sketching sessions. We audio-recorded the post-study interview.}

The entire in-person study took about 45 minutes per participant. Our participants did not receive financial compensation; instead, we provided both pilots and formal study participants with welcome soft drinks free of charge.
% including water, milk, coffee, and tea.
% \marginpar{\tiny\ti{What is ``mike'' for a beverage? What about soft drinks?}\hty{mike-typo of milk.}}  

% \hty{The value of the provided beverages exceeded the local minimum hourly wage (21 RMB, approximately 3 USD), and participants were welcome to take their beverages with them upon completing the study.} 

\subsection{Participants}
We recruited participants via direct e-mail, social media, and word-of-mouth, specifically targeting individuals without design or drawing expertise. Participants were required to be of legal age (18 years in the country in which we conducted our study),
% \marginpar{\tiny\ti{as this was an in-person study, the local laws apply; so please name the Chinese limit here (without saying that it was China)}}
proficient in English, and to lack prior experience or skills in visual design and hand sketching. We excluded anyone from participating in the experiment who reported a level of ``intermediate'' or above in response to the background question ``How proficient or skilled do you believe you are at visual designing or drawing?'' We also excluded participants if they reported using visual design tools or to draw ``more often than once a month.'' 

We had 12 valid participants in total: 5 female and 7 male; 9 aged 18--24, 1 aged 25--34, and 2 aged 45--54. Among them, 7 held a Bachelor's degree, and 5 held a Master's degree. The participants all had limited or no design expertise: 10 did not know visual designing or sketching at all, while 2 considered themselves as beginners. 9 never used visual design tools, 2 used them once a few years, and 1 once a month. 6 participants never hand-sketched, 4 did so once a year, and 2 once a few years. %In addition, 10 participants were \todo{unfamiliar} with semantic associations, and 2 had only heard the term. 
``Semantic association'' is also an unfamiliar term for participants: 10 participants had never encountered this term before, and 2 had only heard this term but did not know its meaning.

% We excluded participants with design expertise based on their responses to specific questions. Participants will be excluded if they select any of the following options:
% How proficient or skilled do you believe you are at visual designing or drawing?
% Exclusion criteria: Intermediate, Advanced, or Expert.
% How frequently do you use visual design tools (e.g., Adobe Photoshop, Adobe Illustrator, Figma, Procreate…)?
% Exclusion criteria: More often than once a month.
% How frequently do you draw?
% Exclusion criteria: More often than once a month.

\subsection{Results}
\label{sec:evaluation-results}
From the evaluation workshops, we collected 24 pie charts with 168 \sr pattern designs from 12 participants. In addition, we gathered participants' subjective ratings and interview recordings, in which they provided feedback on our design methodology. All \sr pattern designs are shown in \hyperref[fig:p1-con]{Figures~}\ref{fig:p1-con}--\ref{fig:p12-ab} in \autoref{appendix:all-designs-workshop2-formal-experiment}. We followed the same procedural process as \autoref{sec:workshop1-data-analysis} and processed the sketches and interview recordings. \hty{We include all original and corrected transcriptions in our OSF repository (\href{https://osf.io/9h5nd/}{\texttt{osf\discretionary{}{.}{.}io\discretionary{/}{}{/}9h5nd\discretionary{/}{}{/}}}).}

\subsubsection{Feedback on our \hty{design methodology}}
\begin{figure}
    \centering
    \includegraphics[width=1\linewidth, trim={0 0 0 42}, clip]{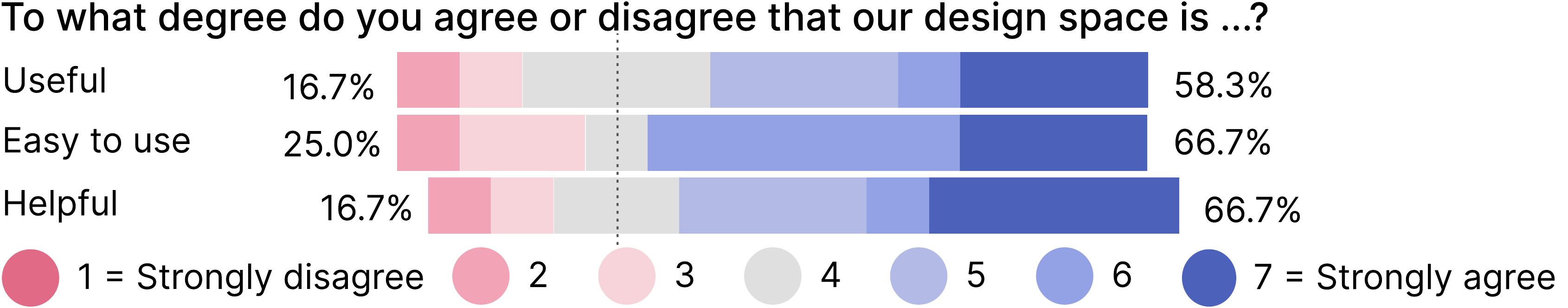}
    \caption{\hty{Rating of how useful, easy to use, and helpful our design methodology is}, for the concrete concept set. \hty{The percentages on the left represent negative scores (1--3), and the percentages on the right represent positive scores (5--7). }} 
    % \ti{[Check the meaning of the Likert values: strongly disagree to strongly agree does not seem to measure ``usefulness/ease of use/helpfulness''. Or was the question phrased as ``Do you agree that the concepts was useful/not useful/... to design?''? And in case you re-do the plot, could you round all \% numbers to 1 digit after the decimal?]}
    \label{fig:evaluation-workshop-rating-concrete}
\end{figure}

\begin{figure}
    \centering
    \includegraphics[width=1\linewidth, trim={0 0 0 41}, clip]{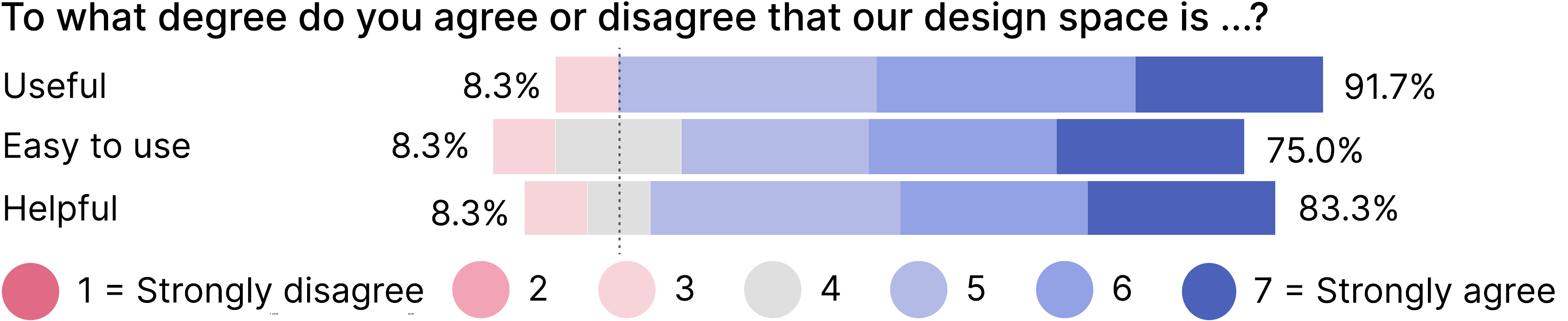}
    \caption{\hty{Rating of how useful, easy to use, and helpful our design methodology is}, for the abstract concept set. \hty{The percentages on the left represent negative scores (1--3), and the percentages on the right represent positive scores (5--7). }}
    % \ti{[Check the meaning of the Likert values: strongly disagree to strongly agree does not seem to measure ``usefulness/ease of use/helpfulness''. Or was the question phrased as ``Do you agree that the concepts was useful/not useful/... to design?''? And in case you re-do the plot, could you round all \% numbers to 1 digit after the decimal?]}
    \label{fig:evaluation-workshop-rating-abstract}
\end{figure}

After each sketching session, we asked our participants to rate the \hty{design methodology} in terms of usefulness, ease of use, and helpfulness on a 7-point Likert scale (as we summarize in \hyperref[fig:evaluation-workshop-rating-concrete]{Figures~}\ref{fig:evaluation-workshop-rating-concrete} and \ref{fig:evaluation-workshop-rating-abstract}) and to explain their reasons and their feedback on our \hty{design methodology}. 

\textbf{Usefulness:}
Overall, participants indicated that the \hty{design methodology} is useful (\textit{M} $=$ 5.29, \textit{SD} $=$ 1.43). The usefulness of the abstract set was rated (\textit{M} $=$ 5.67, \textit{SD} $=$ 1.15), while that of the concrete set was rated (\textit{M} $=$ 4.92, \textit{SD} $=$ 1.62). This result points to a trend that participants think our \hty{design methodology} is more useful for abstract concept sets than concrete concept sets.

Participants thought our \hty{design methodology} is highly inspiring for generating ideas and that it can guide their designs. Especially for the abstract concept set, 11/12 participants expressed this feeling. For example, \emph{``The combination of abstract and concrete is well covered, so it is useful when designing''} (P10 for abstract). For the concrete concept set, 6/12 participants also thought the \hty{design methodology} is useful in helping them generate ideas, and to structure their designs.

However, 5/12 participants mentioned that it is difficult to show concrete concepts using abstract representations. Because they already had own ideas on these concepts in their mind (\ie, on ball games) it was harder for them to reference our \hty{design methodology}. For example, P11 explained, \emph{``This set is something we see every day, and we already have a relatively fixed understanding of their shapes. So, the abstract approach wasn't very helpful.''} This issue was reflected in their scores: one participant rated the usefulness as 2, which shows that they did not think our \hty{design methodology} is useful. In contrast, most gave it a 4 or 5, showing that they agree on the usefulness of our \hty{design methodology}. This observation suggests that, despite there being challenges, our \hty{design methodology} still provided valuable inspiration and guidance for most participants.

\textbf{Ease of use:}
Overall, participants rated the ease of use at (\textit{M} $=$ 5.33, \textit{SD} $=$ 1.52). For the abstract set, the rating was (\textit{M} $=$ 5.42, \textit{SD} $=$ 1.31), while the concrete set received a rating of (\textit{M} $=$ 5.25, \textit{SD} $=$ 1.76).  In terms of ease of use, participants thus have similar feeling for the abstract and the concrete concept sets.

Participants felt our \hty{design methodology} was simple, easy to understand, and memorable. For example,  P3 mentioned, \emph{``The methods you introduced are clear, simple, and easy for everyone to understand''} (P3 for concrete). P12 also noted: \emph{``The association method, for example, is simple to understand''} (P2 for abstract).

Participants also reported challenges in terms of ease to use. For the concrete concept set, 4/12 participants gave a score of 3--4 as they felt that, while the \hty{design methodology} provided methods for conceptualizing and visualizing concepts, it lacked guidance on abstracting concrete shapes into patterns, which made it difficult to design patterns. For example, P9 noted, \emph{``You're asking for abstract patterns, which can't be drawn directly. So, there's a need for transformation or professional processing, and abstracting it is a bit harder.''}

For the abstract concept set, 5/12 participants mentioned it was challenging to apply the \hty{design methodology} to new abstract concepts. They felt that the methods and examples provided were based on specific abstract concepts (the concepts in our first workshop) and might not easily transfer to other new concepts. P1 commented, \emph{``It's just that this abstract concept is hard to capture with just one or two methods... It could be that everyone has a different understanding''} (P1 for abstract).

\textbf{Helpfulness:}
Overall, participants thought our design is helpful (\textit{M} $=$ 5.33, \textit{SD} $=$ 1.46). The abstract set was rated (\textit{M} $=$ 5.50, \textit{SD} $=$ 1.24) about the same as the concrete set (\textit{M} $=$ 5.17, \textit{SD} $=$ 1.70)---we almost see no difference between both concept sets. 

8/12 participants found our \hty{design methodology} helpful for inspiring ideas and guiding their designs. For instance, P6 noted, \emph{``The designs I drew were based on the methods you introduced earlier''} (P6 for abstract). P5 added, \emph{``They helped me consider concepts from different angles and draw them out''} (P5 for concrete).

Two participants, however, expressed concerns that the \hty{design methodology} might constrain their ideas, potentially limiting their creativity. For example, P10 mentioned,  \emph{``If we don't use associations (method), each person's thinking might be a bit more detached, but with the associations (method), people's imagination of a concept might be limited to certain aspects. This could be a bit of an influence on the design''} (P10 for abstract).

\subsubsection{Difficulty of designing semantically-resonant patterns}

\begin{figure}
    \centering
    \includegraphics[width=\columnwidth]{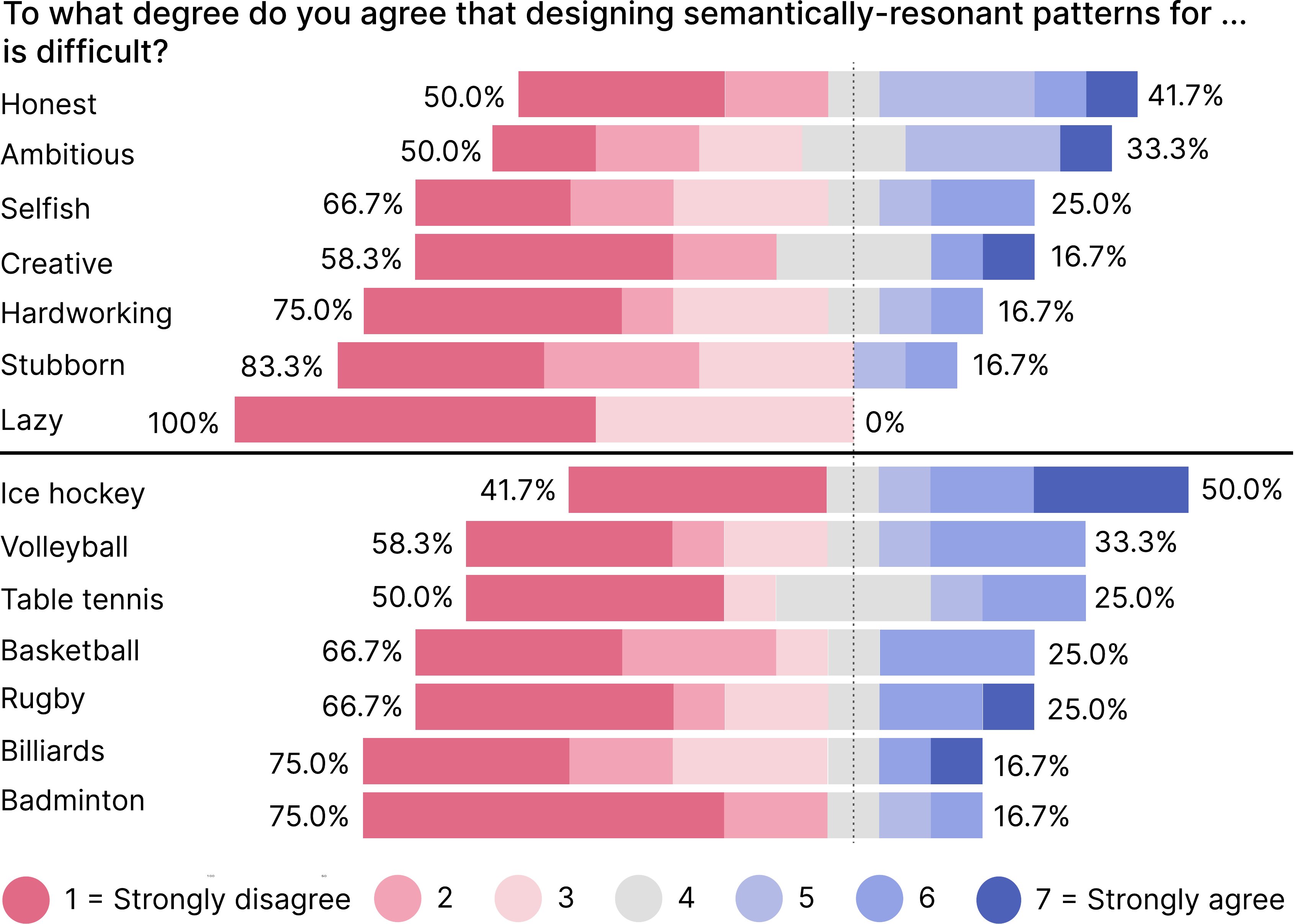}
    \caption{Rating on the difficulty of designing \sr abstract patterns for each concept. The percentages on the left represent scores 1–3, and the percentages on the right represent scores 5–7. }
    % \ti{[Check the meaning of the Likert values: strongly disagree to strongly agree does not seem to measure ``difficulty''. Or was the question phrased as ``Do you agree that the concepts was easy/difficult to design?''? And in case you re-do the plot, could you round all \% numbers to 1 digit after the decimal?]}
    \label{fig:evaluation-workshop-rating-difficult}
\end{figure}

\autoref{fig:evaluation-workshop-rating-difficult} shows our participants' rating on difficulty in creating \sr patterns for each concept.
We can see that, overall, more participants did not think the design procedure was difficult.  Combined with the participants' ratings of usefulness, ease of use, and helpfulness, our \hty{design methodology} thus appears to be meaningful in reducing the challenges users face in design.
From the figure we can see there is a trend that participants think designing concrete concept sets is slightly more difficult than abstract concept sets. This observation is interesting, because during the pilot among the authors, we all had unanimously assumed that designing for an abstract concept set would be much more difficult than for a concrete one. The feedback in the ``usefulness'' rating might explain this difference: non-expert participants feel that it is harder to translate concrete concepts to abstract visual representations.

\subsection{General feedback}
After each drawing session, we had asked participants whether they used the approaches from our \hty{design methodology}. All participants confirmed that they did so. We also asked if they had used methods of their own. While 3/12 participants mentioned they used new methods for some concepts, further clarification showed that these methods were still based on our \hty{design methodology}. We believe the participants gave the initial answer because the specific examples we provided made them think that patterns that are not identical to our examples counted as new methods, even though they were still covered by the \hty{design methodology}. In addition, we received positive feedback on the comprehensiveness of our \hty{design methodology}. P1 said, for instance, \emph{``For people like us without design experience, the design methods you provided are enough for us to create these patterns.''} P10 said \emph{``It covers all basic elements and provides a clear framework for the design process.''} We also received some concerns from participants that a person's cultural background may affect understanding. For example, P8 and P10 commented that \emph{``There are cultural backgrounds to consider''} and \emph{``If these symbols have an internal logic that only a small group understands, it limits the application of these designs.''} However, although these concerns are valid, with our current \hty{design methodology} we aim to help participants to design \sr patterns, whether the patterns they designed are easily understood by others is out of our scope and needs to be investigated in future work.

\section{Discussion}
% From our design walkthrough, we observed key differences in how the methods approached pattern design. Linking a concept to a more concrete concept and extracting an attribute of concept focused on designing for individual abstract concepts, making themselves more generalizable to other concepts. In contrast, comparing concepts to others targeted the entire concept set, requiring the design of each specific concept to account for the relationships and distinctions within the set as a whole. Although we did not evaluate the effectiveness of the patterns created by participants, it is plausible that Method 3 could better convey the overall design logic, enhancing both interpretation and memorability for the entire set. Conversely, Methods 1 and 2, with their emphasis on individual concepts, might be more effective at showcasing the unique characteristics of a single concept. When designing visualizations with patterns, we suggest considering the primary objective: whether to emphasize the overarching structure and coherence of the concept set or to highlight the distinctive attributes of individual concepts. This decision could guide the choice of method to align with the intended communication goals.

Our \hty{design methodology} demonstrated its effectiveness and ease of use for most general participants, showcasing its potential to simplify the process of creating patterns. 
\hty{Since the methodology is grounded in expert guidance and has been shown to be accessible even to non-experts, it is reasonable to assume that experts will also be able to follow it with ease.}
In addition, we believe it could also provide valuable inspiration for designers working on black-and-white visualizations, further investigation is needed to uncover additional factors that influence both the design of such patterns and the perception. Key areas for future exploration include a deeper understanding of how designers' intentions shape their sketching process and the readers' perception workflow. These insights could refine the \hty{design methodology} and enhance its applicability across diverse visualization contexts. We discuss these factors in detail as follows:

\vspace{4pt}
\noindent\textbf{Methodology Evaluation:}
\hty{To evaluate our design methodology, we initially considered comparing conditions with and without its implementation. We determined, however, that such a comparison would not yield meaningful results due to potential confounding factors, particularly given our sample size. In a within-group study design—where each participant engages in sketching tasks under both the ``with design methodology provided'' and ``without design methodology provided (only basic instructions)'' conditions---learning effects would inevitably influence the results. Conversely, in a between-group study design, individual differences, such as variations in sketching abilities, could introduce biases. We could only mitigate these confounding factors with a substantially larger sample size, yet this is unfeasible within our qualitative research approach. A quantitative study approach---as a potential alternative---, however, would not serve our research goals since our primary objective is to gain in-depth insights from our participants’ feedback and by observing their interaction with our design.}

% \marginpar{\tiny\todo{I rephrased this last sentence; in addition, could we add somehow that also a quantitative approach would not be feasible/meaningful either?}}

\vspace{4pt}
\noindent\textbf{Background Diversity:}
As we had already mentioned, a person's cultural background and personal experiences can significantly influence both the design and perception of patterns. During the design process, particularly when using the method of ``linking the target concept to a more concrete concept,'' designers may draw upon culturally specific references or slang. For example, a designer may associate the concept of ``hardworking'' with the slang phrase ``ball of fire'' and create a design inspired by ``fire.'' If the audience is unfamiliar with the slang, however, interpreting the pattern could become confusing, potentially undermining its intended purpose. Similarly, when designing for the ball sports concept set, if a designer uses a subjective metric---such as ranking games based on their popularity among their friends---the underlying logic may be unclear to the audience, making it difficult for readers to interpret the design. This issue was evident in the sketches created by participants, where their \sr pattern designs were often too abstract to identify the relevant items, in particular in the context of abstract concept sets. We thus acknowledge the impact of cultural and personal nuances and emphasize the importance of tailoring designs to the target audience to ensure clarity and effectiveness. 

\vspace{4pt}
\noindent\textbf{Design Experience:}
Another intriguing finding we observed is that individuals without design experience could struggle to translate their ideas into sketches, even when equipped with design strategies. This finding aligns with findings from the work by Lee et al.~\cite{lee:2020:sketch}, which highlighted similar challenges in sketching tasks. While design strategies offer a starting point, they may not fully enable novices to effectively express their ideas in such tasks. Interestingly, participants generally found it harder to sketch for concrete concept sets than for abstract ones. This counter-intuitive result may stem from the inherent familiarity and tangibility of concrete concepts, which often come with predefined visual standards. In contrast, abstract concepts lack fixed representations, potentially giving participants more creative freedom without the pressure of conforming to expected visuals---but only when people are trained in making use of such creative freedom, as designers are but non-designers may not be. Our design methodology focuses on guided people on ideation by the strategies to help them ideation rather than expecting novices to produce complete patterns immediately (as we described it in \autoref{sec:design-space}). For instance, when identifying content by extracting an attribute, novices could begin by writing down the identified attribute before attempting to sketch. If they still found it challenging to translate their ideas into patterns, they could revisit the strategy---exploring more concrete concepts or extracting additional features---to refine their design process iteratively. This structured approach may help bridge the gap between their ideas and final sketches.

% However, based on the outcomes created by our participants in the design workshop, we noticed that our participants' semantical pattern designs were too abstract to identify the relevant items, even for the concrete sets. For example, when we browsed their sketches and tried to find recognizable patterns to fill the concepts illustrated in \autoref{fig:visual-variables-count}, we could only find designs with very high abstract levels that may not inform audiences what the concept is by a first view (especially for ``music'' and ``emotion''). Such observation brings us several interesting questions: To what extent should a pattern be abstract, and how should the balance between the abstract level and the recognizability of the pattern be balanced?   

\section{Limitation and Future Work}
In our work, we focus on the design perspective---specifically, creating \sr patterns---by exploring methods for generating these patterns and evaluating the practicality of the proposed \hty{design methodology}. We do not assess, however, the impact of using \sr patterns in visualizations. For example, the question of how the approach enhances visualization effectiveness or improves the subjective experience of users remains unexplored. Future research can address these gaps by conducting crowd-sourced controlled perception experiments to evaluate how \sr patterns influence chart-reading accuracy and speed across different chart types. Researchers can also apply validated instruments tailored to specific visualization quality criteria, such as the BeauVis scale for aesthetic pleasure \cite{He:2023:BVS} and the PREVis instrument for perceived readability \cite{Cabouat:2025:PPR}.

\hty{In \autoref{sec:evaluation}, we evaluated \ti{a slightly different,} earlier version of our design methodology rather than the final version. Despite this \ti{small difference}, the evaluation demonstrated the effectiveness of our methodology. Given that the final version uses improved organization and terminology (\ti{changes that we made} based on the evaluation), we consider it to be} reasonable to expect even better results based on the final version---which would be worth to be confirmed in further studies.

The findings we reported in \autoref{sec:evaluation-results} show that participants without a design background often use the \hty{design methodology} to generate ideas. Some participants, however, struggled to finalize their abstract pattern designs due to limited sketching skills. To address this limitation and broaden access to \sr patterns, future work on creating systems or libraries seems promising. Such tools may offer pre-configured parameters for digitally generating \sr patterns. In addition, researchers can explore leveraging artificial intelligence, including large language models (LLMs), to collaboratively generate \sr patterns. Such tools may assist non-expert users while enhancing creativity and improving the efficiency of professional designers during the design process.

Building on our findings that participants' designed patterns were often too abstract to be easily understood, we plan to do further investigation on such a phenomenon. Specifically, we aim to determine an optimal level of abstraction for individual patterns to be well perceived and distinguished. Besides, we plan to explore strategies for balancing abstraction with recognizability.

Since \sr patterns visually convey concepts without relying on legends or labels, they work especially well for physical media, such as data embroidery \cite{wannamaker:2019:data, He:2023:DEB}, where embroidery machines handle text elements poorly. Using the \hty{design methodology}, general audiences can design \sr patterns for personal data physicalizations, which opens up exciting future research opportunities. Furthermore, \sr patterns can save space typically allocated to legends and labels, making them highly useful for visualizations, especially in constrained contexts like watch faces or fitness trackers \cite{islam:2024:Visualizing, Islam:2022:Reflections}. Future research can be made on exploring the applications mentioned above and investigating the broader use of \sr patterns across diverse contexts.

\section{Conclusion}
We developed a structured \hty{design methodology} that summarizes the \hty{approaches} for creating \sr abstract patterns and evaluated it. Through workshops with 13 design experts, we contributed to the understanding of the associations between visual variables and the concepts they represent---a fundamental aspect of designing and interpreting visualizations. Specifically, we systematically outlined how design experts form meaningful connections between concepts and black-and-white patterns, a composite variable that incorporates multiple visual elements aside from color. Our contribution complements existing research in the visualization community, which so far has primarily focused on color-concept associations \cite{lin:2013:selecting,schloss:2018:color}. As such, we extend the choices available to visualization designers. 

\hty{Our evaluation with the 12 non-expert participants shows that our \hty{design methodology} effectively supports even individuals with limited design experience in ideating and creating \sr abstract patterns for both abstract and concrete concepts. Our evaluation, however, also reveals that designing \sr patterns is not an easy task for most participants and is often met with hesitation. Prior research \cite{He:2024:DCB} has shown that, \ti{when people are \emph{looking at} pie charts, abstract patterns can improve reading speed and patterns with strong semantic associations can enhance aesthetic appreciation---even if our participants here who \emph{created} the charts were somewhat hesitant.} Thus, our findings suggest that, while novice designers may initially hesitate or struggle with the task \ti{(because it is simply not straightforward)}, our design methodology provides support in easing the design process. Ultimately, once people designed the \sr pattern and incorporated them in visualizations, the visualization \emph{reading results} can improve. By thus lowering barriers and facilitating the embedding of semantic meaning into visual representations, our \hty{design methodology} shows promise in empowering not only expert visualization designers but also the general public to engage in effective and innovative data communication.}

% \marginpar{\tiny\todo{TI: I find that the negative points in the subsubsection above and this summary of the difficulties they had is a valuable additional benefit of the second evaluation: it shows that designing such patterns is not easy for most, and is met with hesitation, yet your previous study at VIS showed that they can be useful for reading speed and intuitive understanding---even if the participants here are hesitant; so the evaluation of the design methodology is essentially demonstrating that even for hesitant people we can ease the design, and then when we have a design the vis reading results will be better; maybe we can add this discussion somewhere (to address the issue of the reviewers that the second eval is meaningless)}} 

% \marginpar{\tiny\todo{maybe here would be the place to discuss the point I mentioned in the margin of section 5.5.2? This would allow you to take this mental step backwards that I always like to do in a conclusion---at the moment it is still largely a summary}}

\section*{Acknowledgments}
We thank the design experts who participated in the design workshop for their valuable input, which helped shape the design strategy and establish the foundation for the \hty{design methodology}. We also thank all participants in the evaluation workshop, including those in the pilot and formal phases.
Lijie Yao is funded by the Research Development Funding, grant number RDF-24-01-062.

\section*{Supplemental Material Pointers}

The pre-registrations for our two experiments can be found at \href{https://osf.io/h62yb}{\texttt{osf\discretionary{}{.}{.}io\discretionary{/}{}{/}h62yb}} and \href{https://osf.io/ystm3}{\texttt{osf\discretionary{}{.}{.}io\discretionary{/}{}{/}ystm3}} seperately. We also share our experiment documents, study results, analysis files, and other material at \href{https://osf.io/9h5nd/}{\texttt{osf\discretionary{}{.}{.}io\discretionary{/}{}{/}9h5nd}}.

\section*{Images/graphs/plots/tables/data license/copyright}
We as authors state that all of our own figures plots, and data tables in this article (\ie, all with the exception of \autoref{fig:expert-design-examples}) are and remain under our own personal copyright, with the permission to be used here. We also make them available under the \href{https://creativecommons.org/licenses/by/4.0/}{Creative Commons At\-tri\-bu\-tion 4.0 International (\ccLogo\,\ccAttribution\ \mbox{CC BY 4.0})} license and share them at \href{https://osf.io/9h5nd/}{\texttt{osf\discretionary{}{.}{.}io\discretionary{/}{}{/}9h5nd}}. The images in \autoref{fig:expert-design-examples} remain under the copyright of their respective authors (\ie, the respective study participants), with the permission to be used here.

%-------------------------------------------------------------------------
% bibtex
\bibliographystyle{abbrv-doi-hyperref}
\bibliography{abbreviations,template}
\flushcolsend
\raggedend

\appendix % You can use the `hideappendix` class option to skip everything after \appendix
\clearpage

\begin{strip} % requires \usepackage{cuted}
\noindent\begin{minipage}{\textwidth}
\makeatletter
\centering%
\sffamily\bfseries\fontsize{15}{16.5}\selectfont
\papertitle \\[.5em]
\large Appendix\\[.75em]
\makeatother
\normalfont\rmfamily\normalsize\noindent\raggedright In this appendix we provide additional tables, plots, and charts that show data beyond the material that we could include in the main paper due to space limitations or because it was not essential for explaining our approach.%\vspace{-.5em}
\end{minipage}
\end{strip}

% \appendix
\section{\hty{Real-world references for concept selection}}
\label{sec:real-world-references-concept-selection}
\begin{figure}[t]
    \centering
    \includegraphics[width=1\columnwidth]{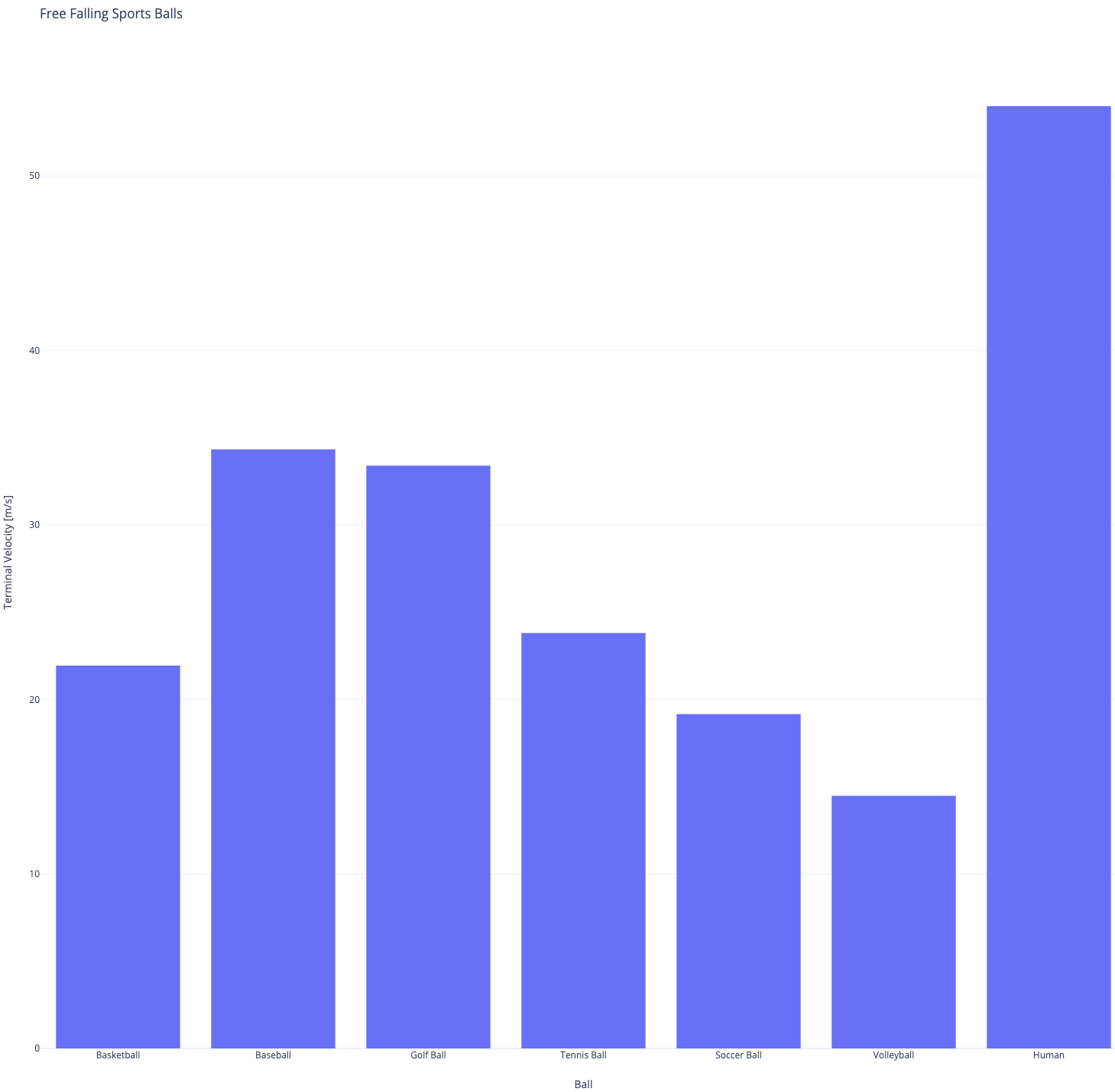}
    \caption{\hty{Screenshot of the real-world visualization example for the ball game concept set. Link: \href{https://chart-studio.plotly.com/~RhettAllain/2211.embed}{\texttt{chart\discretionary{}{-}{-}studio\discretionary{}{.}{.}plotly\discretionary{}{.}{.}com\discretionary{/}{}{/}\~{}RhettAllain\discretionary{/}{}{/}2211\discretionary{}{.}{.}embed}}; website accessed: Feburary 2025.}}
    \label{fig:real-world-example_ball-games}
\end{figure}

\begin{figure}[t]
    \centering
    \includegraphics[width=1\columnwidth]{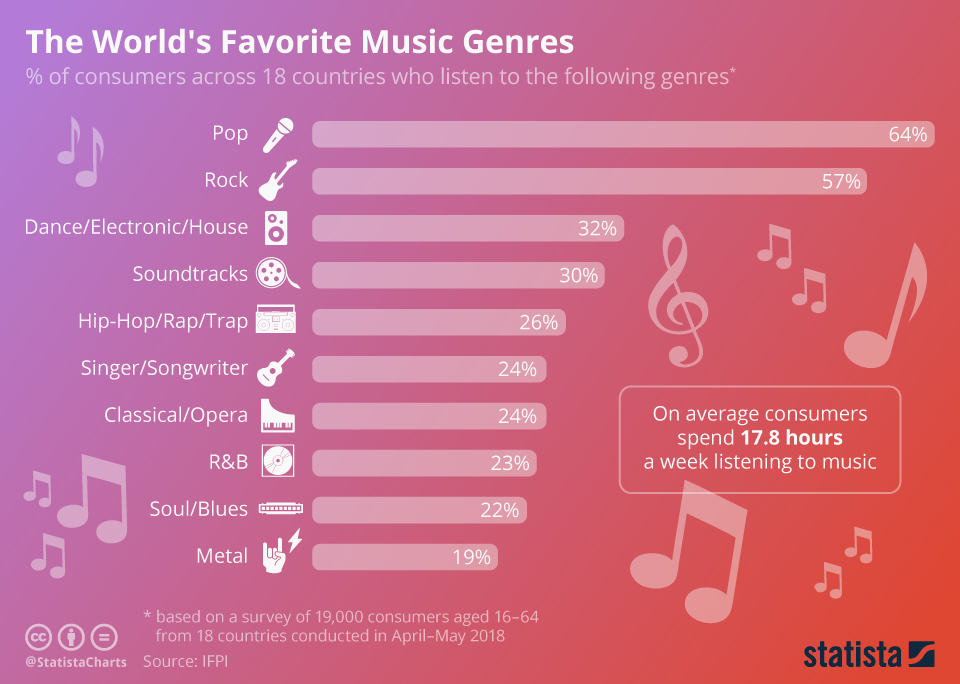}
    \caption{\hty{Screenshot of the real-world visualization example for the music genre concept set. Link: \href{https://www.statista.com/chart/15763/most-popular-music-genres-worldwide/}{\texttt{www\discretionary{}{.}{.}statista\discretionary{}{.}{.}com\discretionary{/}{}{/}chart\discretionary{/}{}{/}15763\discretionary{/}{}{/}most\discretionary{}{-}{-}popular\discretionary{}{-}{-}music\discretionary{}{-}{-}genres\discretionary{}{-}{-}worldwide}}; website accessed: Feburary 2025.}}
    \label{fig:real-world-example_music-genres}
\end{figure}

\begin{figure}[t]
    \centering
    \includegraphics[width=1\columnwidth]{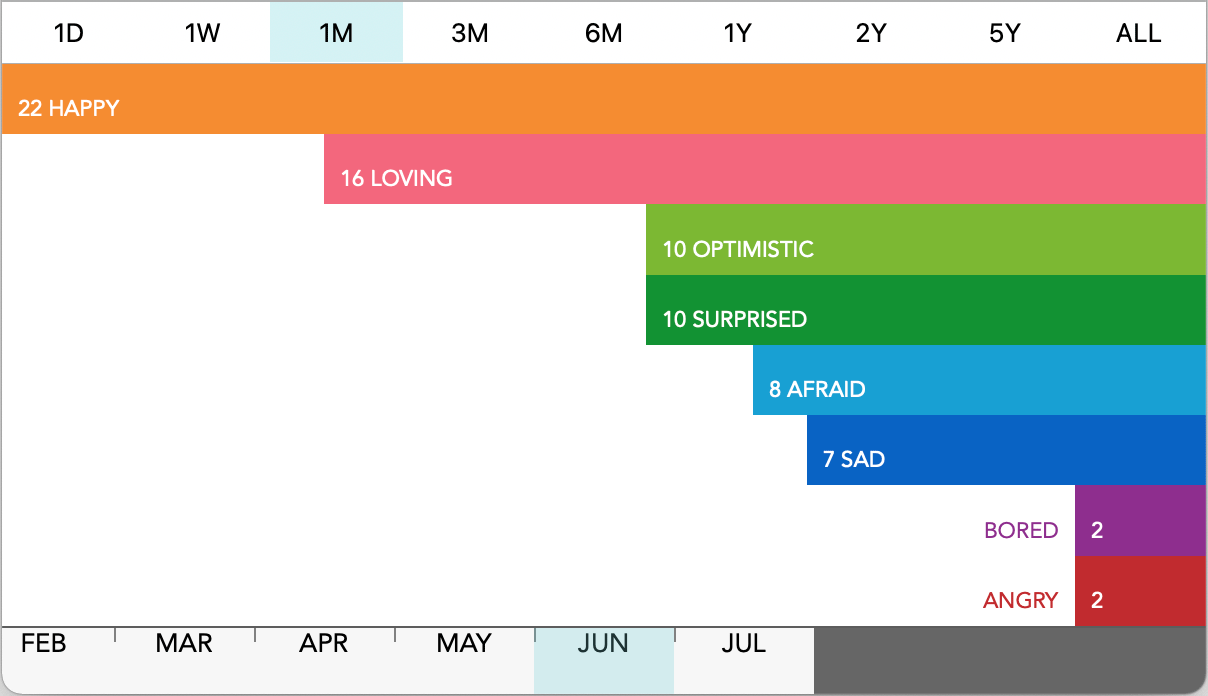}
    \caption{\hty{Screenshot of the real-world visualization example for the emotion concept set. Link: \href{https://support.lifecraft.com/hc/en-us/articles/4404785101079-Emotion-Charts}{\texttt{support\discretionary{}{.}{.}lifecraft\discretionary{}{.}{.}com\discretionary{/}{}{/}hc\discretionary{/}{}{/}en\discretionary{}{-}{-}us\discretionary{/}{}{/}articles\discretionary{/}{}{/}4404785101079\discretionary{}{-}{-}Emotion\discretionary{}{-}{-}Charts}}; website accessed: Feburary 2025.}}
    \label{fig:real-world-example_emotions}
\end{figure}

\begin{figure}[t]
    \centering
    \includegraphics[width=1\columnwidth]{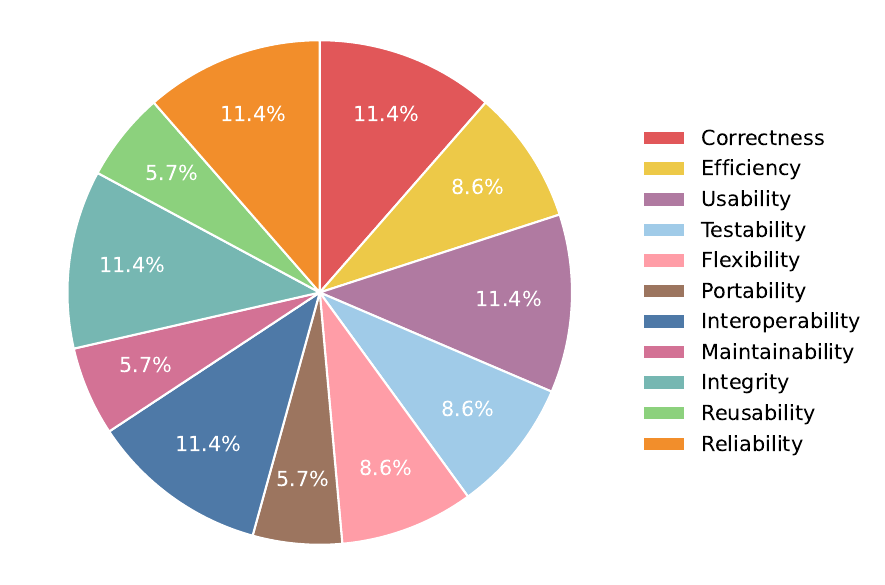}
    \caption{\hty{Real-world visualization example for the quality metric concept set. This is our own re-creation of a pie chart based on the data shown in Figure~5 in Awan et al.'s \cite{Awan:2015:Efficient} paper (because we do not have the copyright for the original figure).} 
    % \todo{TI: I would just get the data from that paper and create a new (and prettier) pie chart with it.} \hty{But I checked the paper, it does not provide the data of this figure. How should we do?} \todo{Simple. See the SVG with the same name of the image. I drew a pie piece on one of the missing data items, and then copied and rotated it and moved it to the other. It matches perfectly. So both missing data items are of equal size. So we can easily calculate the value: 100\% - all existing percentages is 11.5\%, so each of the missing slices has 5.75\%. I assume that these tiny data values that lead to the strange steps of 11.4, 8.6, and 5.7 data. In fact, I think that all 11.4\% values are really a 4, the 8.6\% values are really a 3, and the 5.7\% values are really a 2. That would make 35 as a total, and then the 11.4\% are really 11.429\%, the 8.6\% are really 8.571\%, and the 5.7\% would be 5.714\%. And then everything sums up to 100\% properly. So now you have the data, go and recreate the chart, but with nicer colors. For example Tableau10 from Vega (\url{https://vega.github.io/vega/docs/schemes/}), but with one more color shade addded to it because we have 11 categories. For instance, you could add the 5th color from \#category20b (the darkish beige-green). Also please ensure that the legend is not moved to the top as in the example above (instead top-align it with the pie or vertically center it to the pie, whichever looks better). But do use the same sequence of the data, both in the legend and in the actual pie chart.}
    }
    \label{fig:real-world-example_quality-metrics}
\end{figure}

\hty{We selected our concept sets based on real-world categorical visualization examples, which shows the practical necessity of designing visualizations for these categories. We show screenshots of our reference visualization examples (and mention their sources in the figure captions) for the ball game concept set, music genre concept set and emotion concept set in \hyperref[fig:real-world-example_ball-games]{Figures~}\ref{fig:real-world-example_ball-games}--\ref{fig:real-world-example_emotions}. We show our own re-creation of our reference visualization example (and mention its source in the figure captions) for the quality metric concept set in \autoref{fig:real-world-example_quality-metrics} (because we do not have the copyright for the original figure). These are the concept sets we have considered. We ultimately selected the music genre concept set and the emotion concept set, as described in \autoref{sec:preparation-concept-sets}.}

\section{Semantic-resonant abstract pattern designs by a design expert in a previous study \cite{He:2024:DCB}}
\label{appendix:semantic-association-in-previous-study}

\begin{figure} % htbp are optional placement specifiers (here, top, bottom, page)
	\centering % Centers the figure
	\includegraphics[width=0.8\columnwidth]{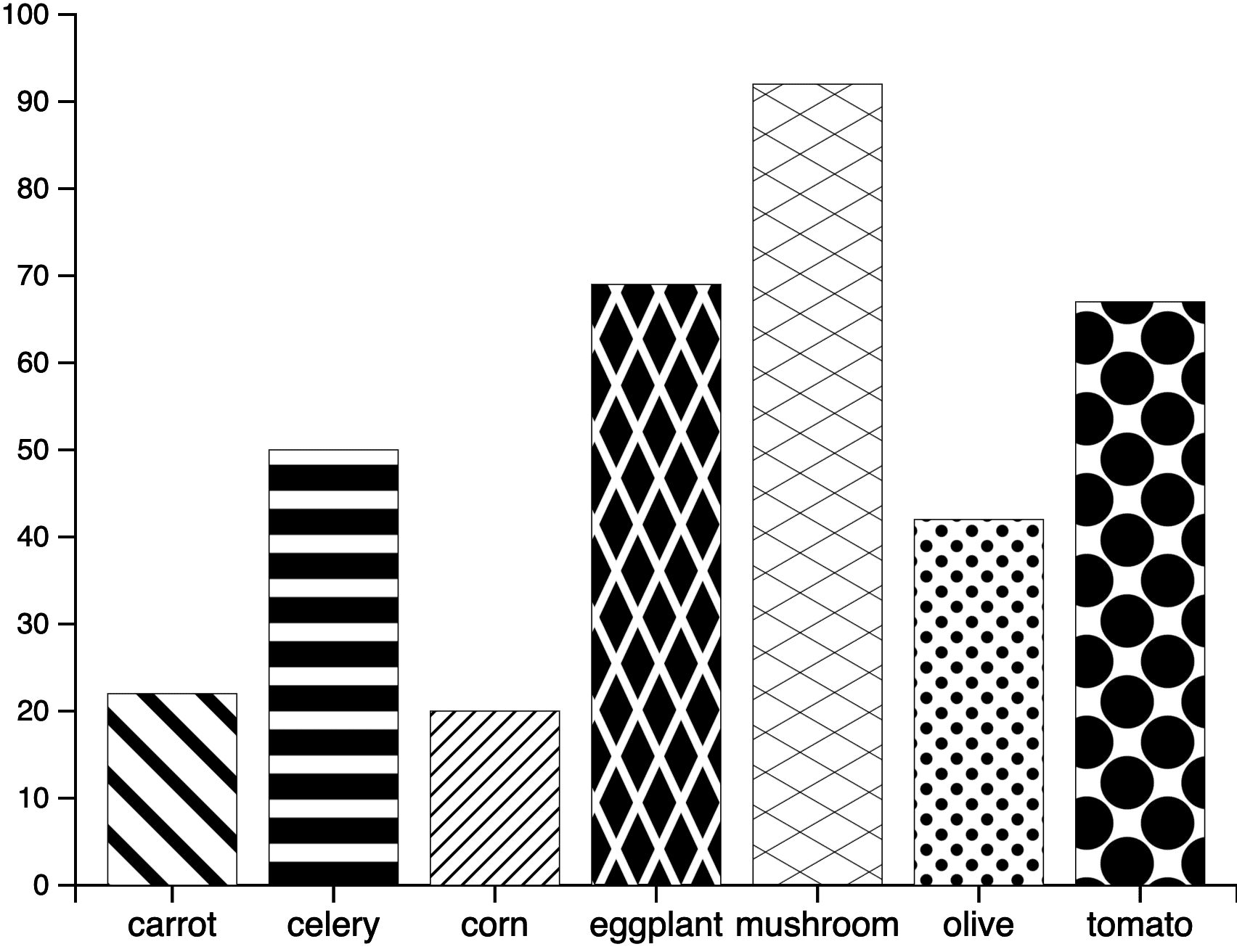}\\
	\caption{An example of semantically-resonant abstract patterns designed for vegetable concept sets, as collected in He et al.'s study \cite{He:2024:DCB} (their Figure~20 as well as part of their Table~2). Here, the designer aimed to create semantic associations between the patterns and the vegetables they represent.}
  \label{fig:BG1}
\end{figure}

Prior research by He et al. \cite{He:2024:DCB} explored the influence of using patterns in visualization by asking designers to create black-and-white patterns for visualizations, focusing on two types: abstract geometric patterns (using simple lines or dots) \inlinevis{-1pt}{1em}{figures/geometrictextures.pdf} and iconic patterns (such as a banana icon for bananas) \inlinevis{-1pt}{1em}{figures/iconictextures.pdf}. When designing abstract geometric patterns, participants attempted to embed semantic meaning into these patterns. One designer primarily used the strategy of creating associations between the patterns and the concepts they represented, as shown in \autoref{fig:BG1}. This design received the highest aesthetic pleasure rating among 14 expert-designed bar charts featuring abstract patterns.

Here is the quote of the design strategies from the designer of \autoref{fig:BG1}: \textit{``I also wanted to elicit visual associations with the geometric texture where possible: olives are small and circular, tomatoes are large and circular; black olives are dark; ripe tomatoes are a deep red; carrots, celery, and stalks of corn are elongated, so line textures seemed appropriate, though the line thickness  and foreground / background choice seemed less deliberate or coherent (some carrots are larger than some celery stalks, and vice versa, while individual corn kernels are quite small, hence the finer texture for corn); eggplants are neither round nor long, but they are dark in color, so the rotated grid pattern with a dark background seemed appropriate; lastly, mushrooms are small and white, but are not circular or elongated, so once again a grid pattern with a white background seemed appropriate, though one that is finer than the eggplant grid. ''}

\section{Iterations on the \hty{design methodology}}
\label{appendix:iteration-design-space}
\begin{figure}[h]
    \centering
    \includegraphics[width=1\linewidth]{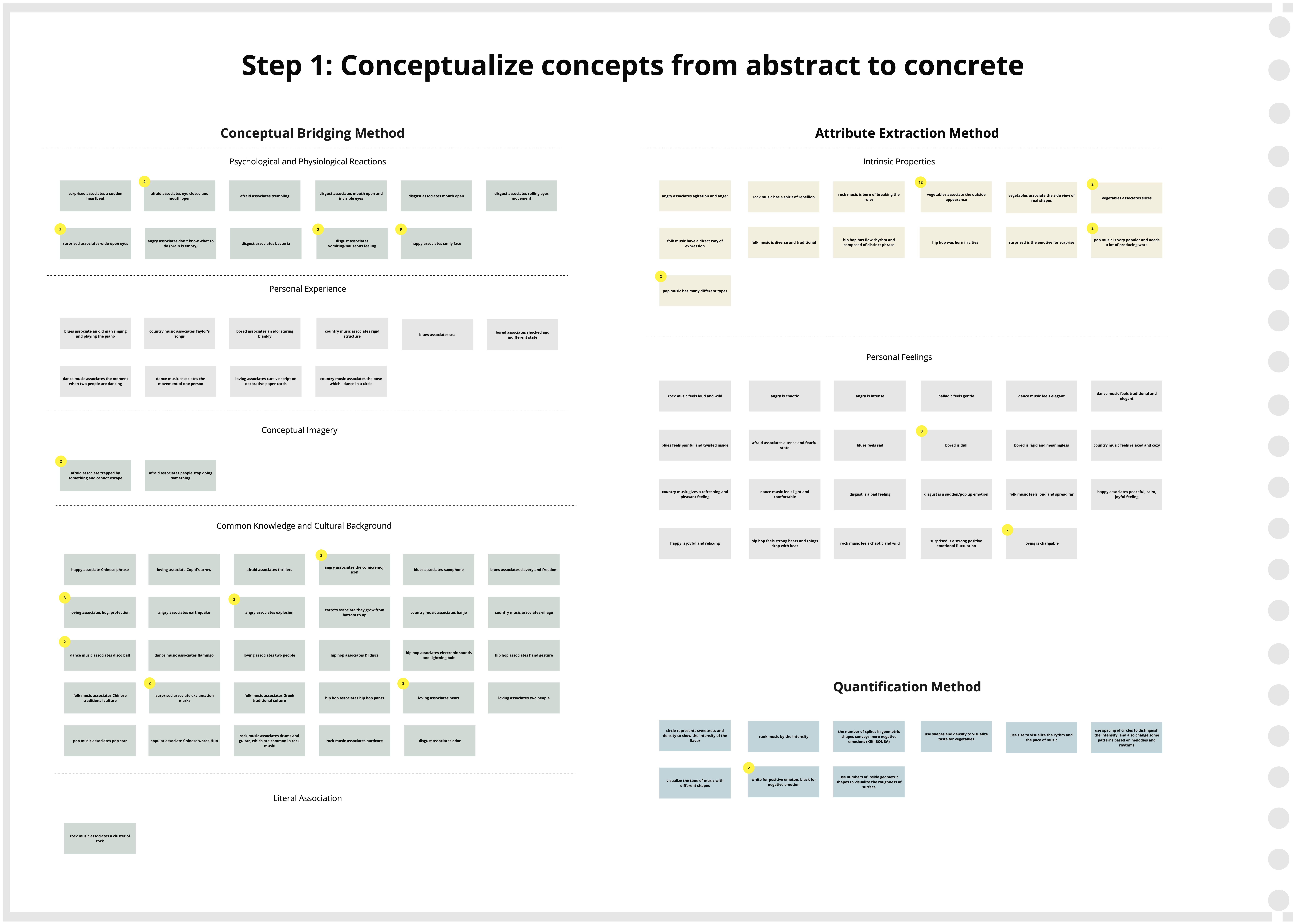}
    \includegraphics[width=1\linewidth]{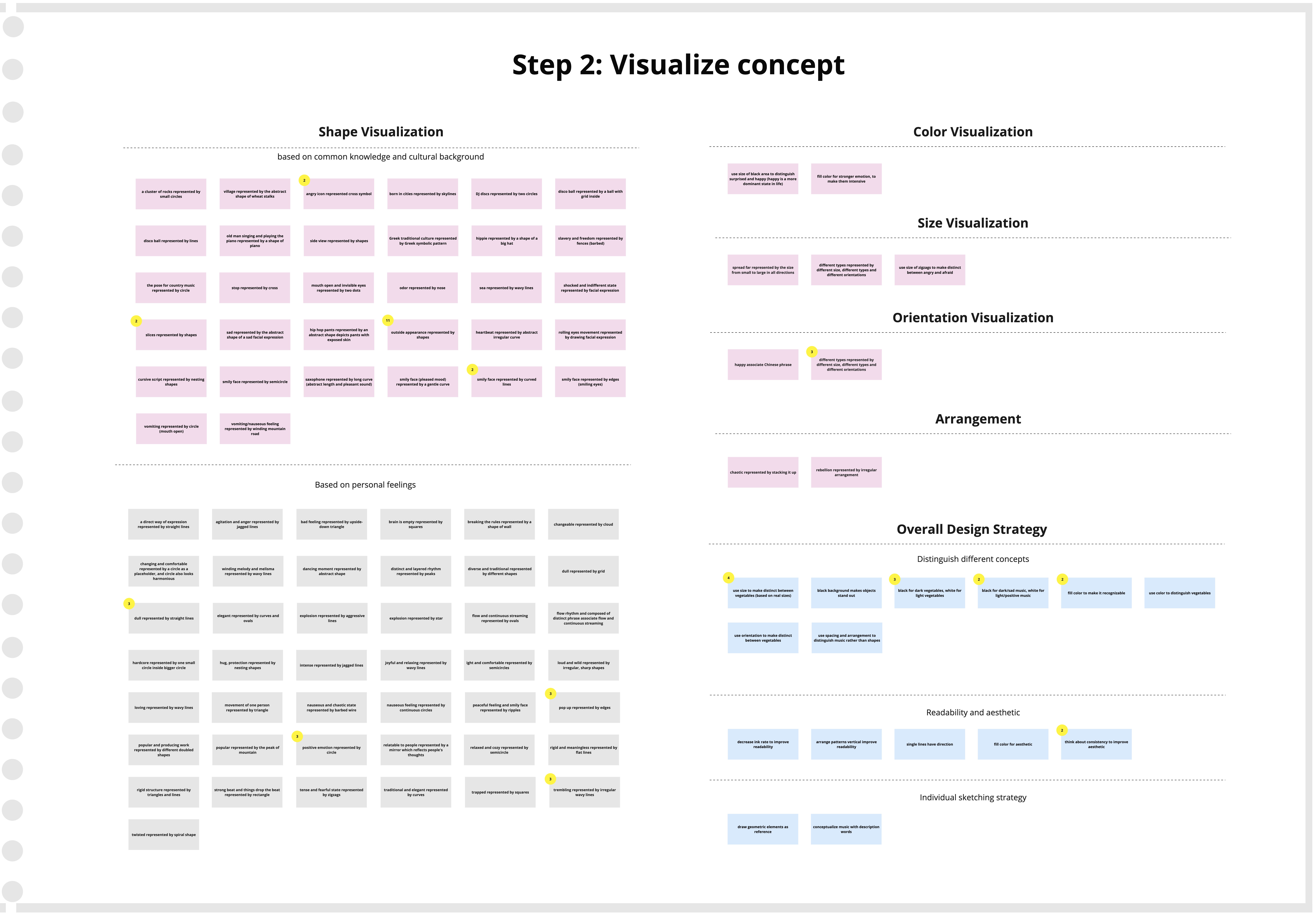}
    \caption{\hty{Design methodology}, version 1.}
    \label{fig:design-space-v1}
\end{figure}

\begin{figure}[h]
    \centering
    \includegraphics[width=1\linewidth]{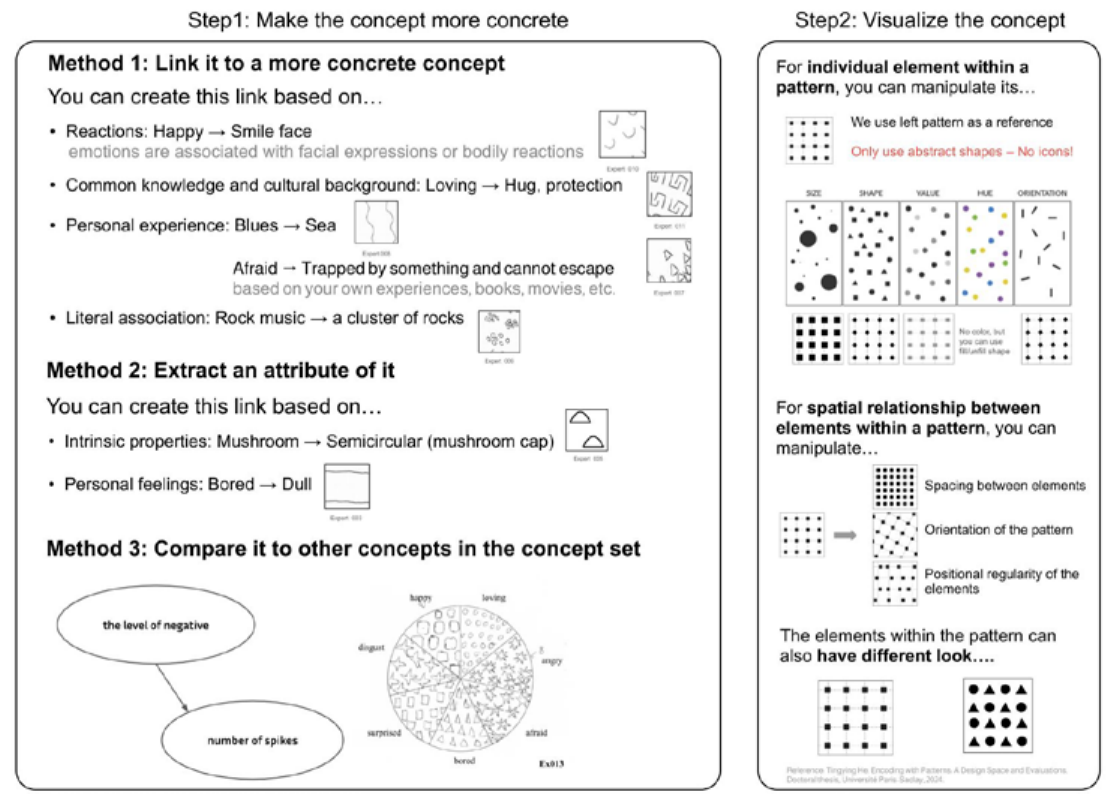}
    \caption{\hty{Design methodology}, version 2.}
    \label{fig:design-space-v2}
\end{figure}

In this section, we present the evolution of our \hty{design methodology} across three main versions: v1 (\autoref{fig:design-space-v1}), v2 (\autoref{fig:design-space-v2}), and v3 (\autoref{fig:design-space}). To be specific, we developed v1 immediately after analyzing the results of the design workshop with experts (\autoref{sec:workshop}), v2 is an iteration over v1 and is what we evaluated during the user study (\autoref{sec:evaluation}), and v3 is the final version which we had presented and discussed in detail in \autoref{sec:design-space}. \hty{Please note that this appendix \emph{\textbf{does not present the final design methodology}} (v3), which is already part of the main paper as \autoref{sec:design-space}.}

The differences between the versions only lie in the structure and terminology. Both v1 and v2 included all the \hty{approaches} we described in the final version, but they were poorly organized. In the final version, we reorganized these \hty{approaches} into a more logical structure and chose better words to describe them.

Below we outline the details of this evolution. First, we introduce v1 and explain its initial structure. Next, we describe the transition to v2 and explain how it corresponds to both v1 and v3. Finally, we discuss the transition to v3, focusing on the changes made from v2 and the reasoning behind them. Since v3 is already detailed in \autoref{sec:evaluation}, we do not explain it here again but instead concentrate on the modifications and improvements from v2 to v3.

\subsection{Initial version (v1)}

We created the initial version based on the open coding and qualitative analysis of the design strategies we collected from design experts. This version has a two-step structure similar to the final version, and a set of conceptual methods within each step. 

\textbf{Step 1: Conceptualizing abstract concepts.}
This step focused on making abstract concepts more concrete and included the following methods:
\begin{enumerate} 
\item Conceptual bridging: Linking the target concept to a more concrete one using: Psychological and physiological reactions/Personal experiences/Conceptual imagery/Common knowledge and cultural background/Literal associations. 
\item Attribute extraction: Identifying features of the target concept, categorized into: Intrinsic properties of the concept/Personal feelings to the concept.
\item Quantification: Unlike the previous two methods, which focus on designing for individual target concepts, this method treats the entire concept set as a whole. The designer transforms the categorical value set into an orderable value set. Since all concepts within the set share a common theme, the designer extracts a characteristic that is shared by all concepts and can be ordered or even quantified. Based on this characteristic, the designer arranges the concepts in a specific order. In the next step, the designer selects an ordered visual variable to represent these concepts.
\end{enumerate}

\textbf{Step 2: Visualizing the concept.}
In this step, designers transform the refined concept into an abstract pattern. They can use different visual variables to create the semantic association. We followed the framework of He \cite{He:2024:Encoding} to categorize the visual variables.

\subsection{Transition to v2}
When designing the evaluation workshop, we need to think about how to explain these methods to the non-experts participants, in this process we identified some minor structural and logical issues in v1. To address this, we introduced minor refinements in v2, focusing on terminology and structural clarity.

\textbf{Step 1.} We renamed this step to ``Make the concept more concrete'' with revised method categories: 
\begin{enumerate}
\item Link to a more concrete concept: Similar to ``conceptual bridging'' in v1, but with revised the categories what this link can be based on. The new categories under this method are as following:
``reactions''(refined from ``psychological and physiological reactions'' in v1, corresponding to ``Step 1: Human Reaction $\rightarrow$ Behavioral reaction'' in v3),
``common knowledge and cultural background'' (same in v1, corresponding to ``Step 1: Meaning $\rightarrow$ Shared understanding'' in v3),
``personal experience'' (we combined ``common knowledge and cultural background'' and ``conceptual imaginary'' in v1 to this category, corresponding to ``Step 1: Meaning $\rightarrow$ Personal experience'' in v3).
``literal association'' (same in v1 and ``Step 1: Literal association'' in v3).
\item Extract an attribute of the target concept. Same as attribute extraction method in v1.  Within this method,  ``intrinsic properties'' corresponding to ``Step 1: Features'' in v3, ``personal feelings'' corresponding to ``Step 1: Human reaction $\rightarrow$ Emotional reaction'' in v3. 
\item Compare the target concept to other concepts in the concept set. Same as the quantification method in v1. Corresponding to ``Step 2: Basic patterns $\rightarrow$ Ordinal scale'' in v3. We explain why we moved this method from Step 1 to Step 2 in next subsection. 
\end{enumerate}

\textbf{Step 2.} In v2, for Step 2, we retained the structure and methods from v1 but just refined the language and provided examples for clarity.

\subsection{Transition to v3 (the final version)}
After the evaluation workshop (\autoref{sec:evaluation}), when writing paper, one author noted that the structure of \hty{design methodology} (v2) still lacked some logical coherence. This author reanalyzed the interview data, experimented with reorganizing the methods, and proposed a revised organization based on v2. Through collaborative discussions we then finally developed v3. We made the following changes from v2 to v3:
\begin{enumerate}
% \item We renamed conceptual ``methods'' to conceptual ``pathways.'' to emphasize their role as inspiration on how people can think to generate design ideas, rather than as prescriptive guidelines.
\item) We refined the two step names as ``Step 1: Identifying the content to be visualized'' and ``Step 2: Visualizing the refined concept as a pattern.'' Because these new names can better and precisely describe what design experts exactly did during each step. 
\item We reorganized Step 1. 
\begin{enumerate}
\item Since all specific \hty{approaches} under Step 1 involve conceptual bridging or linking, we no longer use ``concept bridging'' or ``link the target concept to a more concrete concept'' as a specific \hty{approach} name, as was done in v1 and v2. Instead, we interpret this specific \hty{approach} as identifying the content to visualize based on the meaning of the target concept. 
\item ``literal association'' was classified under conceptual bridging method before, but we identified ``literal association'' as distinct from connections based on the explicit meaning of the target concept and, therefore, separated it into a new \hty{approach}, just named ``literal association''.
\item In v2, ``personal feelings'' and ``reactions'' were categorized under separate methods. However, since both of them stem from human reactions, we combined them into a single new \hty{approach} called ``Human reaction'' in v3.
\item The last method (called ``quantification'' in v1 and ``compare it to other concepts in the concept set'' in v2) in fact describe this process: extract a common feature of concepts from the concept set (so it should belong to ``Feature'' \hty{approach}), order them and use a visual variable suitable for ordinal data to visualize them (which is more about visualizing, should be in Step 2). Therefore, we consider extract common feature as part of ``Feature'' \hty{approach} and do not list it as a separate \hty{approach} in Step 1. We also add a ``Ordinal scale'' \hty{approach} in Step 2 for this idea of creating an ordinal scale.
\end{enumerate}
\item The visual variables used in Step 2 remained unchanged from v1 and v2. However, upon reviewing the pattern designs again, we observed that it is important to point out that some patterns have more basic look, resembled those more commonly found in current charts (characterized by repetitive shapes), while others were more complex. We further categorized these patterns based on their use or not use of relational variables into ``complex patterns'' and ``basic patterns''
\end{enumerate}

This revised version of v3 addresses the structural and terminological inconsistencies identified in earlier iterations, providing a more coherent and precise framework for the \hty{design methodology}. We evaluated v2, which included the same \hty{approach}s as v3 but with a less logical organization. As we show in \autoref{sec:evaluation}, v2 already demonstrated its effectiveness. Therefore, we have reason to argue that with its improved structure and terminology, v3 should perform even better.

\section{All designs generated by the visualization experts in the design workshops}
\label{appendix:all-designs-workshop1}

In \hyperref[fig:ex001-ve]{Figures~}\ref{fig:ex001-ve}--\ref{fig:ex013-emo} we show the 39 designs we collected from 13 visualization designers in the design workshop. The collection comprised 13 pie charts designed for vegetable concept set (\hyperref[fig:ex001-ve]{Figures~}\ref{fig:ex001-ve}--\ref{fig:ex013-ve}), 13 pie charts designed for music concept set (\hyperref[fig:ex001-mus]{Figures~}\ref{fig:ex001-mus}--\ref{fig:ex013-mus}) and 13 pie charts designed for emotion concept set (\hyperref[fig:ex001-emo]{Figures~}\ref{fig:ex001-emo}--\ref{fig:ex013-emo}).

The original sketching template did not provide labels for each concept. To improve readability, we added labels to the sketches of participants who did not include labels. All design results of in-person studies were drawn on A4 paper and scanned to generate PDF files. 
For remote participants who were unable to print the design template, we allowed the use of a digital version of the template on their iPad (Ex005) or hand-drawing the blank pie chart on A4 papers (Ex006). Participants were instructed to ensure that the charts matched the dimensions of the printed version of the design template.

% Nonetheless, you can find the original sketches in our OSF repository at (\href{https://osf.io/9h5nd}{\texttt{osf\discretionary{}{.}{.}io\discretionary{/}{}{/}9h5nd}}) and you can look at them with a browser such as Chrome, Microsoft Edge, or Firefox.

% We include all these images (here and also the images in the main paper such as \todo(the teaser (\autoref{fig:teaser}) and Tables \ref{tab:exp2-bar-geo}--\ref{tab:exp2-map-icon})) as pixel images on purpose because the SVG vector version relies on tiled texture samples, which---when converted to PDF for the inclusion in the paper---lead to unfortunate errors in the display in all PDF readers we tested. Likely this effect is due to numeric issues that affect the exact positions where the texture tiles meet. Nonetheless, you can find the original SVG images in our OSF repository at \href{https://osf.io/n5zut/}{\texttt{osf.io/n5zut}} and you can look at them with a browser such as Chrome, Microsoft Edge, or Firefox.

\section{All designs generated by the authors in the pilot studies of our evaluation workshops}
\label{appendix:all-designs-workshop2-pilot-authors}

In \hyperref[fig:pil1-con]{Figures~}\ref{fig:pil1-con}--\ref{fig:pil4-ab} we show the 8 designs we collected from 4 authors in the evaluation workshop pilot. The collection comprised 4 pie charts designed for ball sports concept set (\hyperref[fig:pil1-con]{Figures~}\ref{fig:pil1-con}--\ref{fig:pil4-con}) and 4 pie charts designed for personality concept set (\hyperref[fig:pil1-ab]{Figures~}\ref{fig:pil1-ab}--\ref{fig:pil4-ab}). For remote participants who were unable to print the design template, they drew the template by hand on A4 paper.

\section{All designs generated by the non-expert participants in the pilot studies of our evaluation workshops}
\label{appendix:all-designs-workshop2-pilot-participants}

In \hyperref[fig:sz1-con]{Figures~}\ref{fig:sz1-con}--\ref{fig:sz4-ab} we show the 8 designs we collected from 4 non-expert participants in the evaluation workshop pilot. The collection comprised 4 pie charts designed for ball sports concept set (\hyperref[fig:sz1-con]{Figures~}\ref{fig:sz1-con}--\ref{fig:sz4-con}) and 4 pie charts designed for personality concept set (\hyperref[fig:sz1-ab]{Figures~}\ref{fig:sz1-ab}--\ref{fig:sz4-ab}). 

\section{All designs generated by the non-expert participants in our evaluation workshops}
\label{appendix:all-designs-workshop2-formal-experiment}
In \hyperref[fig:p1-con]{Figures~}\ref{fig:p1-con}--\ref{fig:p12-ab} we show the 8 designs we collected from 12 non-expert participants in the evaluation workshop. The collection comprised 12 pie charts designed for ball sports concept set (\hyperref[fig:p1-con]{Figures~}\ref{fig:p1-con}--\ref{fig:p12-con}) and 12 pie charts designed for personality concept set (\hyperref[fig:p1-ab]{Figures~}\ref{fig:p1-ab}--\ref{fig:p12-ab}). 

\section*{Images license/copyright}
\noindent We as authors state that all of our own figures in this appendix (\ie, \hty{the screenshots
% \marginpar{\todo{Please make sure to upload all the new figures, including \autoref{fig:real-world-example_ball-games}--\ref{fig:real-world-example_quality-metrics}, and ensure that all figures are correctly named on OSF} \hty{will do it}} 
in \hyperref[fig:real-world-example_ball-games]{Figures~}\ref{fig:real-world-example_ball-games}--\ref{fig:real-world-example_emotions}, our own depiction of the data from Figure 5 in Awan et al.'s \cite{Awan:2015:Efficient} paper in our \autoref{fig:real-world-example_quality-metrics}, as well as} \hyperref[fig:design-space-v1]{Figures~}\ref{fig:design-space-v1} and \ref{fig:design-space-v2} and our own pattern designs in \hyperref[fig:pil1-con]{Figures~}\ref{fig:pil1-con}--\ref{fig:pil4-ab}) are and remain under our own personal copyright, with the permission to be used here. We also make them available under the \href{https://creativecommons.org/licenses/by/4.0/}{Creative Commons At\-tri\-bu\-tion 4.0 International (\ccLogo\,\ccAttribution\ \mbox{CC BY 4.0})} license and share them at \href{https://osf.io/9h5nd/}{\texttt{osf\discretionary{}{.}{.}io\discretionary{/}{}{/}9h5nd}}. \autoref{fig:BG1} is \textcopyright\ He et al. \cite{He:2024:DCB}, which we use under the \href{https://creativecommons.org/licenses/by/4.0/}{Creative Commons At\-tri\-bu\-tion 4.0 International (\ccLogo\,\ccAttribution\ \mbox{CC BY 4.0})} license. All remaining figures, \ie, \hyperref[fig:ex001-ve]{Figures~}\ref{fig:ex001-ve}--\ref{fig:ex013-emo} and \ref{fig:sz1-con}--\ref{fig:p12-ab}, remain under the copyright of their respective authors (\ie, the respective study participants), with the permission to be used here.

\newlength{\appendixfigurewidth}
\setlength{\appendixfigurewidth}{.85\columnwidth}

%vegetable concept set
\begin{figure}[t]
    \centering
    \includegraphics[width=\appendixfigurewidth]{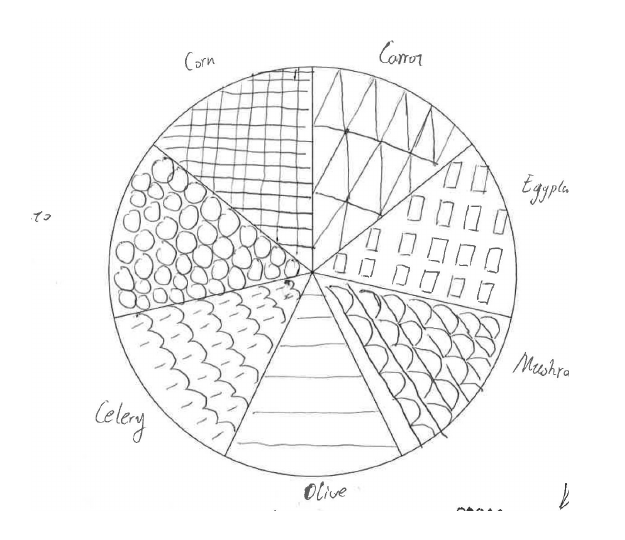}
    \caption{A semantically-resonant pattern design (Ex1) for vegetable concept set collected in our design workshop.}
    \label{fig:ex001-ve}
\end{figure}

\begin{figure}[t]
    \centering
    \includegraphics[width=\appendixfigurewidth]{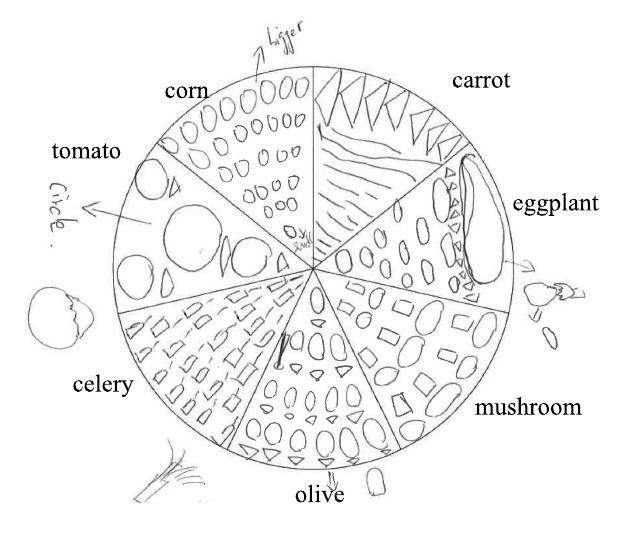}
    \caption{A semantically-resonant pattern design (Ex2) for vegetable concept set collected in our design workshop.}
    \label{fig:ex002-ve}
\end{figure}

\begin{figure}[t]
    \centering
    \includegraphics[width=\appendixfigurewidth]{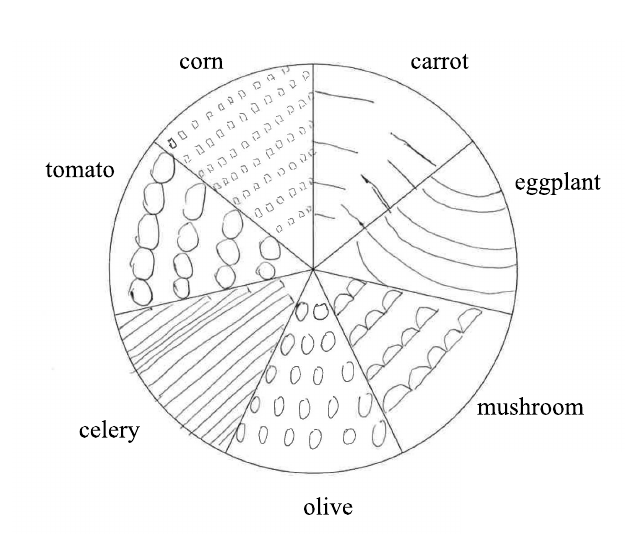}
    \caption{A semantically-resonant pattern design (Ex3) for vegetable concept set collected in our design workshop.}
    \label{fig:ex003-ve}
\end{figure}

\begin{figure}[t]
    \centering
    \includegraphics[width=\appendixfigurewidth]{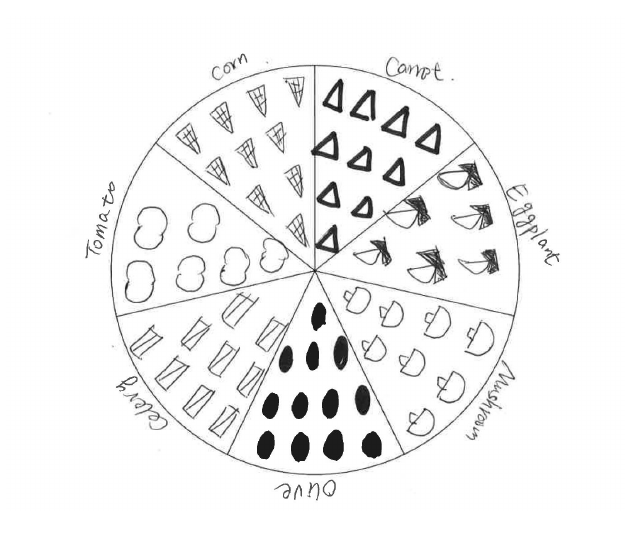}
    \caption{A semantically-resonant pattern design (Ex4) for vegetable concept set collected in our design workshop.}
    \label{fig:ex004-ve}
\end{figure}

\begin{figure}[t]
    \centering
    \includegraphics[width=\appendixfigurewidth]{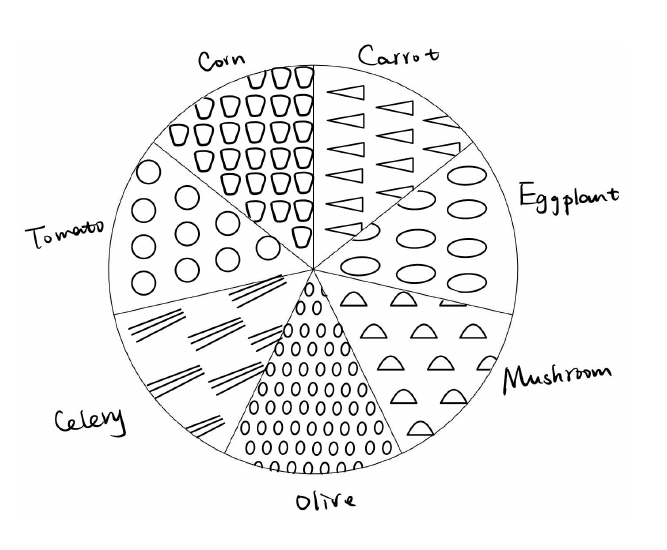}
    \caption{A semantically-resonant pattern design (Ex5) for vegetable concept set collected in our design workshop.}
    \label{fig:ex005-ve}
\end{figure}

\begin{figure}[t]
    \centering
    \includegraphics[width=\appendixfigurewidth]{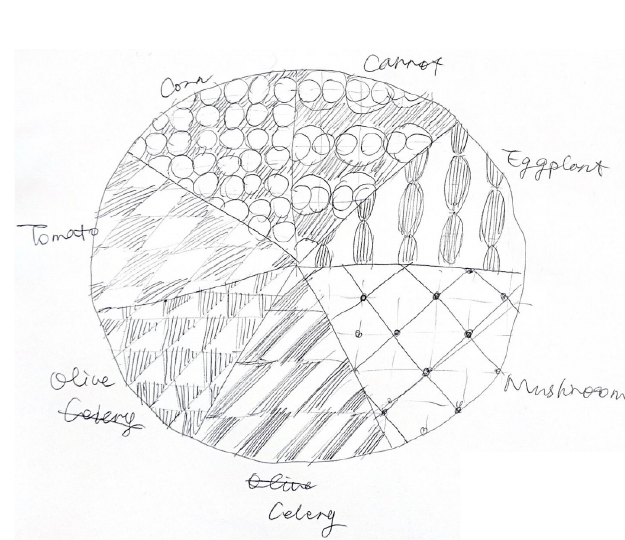}
    \caption{A semantically-resonant pattern design (Ex6) for vegetable concept set collected in our design workshop.}
    \label{fig:ex006-ve}
\end{figure}

\begin{figure}[t]
    \centering
    \includegraphics[width=\appendixfigurewidth]{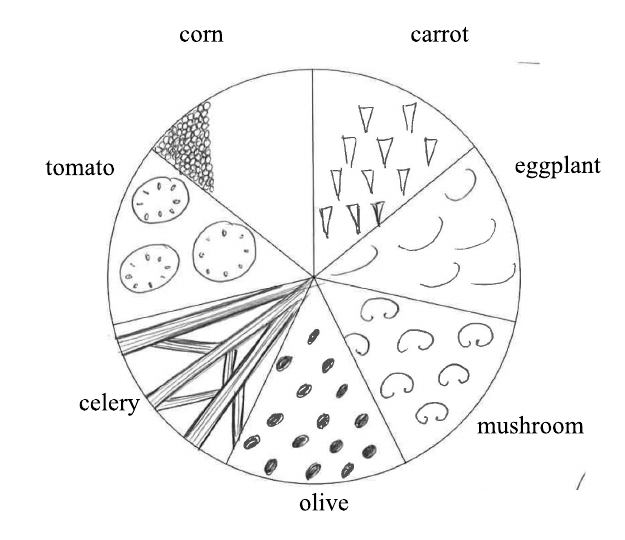}
    \caption{A semantically-resonant pattern design (Ex7) for vegetable concept set collected in our design workshop.}
    \label{fig:ex007-ve}
\end{figure}

\begin{figure}[t]
    \centering
    \includegraphics[width=\appendixfigurewidth]{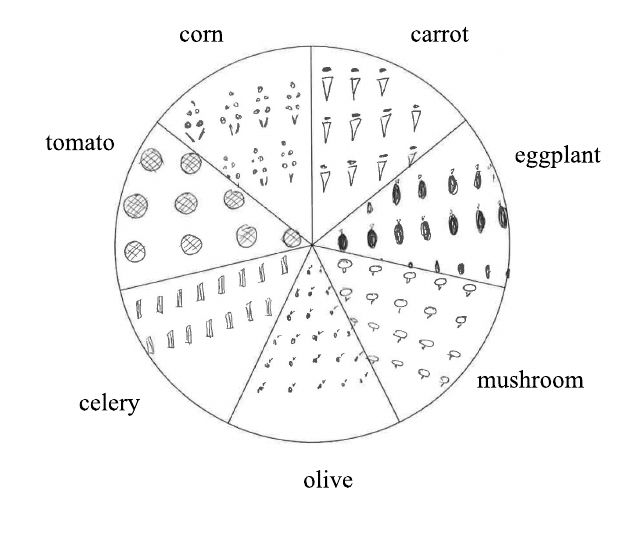}
    \caption{A semantically-resonant pattern design (Ex8) for vegetable concept set collected in our design workshop.}
    \label{fig:ex008-ve}
\end{figure}

\begin{figure}[t]
    \centering
    \includegraphics[width=\appendixfigurewidth]{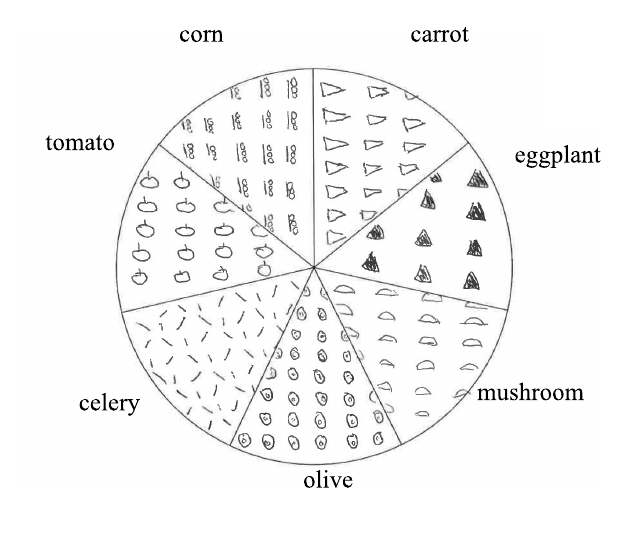}
    \caption{A semantically-resonant pattern design (Ex9) for vegetable concept set collected in our design workshop.}
    \label{fig:ex009-ve}
\end{figure}

\begin{figure}[t]
    \centering
    \includegraphics[width=\appendixfigurewidth]{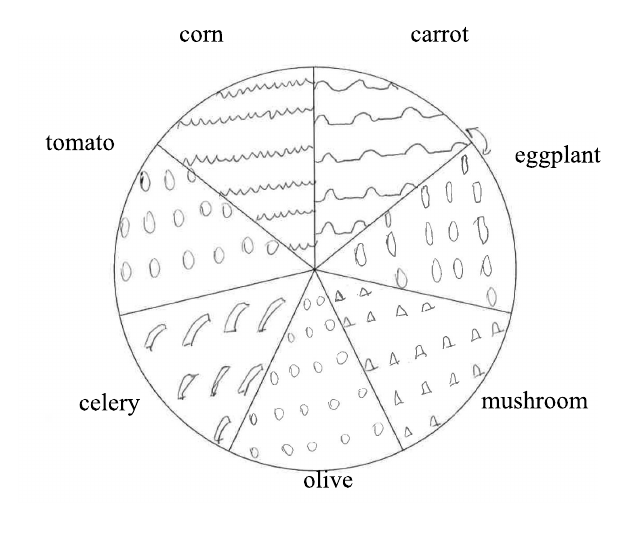}
    \caption{A semantically-resonant pattern design (Ex10) for vegetable concept set collected in our design workshop.}
    \label{fig:ex010-ve}
\end{figure}

\begin{figure}[t]
    \centering
    \includegraphics[width=\appendixfigurewidth]{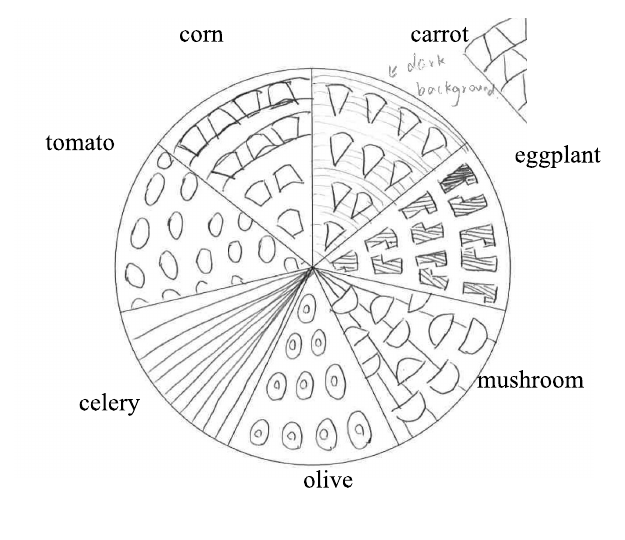}
    \caption{A semantically-resonant pattern design (Ex11) for vegetable concept set collected in our design workshop.}
    \label{fig:ex011-ve}
\end{figure}

\begin{figure}[t]
    \centering
    \includegraphics[width=\appendixfigurewidth]{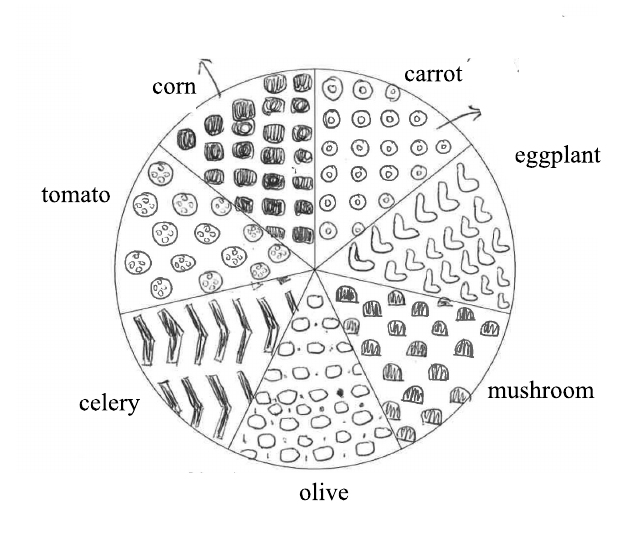}
    \caption{A semantically-resonant pattern design (Ex12) for vegetable concept set collected in our design workshop.}
    \label{fig:ex012-ve}
\end{figure}

\begin{figure}[t]
    \centering
    \includegraphics[width=\appendixfigurewidth]{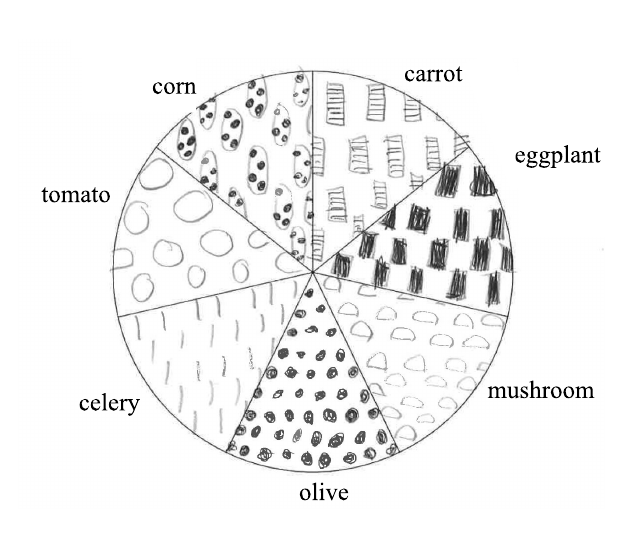}
    \caption{A semantically-resonant pattern design (Ex13) for vegetable concept set collected in our design workshop.}
    \label{fig:ex013-ve}
\end{figure}

%music concept set
\begin{figure}[t]
    \centering
    \includegraphics[width=\appendixfigurewidth]{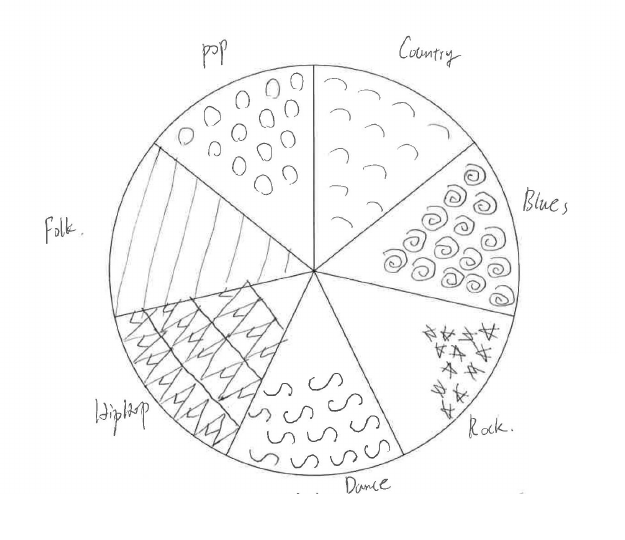}
    \caption{A semantically-resonant pattern design (Ex1) for music concept set collected in our design workshop.}
    \label{fig:ex001-mus}
\end{figure}

\begin{figure}[t]
    \centering
    \includegraphics[width=\appendixfigurewidth]{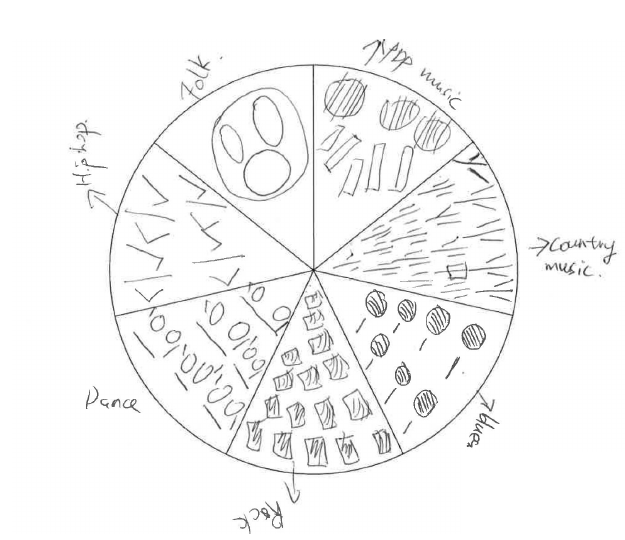}
    \caption{A semantically-resonant pattern design (Ex2) for music concept set collected in our design workshop.}
    \label{fig:ex002-mus}
\end{figure}

\begin{figure}[t]
    \centering
    \includegraphics[width=\appendixfigurewidth]{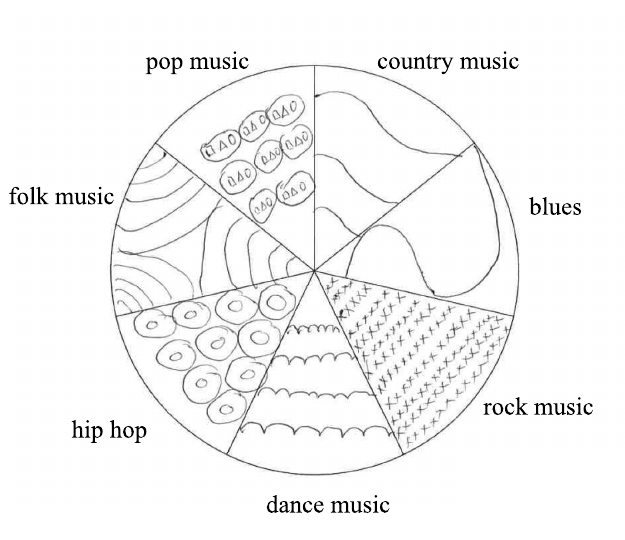}
    \caption{A semantically-resonant pattern design (Ex3) for music concept set collected in our design workshop.}
    \label{fig:ex003-mus}
\end{figure}

\begin{figure}[t]
    \centering
    \includegraphics[width=\appendixfigurewidth]{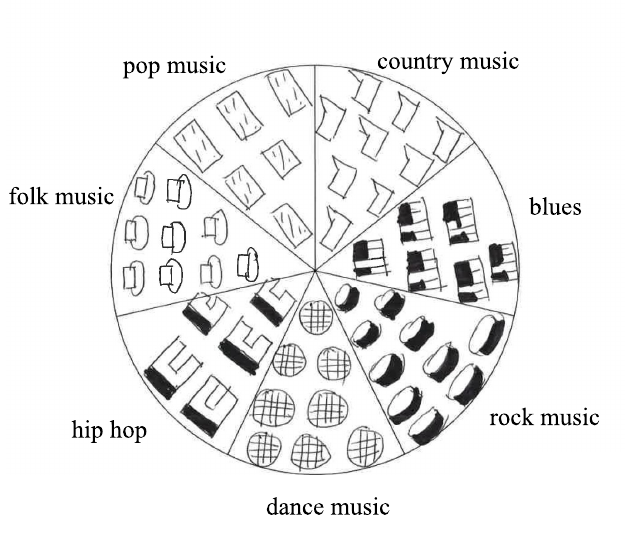}
    \caption{A semantically-resonant pattern design (Ex4) for music concept set collected in our design workshop.}
    \label{fig:ex004-mus}
\end{figure}

\begin{figure}[t]
    \centering
    \includegraphics[width=\appendixfigurewidth]{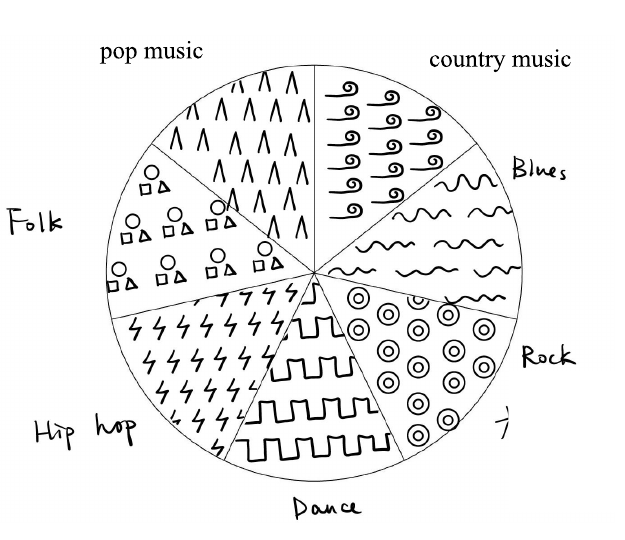}
    \caption{A semantically-resonant pattern design (Ex5) for music concept set collected in our design workshop.}
    \label{fig:ex005-mus}
\end{figure}

\begin{figure}[t]
    \centering
    \includegraphics[width=\appendixfigurewidth]{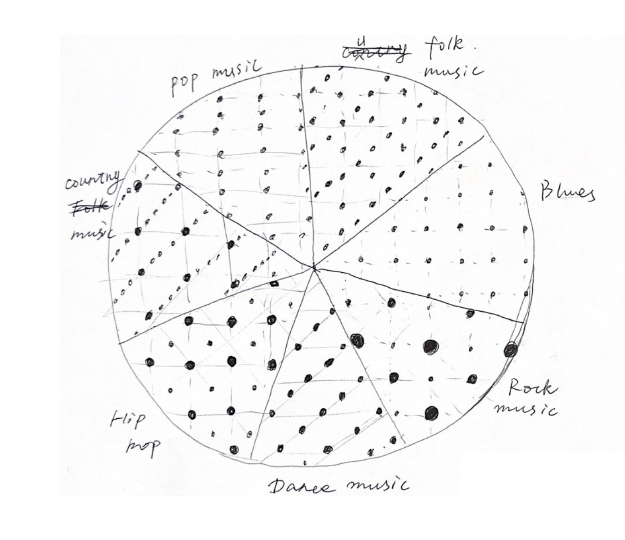}
    \caption{A semantically-resonant pattern design (Ex6) for music concept set collected in our design workshop.}
    \label{fig:ex006-mus}
\end{figure}

\begin{figure}[t]
    \centering
    \includegraphics[width=\appendixfigurewidth]{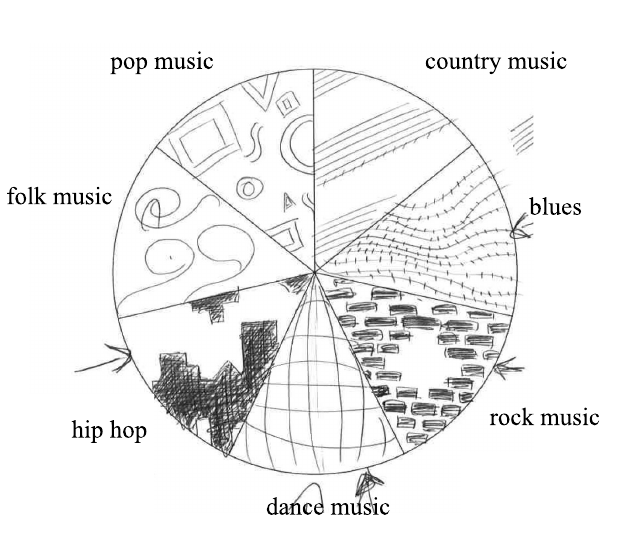}
    \caption{A semantically-resonant pattern design (Ex7) for music concept set collected in our design workshop.}
    \label{fig:ex007-mus}
\end{figure}

\begin{figure}[t]
    \centering
    \includegraphics[width=\appendixfigurewidth]{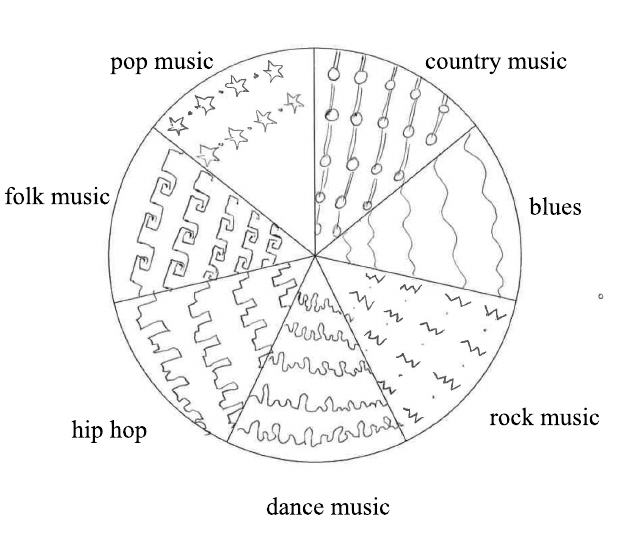}
    \caption{A semantically-resonant pattern design (Ex8) for music concept set collected in our design workshop.}
    \label{fig:ex008-mus}
\end{figure}

\begin{figure}[t]
    \centering
    \includegraphics[width=\appendixfigurewidth]{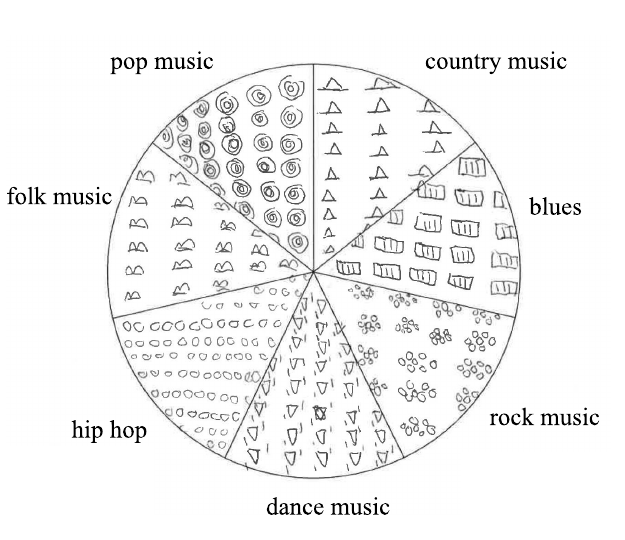}
    \caption{A semantically-resonant pattern design (Ex9) for music concept set collected in our design workshop.}
    \label{fig:ex009-mus}
\end{figure}

\begin{figure}[t]
    \centering
    \includegraphics[width=\appendixfigurewidth]{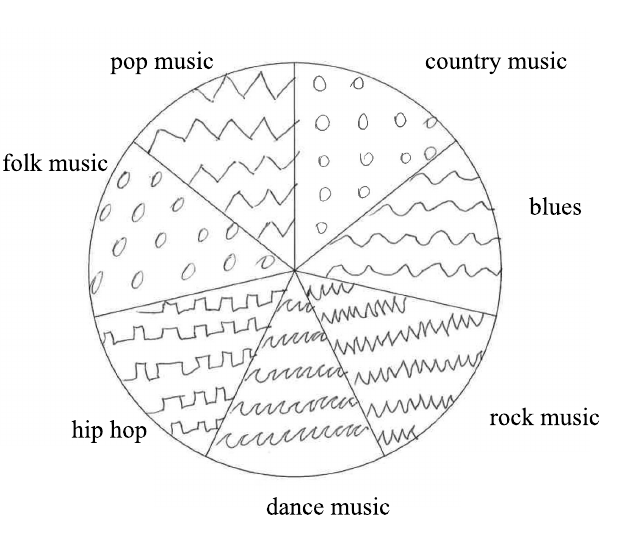}
    \caption{A semantically-resonant pattern design (Ex10) for music concept set collected in our design workshop.}
    \label{fig:ex010-mus}
\end{figure}

\begin{figure}[t]
    \centering
    \includegraphics[width=\appendixfigurewidth]{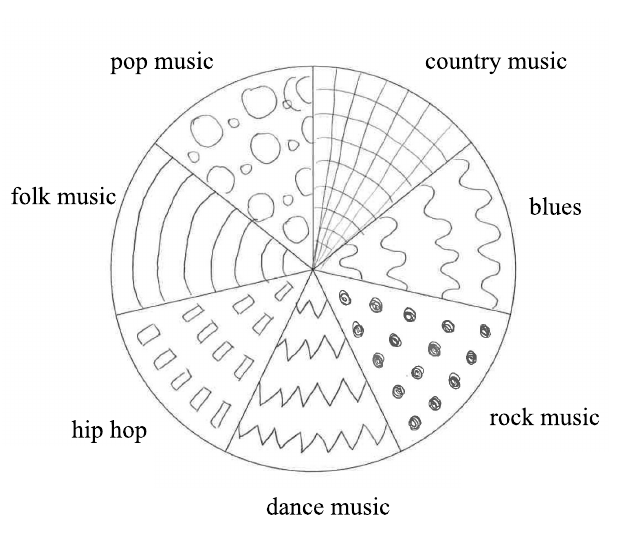}
    \caption{A semantically-resonant pattern design (Ex11) for music concept set collected in our design workshop.}
    \label{fig:ex011-mus}
\end{figure}

\begin{figure}[t]
    \centering
    \includegraphics[width=\appendixfigurewidth]{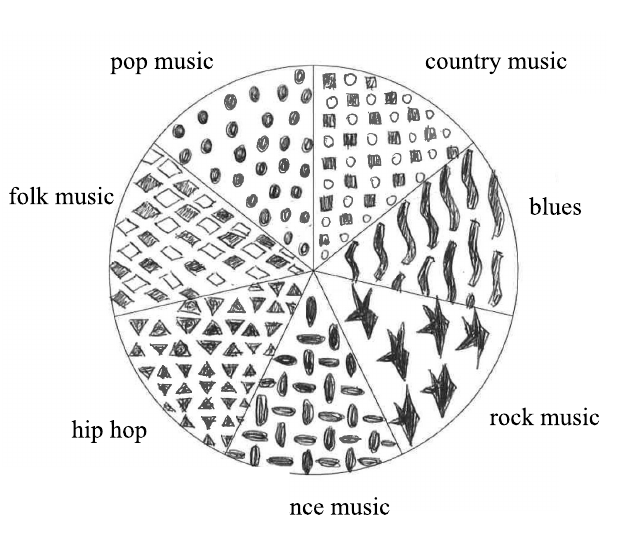}
    \caption{A semantically-resonant pattern design (Ex12) for music concept set collected in our design workshop.}
    \label{fig:ex012-mus}
\end{figure}

\begin{figure}[t]
    \centering
    \includegraphics[width=\appendixfigurewidth]{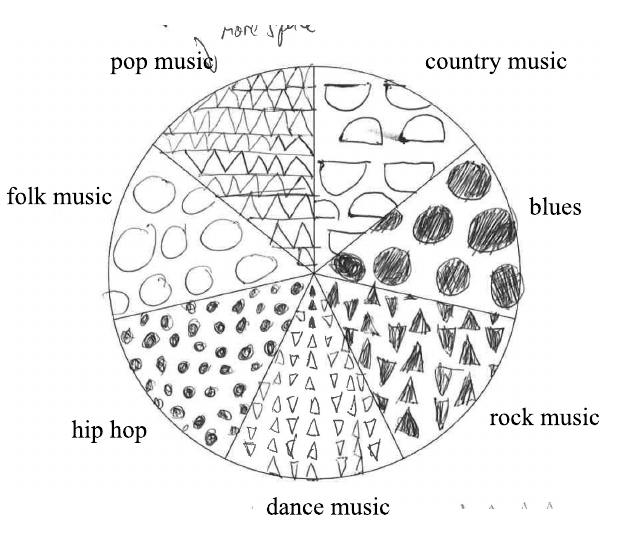}
    \caption{A semantically-resonant pattern design (Ex13) for music concept set collected in our design workshop.}
    \label{fig:ex013-mus}
\end{figure}

%emotion concept set
\begin{figure}[t]
    \centering
    \includegraphics[width=\appendixfigurewidth]{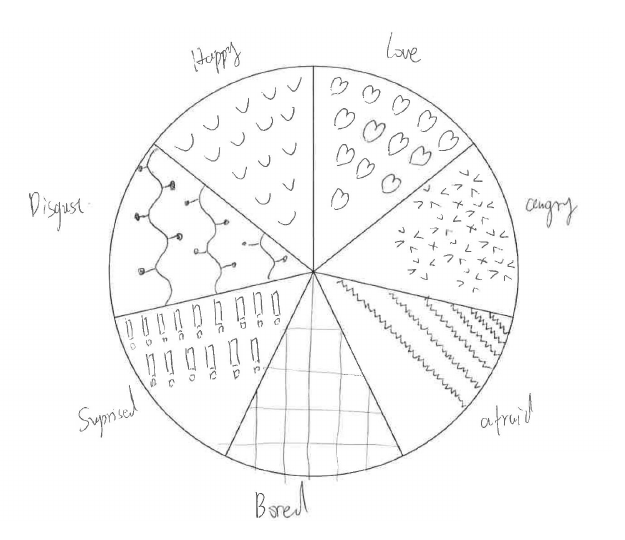}
    \caption{A semantically-resonant pattern design (Ex1) for emotion concept set collected in our design workshop.}
    \label{fig:ex001-emo}
\end{figure}

\begin{figure}[t]
    \centering
    \includegraphics[width=\appendixfigurewidth]{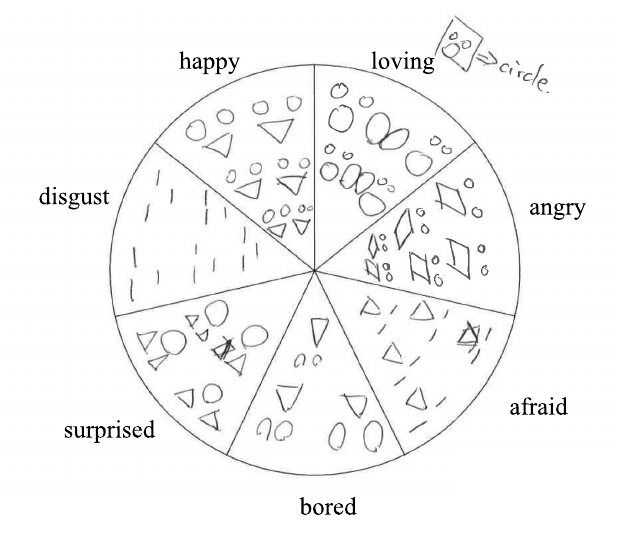}
    \caption{A semantically-resonant pattern design (Ex2) for emotion concept set collected in our design workshop.}
    \label{fig:ex002-emo}
\end{figure}

\begin{figure}[t]
    \centering
    \includegraphics[width=\appendixfigurewidth]{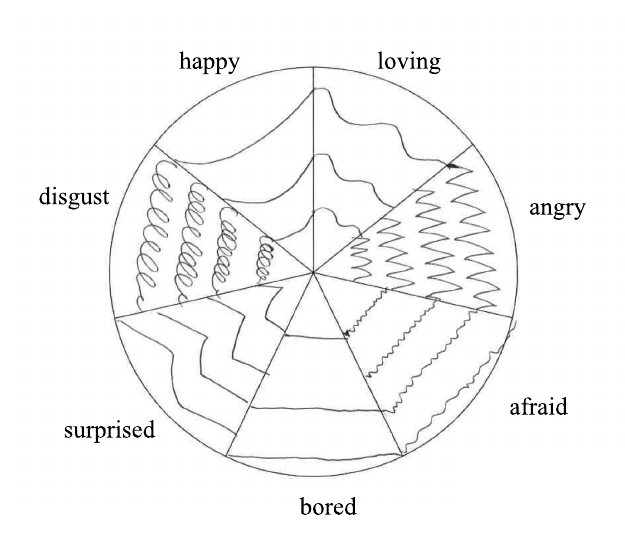}
    \caption{A semantically-resonant pattern design (Ex3) for emotion concept set collected in our design workshop.}
    \label{fig:ex003-emo}
\end{figure}

\begin{figure}[t]
    \centering
    \includegraphics[width=\appendixfigurewidth]{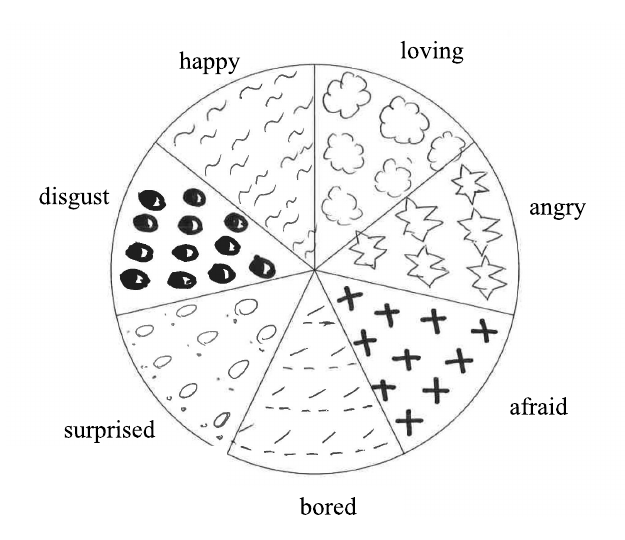}
    \caption{A semantically-resonant pattern design (Ex4) for emotion concept set collected in our design workshop.}
    \label{fig:ex004-emo}
\end{figure}

\begin{figure}[t]
    \centering
    \includegraphics[width=\appendixfigurewidth]{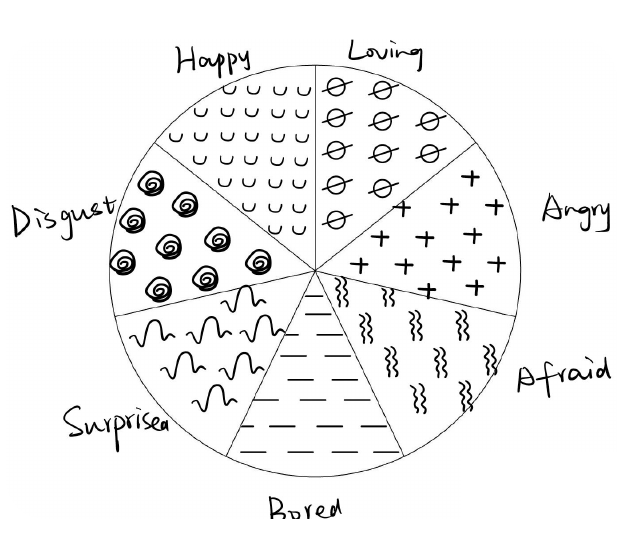}
    \caption{A semantically-resonant pattern design (Ex5) for emotion concept set collected in our design workshop.}
    \label{fig:ex005-emo}
\end{figure}

\begin{figure}[t]
    \centering
    \includegraphics[width=\appendixfigurewidth]{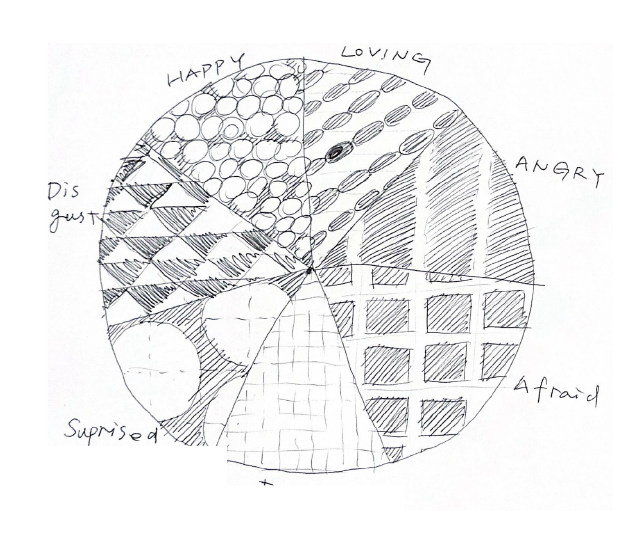}
    \caption{A semantically-resonant pattern design (Ex6) for emotion concept set collected in our design workshop.}
    \label{fig:ex006-emo}
\end{figure}

\begin{figure}[t]
    \centering
    \includegraphics[width=\appendixfigurewidth]{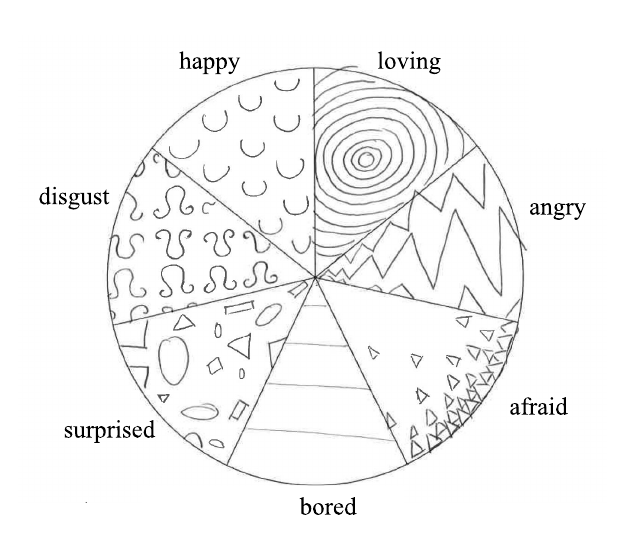}
    \caption{A semantically-resonant pattern design (Ex7) for emotion concept set collected in our design workshop.}
    \label{fig:ex007-emo}
\end{figure}

\begin{figure}[t]
    \centering
    \includegraphics[width=\appendixfigurewidth]{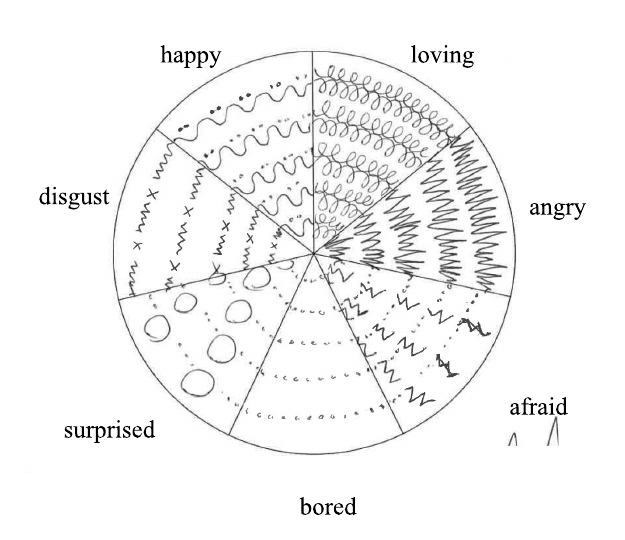}
    \caption{A semantically-resonant pattern design (Ex8) for emotion concept set collected in our design workshop.}
    \label{fig:ex008-emo}
\end{figure}

\begin{figure}[t]
    \centering
    \includegraphics[width=\appendixfigurewidth]{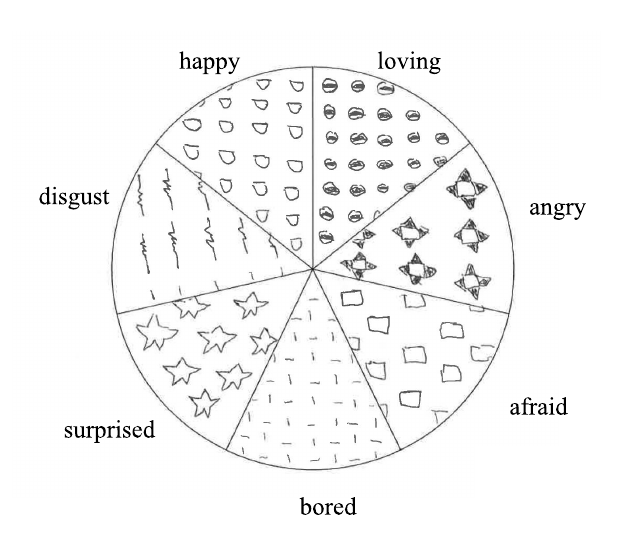}
    \caption{A semantically-resonant pattern design (Ex9) for emotion concept set collected in our design workshop.}
    \label{fig:ex009-emo}
\end{figure}

%\clearpage
%
\begin{figure}[t]
    \centering
    \includegraphics[width=\appendixfigurewidth]{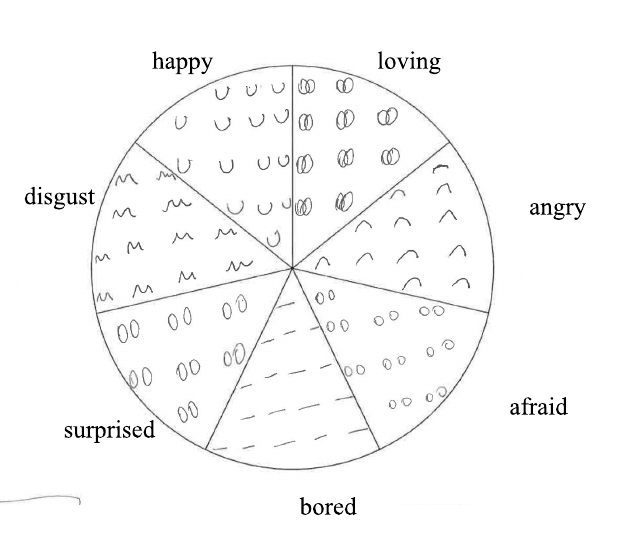}
    \caption{A semantically-resonant pattern design (Ex10) for emotion concept set collected in our design workshop.}
    \label{fig:ex010-emo}
\end{figure}

\begin{figure}[t]
    \centering
    \includegraphics[width=\appendixfigurewidth]{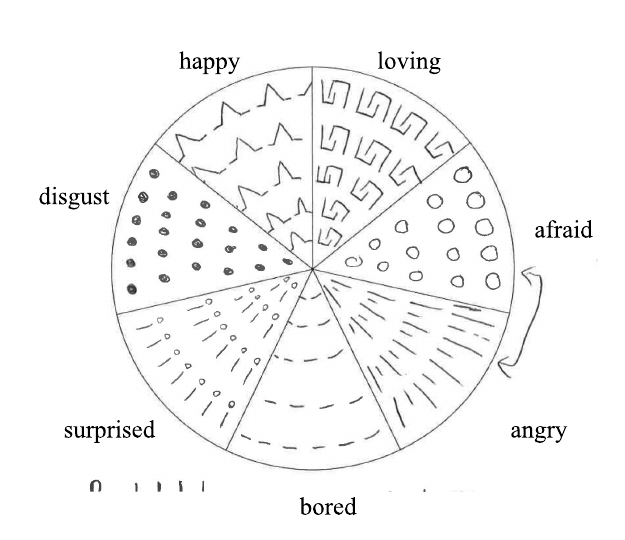}
    \caption{A semantically-resonant pattern design (Ex11) for emotion concept set collected in our design workshop.}
    \label{fig:ex011-emo}
\end{figure}

\begin{figure}
    \centering
    \includegraphics[width=\appendixfigurewidth]{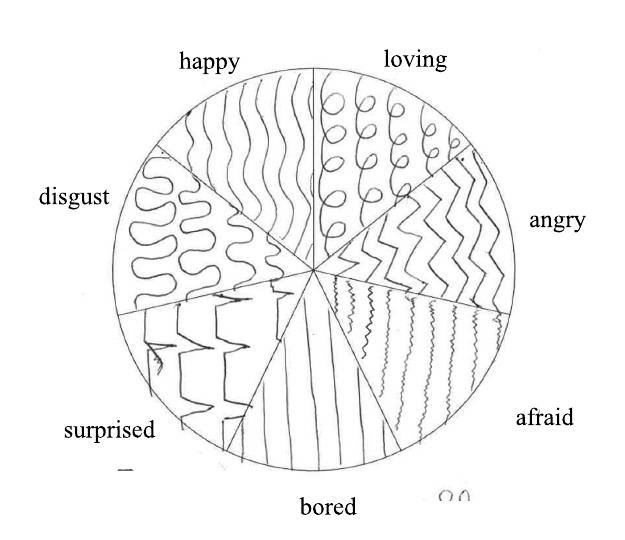}
    \caption{A semantically-resonant pattern design (Ex12) for emotion concept set collected in our design workshop.}
    \label{fig:ex012-emo}
\end{figure}

\begin{figure}
    \centering
    \includegraphics[width=\appendixfigurewidth]{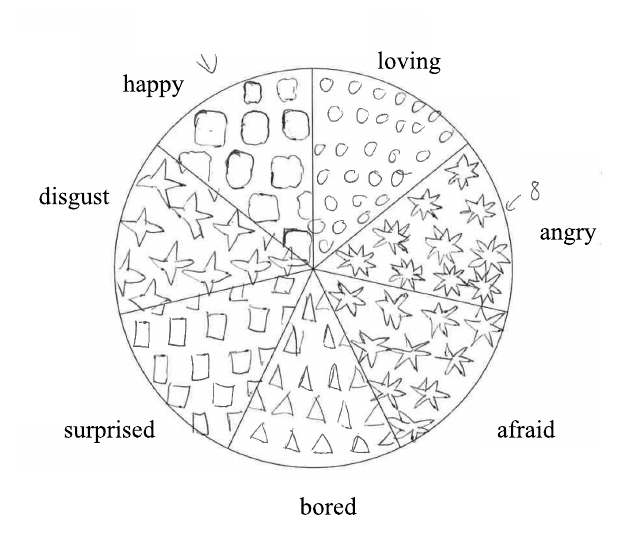}
    \caption{A semantically-resonant pattern design (Ex13) for emotion concept set collected in our design workshop.}
    \label{fig:ex013-emo}
\end{figure}

%ball sports concept set
\begin{figure}[t]
    \centering
    \includegraphics[width=\appendixfigurewidth]{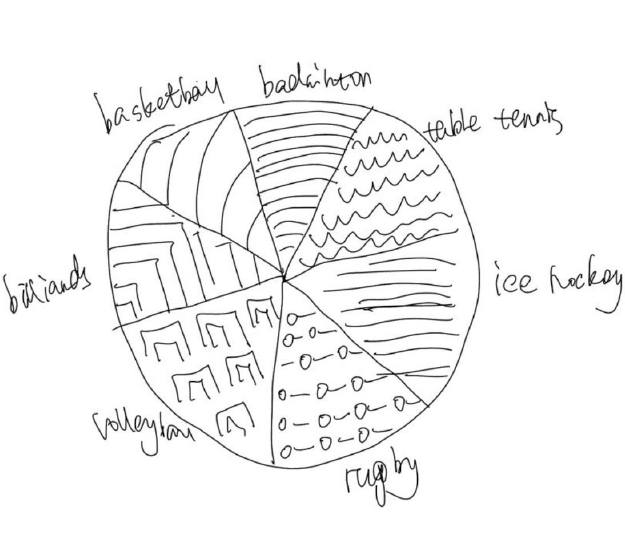}
    \caption{A semantically-resonant pattern design by the author (Au1) for ball sports concept set collected in our evaluation workshop pilot.}
    \label{fig:pil1-con}
\end{figure}

\clearpage

\begin{figure}[t]
    \centering
    \includegraphics[width=\appendixfigurewidth]{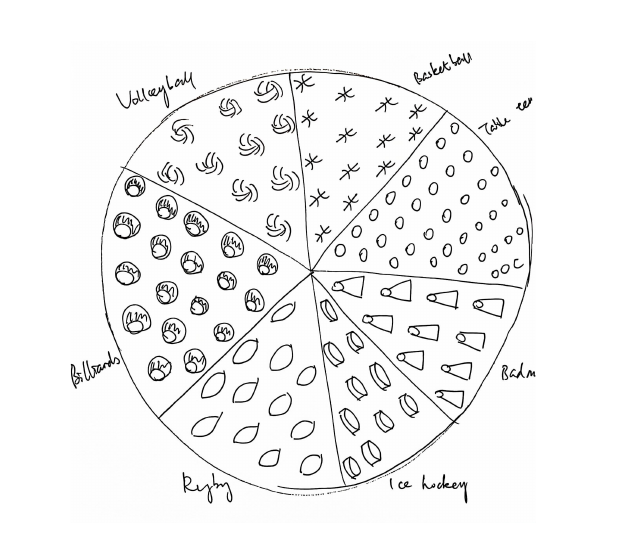}
    \caption{A semantically-resonant pattern design by the author (Au2) for ball sports concept set collected in our evaluation workshop pilot.}
    \label{fig:pil2-con}
\end{figure}

\begin{figure}[t]
    \centering
    \includegraphics[width=\appendixfigurewidth]{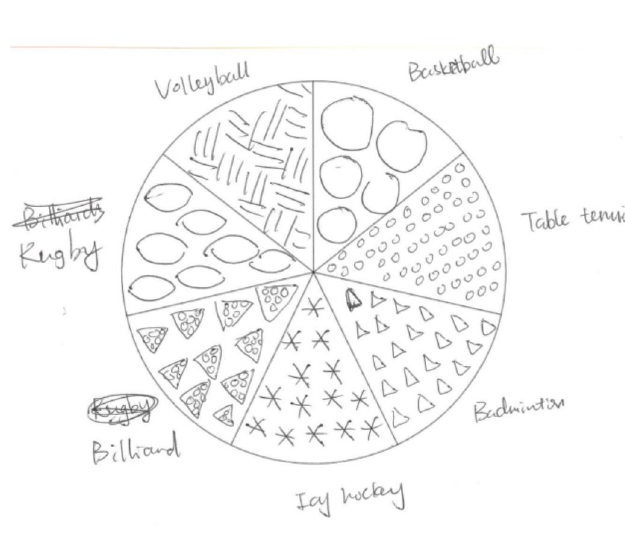}
    \caption{A semantically-resonant pattern design by the author (Au3) for ball sports concept set collected in our evaluation workshop pilot.}
    \label{fig:pil3-con}
\end{figure}

\begin{figure}[t]
    \centering
		\includegraphics[width=\appendixfigurewidth]{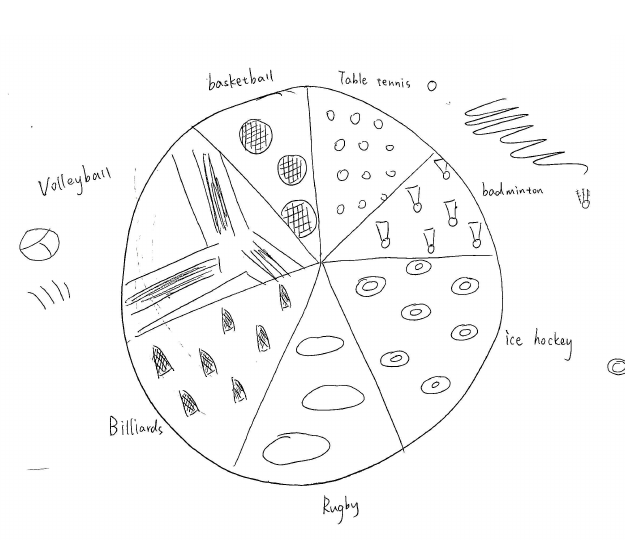}
    \caption{A semantically-resonant pattern design by the author (Au4) for ball sports concept set collected in our evaluation workshop pilot.}
    \label{fig:pil4-con}
\end{figure}

%personality concept set
\begin{figure}[t]
    \centering
		\includegraphics[width=\appendixfigurewidth]{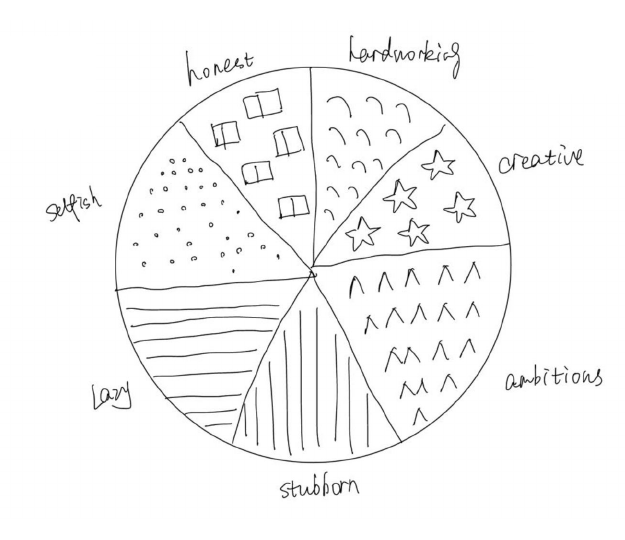}
    \caption{A semantically-resonant pattern design by the author (Au1) for personality concept set collected in our evaluation workshop pilot.}
    \label{fig:pil1-ab}
\end{figure}

\begin{figure}[t]
    \centering
    \includegraphics[width=\appendixfigurewidth]{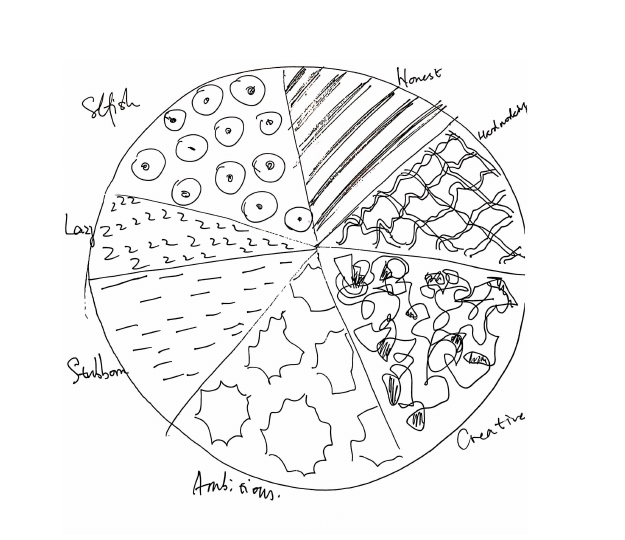}
    \caption{A semantically-resonant pattern design by the author (Au2) for personality concept set collected in our evaluation workshop pilot.}
    \label{fig:pil2-ab}
\end{figure}

\begin{figure}[htb]
    \centering
    \includegraphics[width=\appendixfigurewidth]{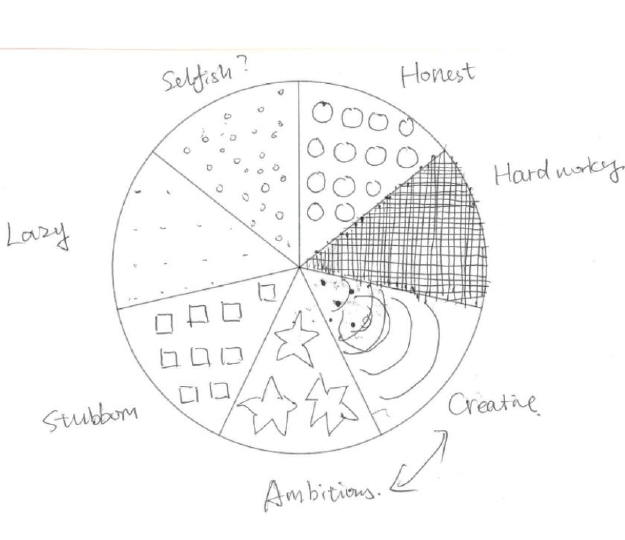}
    \caption{A semantically-resonant pattern design by the author (Au3) for personality concept set collected in our evaluation workshop pilot.}
    \label{fig:pil3-ab}
\end{figure}

\begin{figure}[htb]
    \centering
    \includegraphics[width=\appendixfigurewidth]{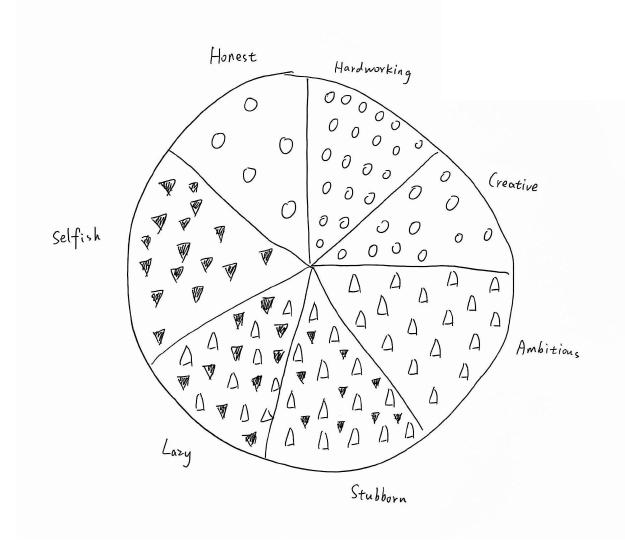}
    \caption{A semantically-resonant pattern design by the author (Au4) for personality concept set collected in our evaluation workshop pilot.}
    \label{fig:pil4-ab}
\end{figure}

%ball sports concept set
\begin{figure}[htb]
    \centering
    \includegraphics[width=\appendixfigurewidth]{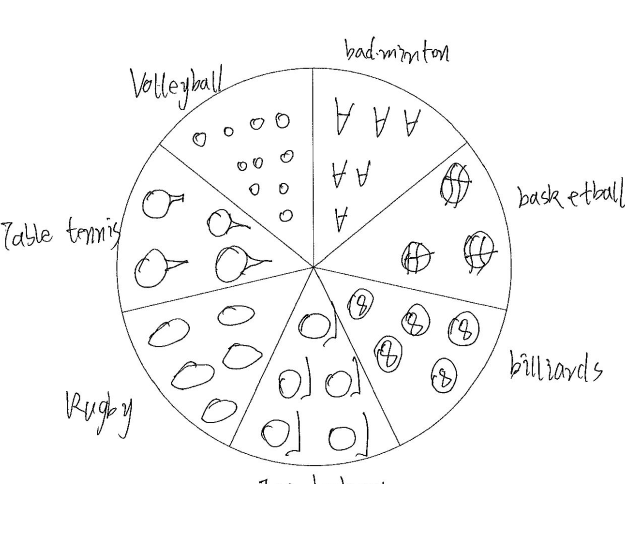}
    \caption{A semantically-resonant pattern design by the non-expert participants (Pi1) for ball sports concept set collected in our evaluation workshop pilot.}
    \label{fig:sz1-con}
\end{figure}

\begin{figure}[htb]
    \centering
    \includegraphics[width=\appendixfigurewidth]{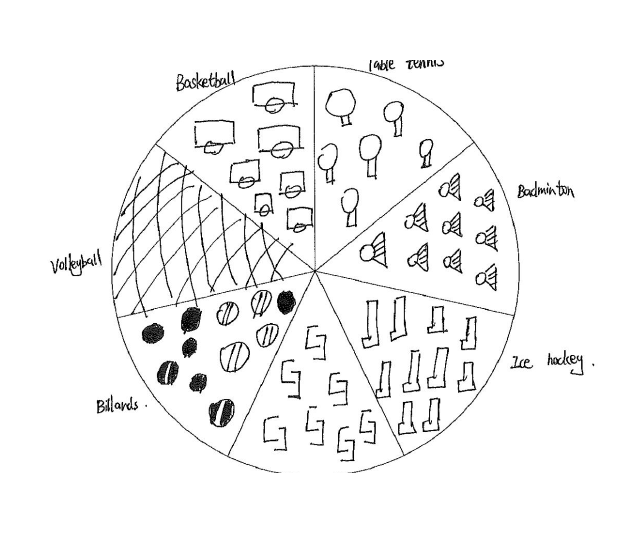}
    \caption{A semantically-resonant pattern design by the non-expert participants (Pi2) for ball sports concept set collected in our evaluation workshop pilot.}
    \label{fig:sz2-con}
\end{figure}

\begin{figure}[t]
    \centering
    \includegraphics[width=\appendixfigurewidth]{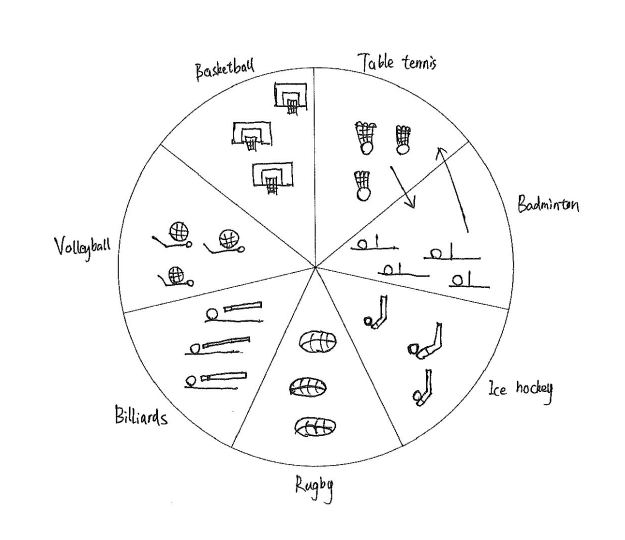}
    \caption{A semantically-resonant pattern design by the non-expert participants (Pi3) for ball sports concept set collected in our evaluation workshop pilot.}
    \label{fig:sz3-con}
\end{figure}

\begin{figure}[t]
    \centering
    \includegraphics[width=\appendixfigurewidth]{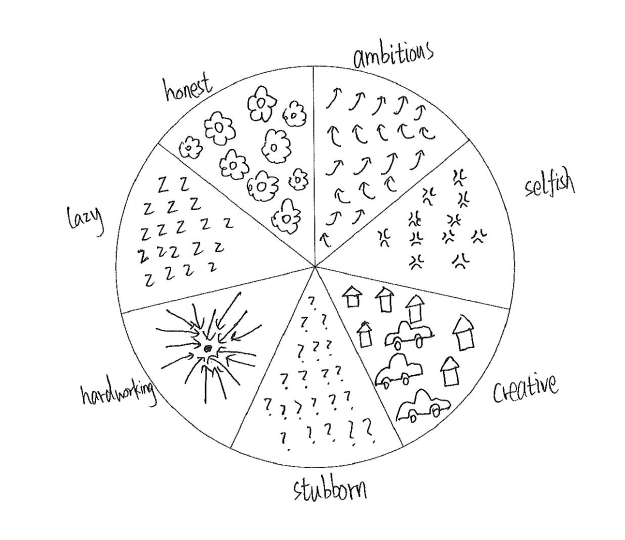}
    \caption{A semantically-resonant pattern design by the non-expert participants (Pi4) for ball sports concept set collected in our evaluation workshop pilot.}
    \label{fig:sz4-con}
\end{figure}

%personality concept set
\begin{figure}[t]
    \centering
    \includegraphics[width=\appendixfigurewidth]{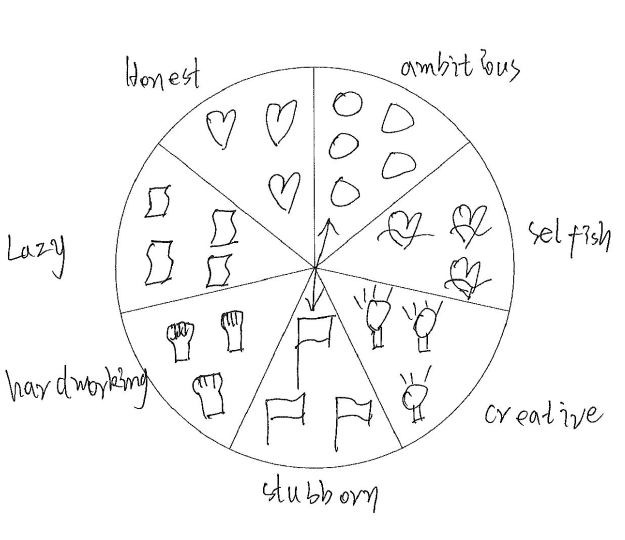}
    \caption{A semantically-resonant pattern design by the non-expert participants (Pi1) for personality concept set collected in our evaluation workshop pilot.}
    \label{fig:sz1-ab}
\end{figure}

\begin{figure}[t]
    \centering
    \includegraphics[width=\appendixfigurewidth]{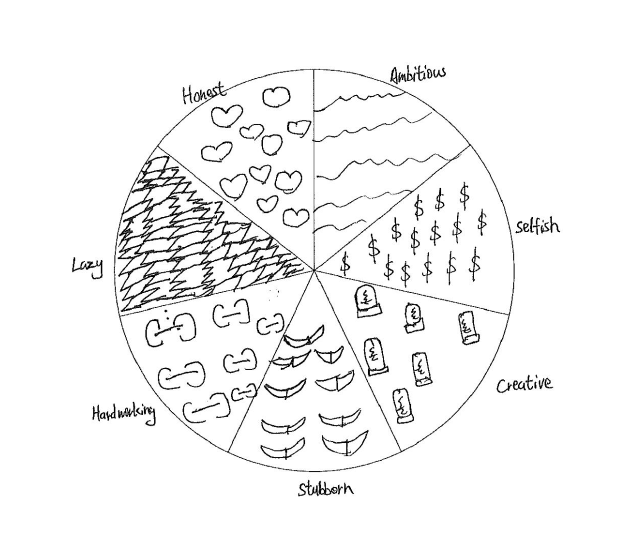}
    \caption{A semantically-resonant pattern design by the non-expert participants (Pi2) for personality concept set collected in our evaluation workshop pilot.}
    \label{fig:sz2-ab}
\end{figure}

\begin{figure}[t]
    \centering
    \includegraphics[width=\appendixfigurewidth]{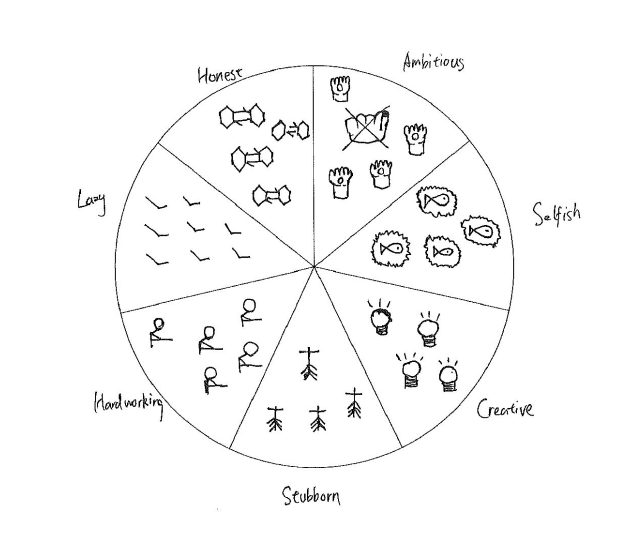}
    \caption{A semantically-resonant pattern design by the non-expert participants (Pi3) for personality concept set collected in our evaluation workshop pilot.}
    \label{fig:sz3-ab}
\end{figure}

\begin{figure}[t]
    \centering
    \includegraphics[width=\appendixfigurewidth]{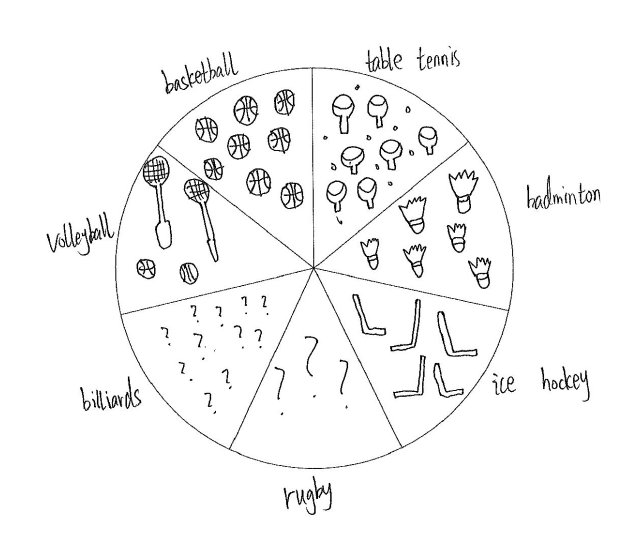}
    \caption{A semantically-resonant pattern design by the non-expert participants (Pi4) for personality concept set collected in our evaluation workshop pilot.}
    \label{fig:sz4-ab}
\end{figure}

%ball sports concept set
\begin{figure}[t]
    \centering
    \includegraphics[width=\appendixfigurewidth]{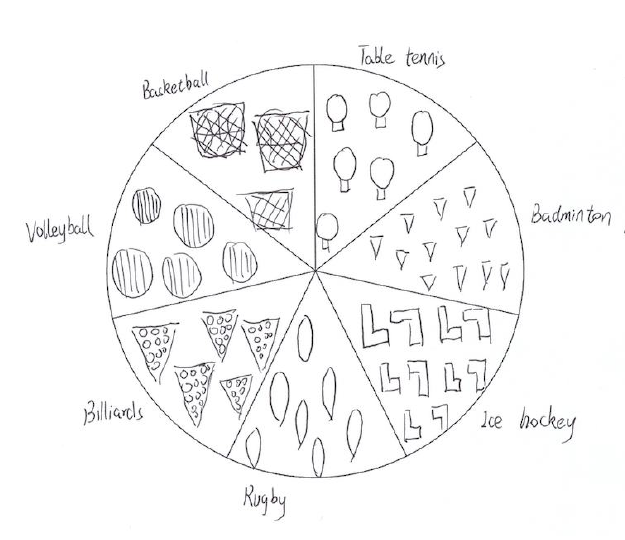}
    \caption{A semantically-resonant pattern design by the non-expert participants (P1) for ball sports concept set collected in our evaluation workshop.}
    \label{fig:p1-con}
\end{figure}

\begin{figure}[t]
    \centering
    \includegraphics[width=\appendixfigurewidth]{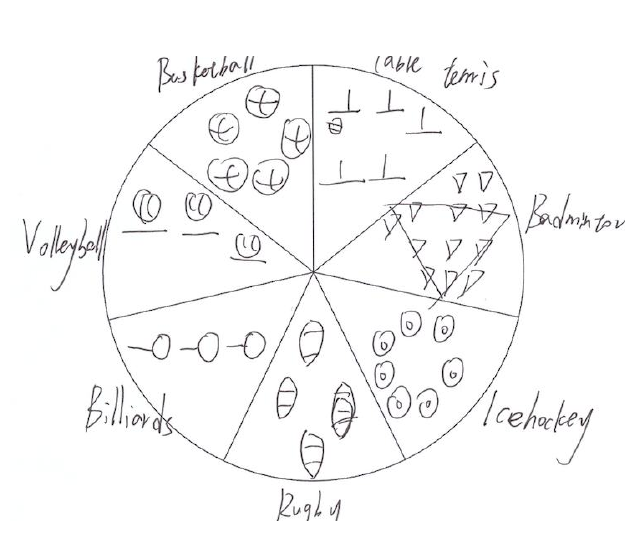}
    \caption{A semantically-resonant pattern design by the non-expert participants (P2) for ball sports concept set collected in our evaluation workshop.}
    \label{fig:p2-con}
\end{figure}

\begin{figure}[t]
    \centering
    \includegraphics[width=\appendixfigurewidth]{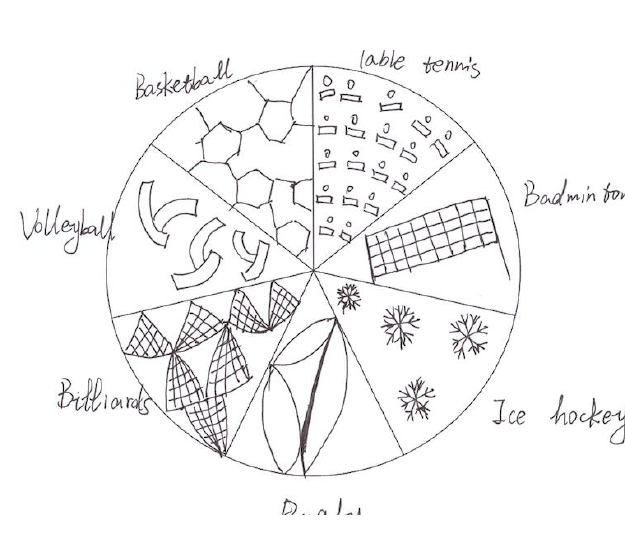}
    \caption{A semantically-resonant pattern design by the non-expert participants (P3) for ball sports concept set collected in our evaluation workshop.}
    \label{fig:p3-con}
\end{figure}

\begin{figure}[t]
    \centering
    \includegraphics[width=\appendixfigurewidth]{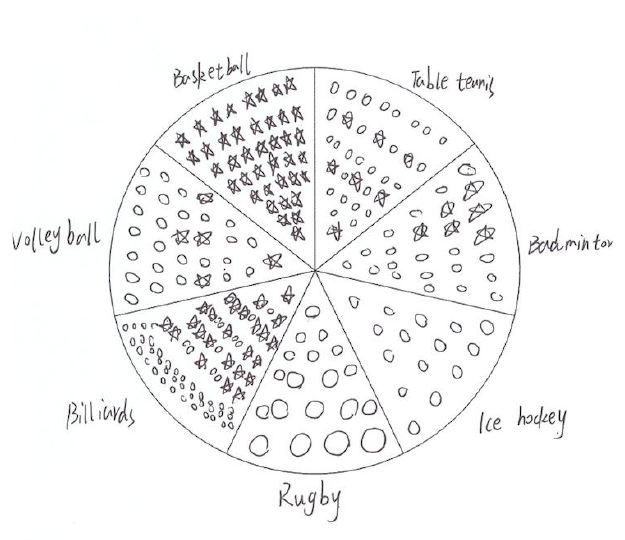}
    \caption{A semantically-resonant pattern design by the non-expert participants (P4) for ball sports concept set collected in our evaluation workshop.}
    \label{fig:p4-con}
\end{figure}

\begin{figure}[t]
    \centering
    \includegraphics[width=\appendixfigurewidth]{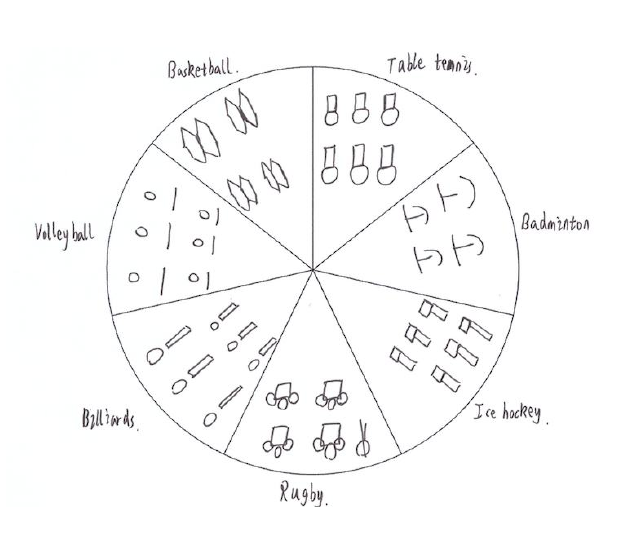}
    \caption{A semantically-resonant pattern design by the non-expert participants (P5) for ball sports concept set collected in our evaluation workshop.}
    \label{fig:p5-con}
\end{figure}

\begin{figure}[t]
    \centering
    \includegraphics[width=\appendixfigurewidth]{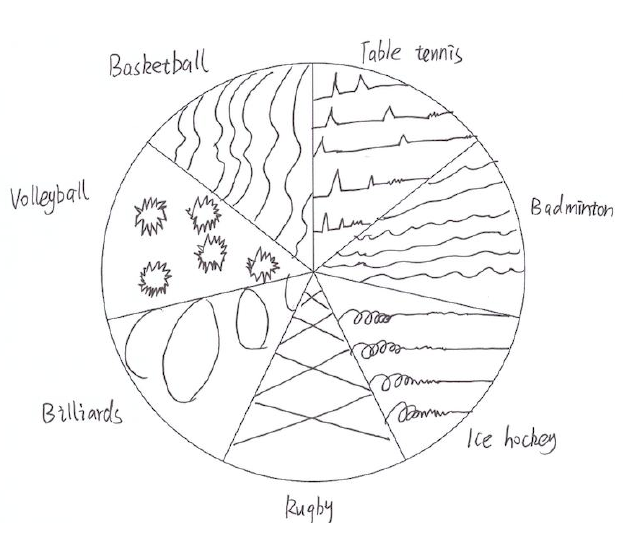}
    \caption{A semantically-resonant pattern design by the non-expert participants (P6) for ball sports concept set collected in our evaluation workshop.}
    \label{fig:p6-con}
\end{figure}

\begin{figure}[t]
    \centering
    \includegraphics[width=\appendixfigurewidth]{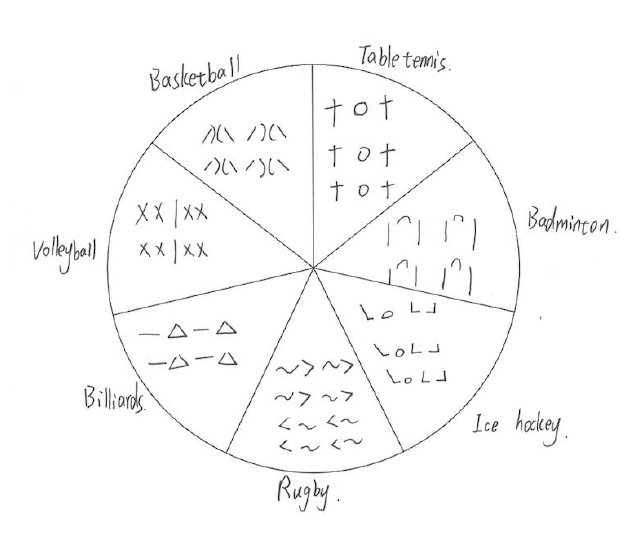}
    \caption{A semantically-resonant pattern design by the non-expert participants (P7) for ball sports concept set collected in our evaluation workshop.}
    \label{fig:p7-con}
\end{figure}

\begin{figure}[t]
    \centering
    \includegraphics[width=\appendixfigurewidth]{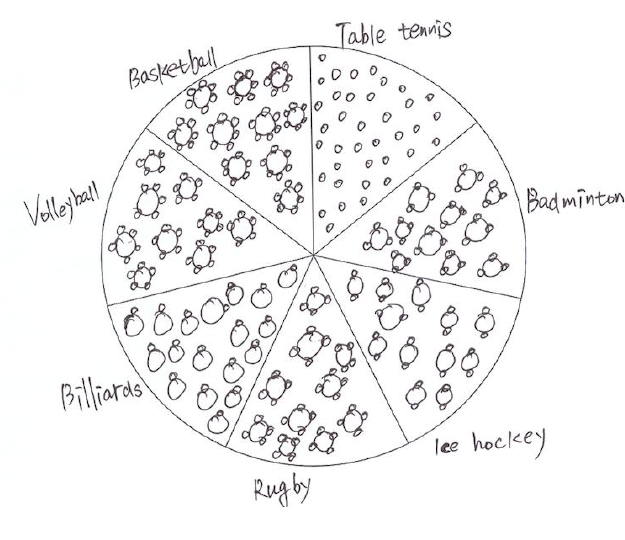}
    \caption{A semantically-resonant pattern design by the non-expert participants (P8) for ball sports concept set collected in our evaluation workshop.}
    \label{fig:p8-con}
\end{figure}

\begin{figure}[t]
    \centering
    \includegraphics[width=\appendixfigurewidth]{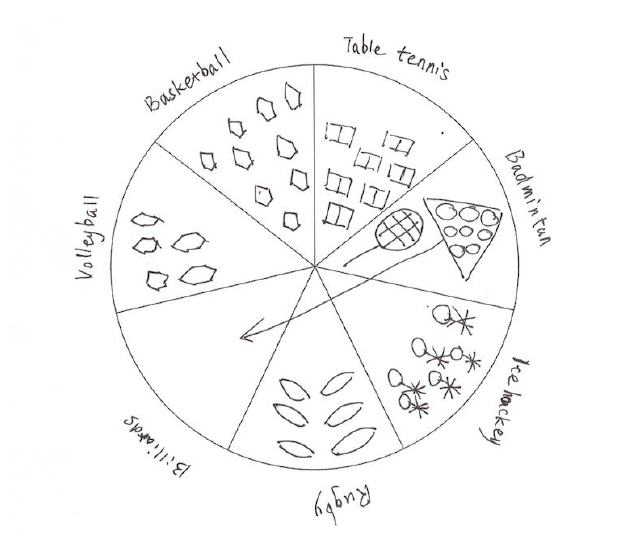}
    \caption{A semantically-resonant pattern design by the non-expert participants (P9) for ball sports concept set collected in our evaluation workshop.}
    \label{fig:p9-con}
\end{figure}

\begin{figure}[t]
    \centering
    \includegraphics[width=\appendixfigurewidth]{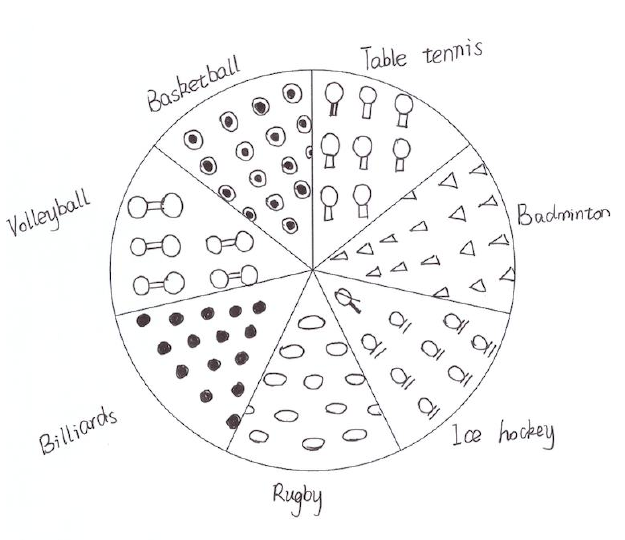}
    \caption{A semantically-resonant pattern design by the non-expert participants (P10) for ball sports concept set collected in our evaluation workshop.}
    \label{fig:p10-con}
\end{figure}

\begin{figure}[t]
    \centering
    \includegraphics[width=\appendixfigurewidth]{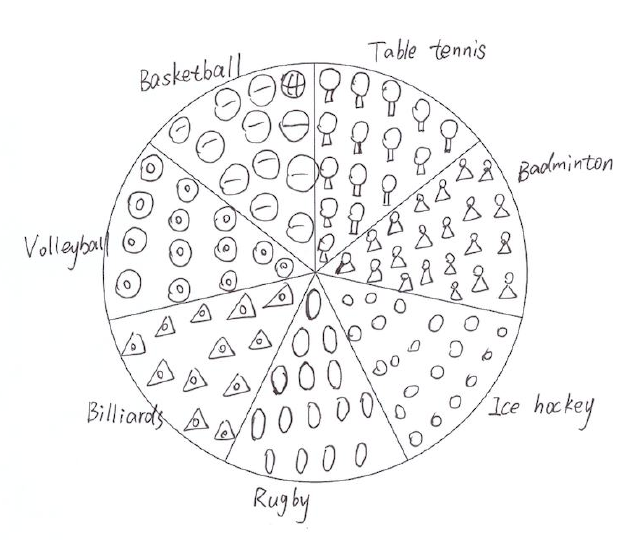}
    \caption{A semantically-resonant pattern design by the non-expert participants (P11) for ball sports concept set collected in our evaluation workshop.}
    \label{fig:p11-con}
\end{figure}

\begin{figure}[t]
    \centering
    \includegraphics[width=\appendixfigurewidth]{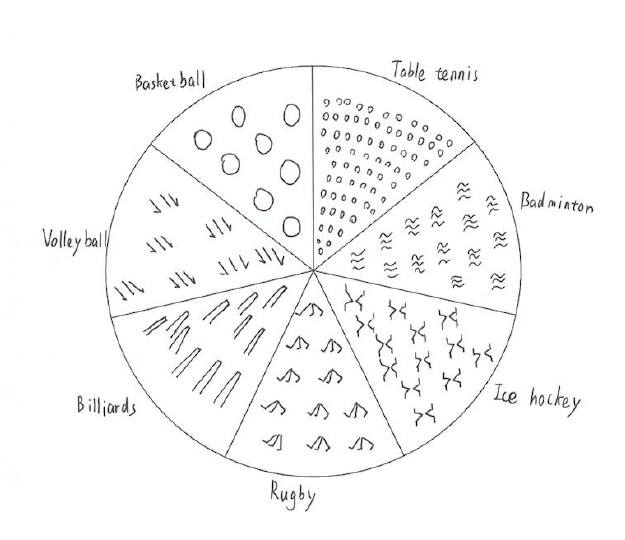}
    \caption{A semantically-resonant pattern design by the non-expert participants (P12) for ball sports concept set collected in our evaluation workshop.}
    \label{fig:p12-con}
\end{figure}

%personality concept set
\begin{figure}[t]
    \centering
    \includegraphics[width=\appendixfigurewidth]{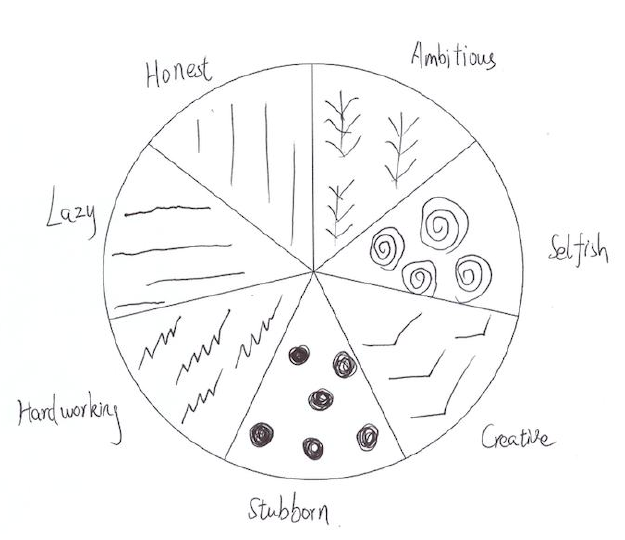}
    \caption{A semantically-resonant pattern design by the non-expert participants (P1) for personality concept set collected in our evaluation workshop.}
    \label{fig:p1-ab}
\end{figure}

\begin{figure}[t]
    \centering
    \includegraphics[width=\appendixfigurewidth]{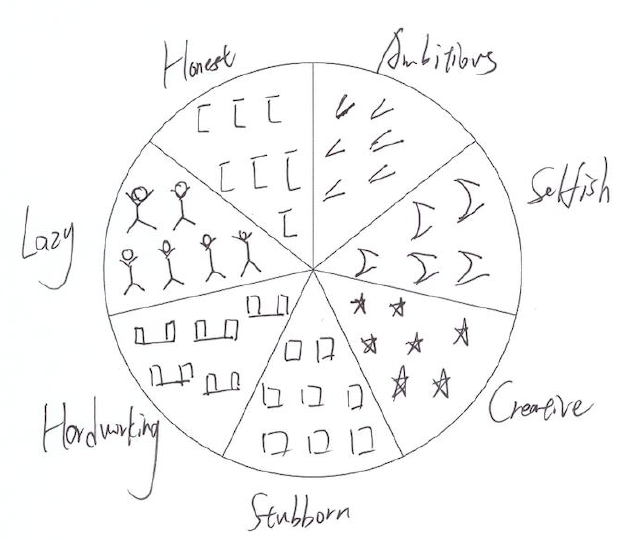}
    \caption{A semantically-resonant pattern design by the non-expert participants (P2) for personality concept set collected in our evaluation workshop.}
    \label{fig:p2-ab}
\end{figure}

\begin{figure}[t]
    \centering
    \includegraphics[width=\appendixfigurewidth]{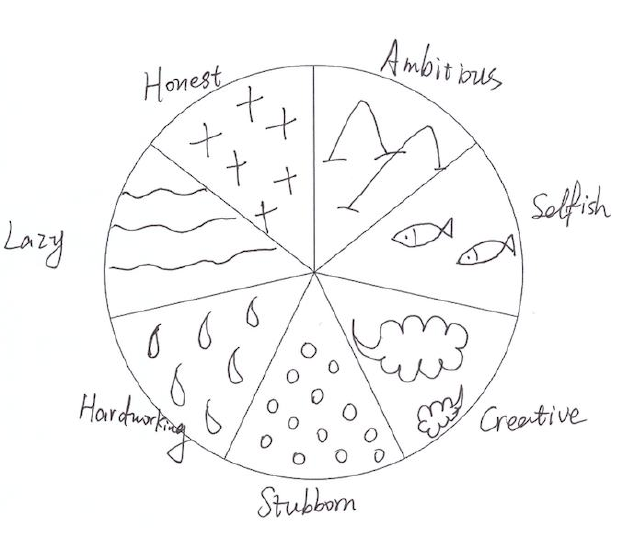}
    \caption{A semantically-resonant pattern design by the non-expert participants (P3) for personality concept set collected in our evaluation workshop.}
    \label{fig:p3-ab}
\end{figure}

\begin{figure}[t]
    \centering
    \includegraphics[width=\appendixfigurewidth]{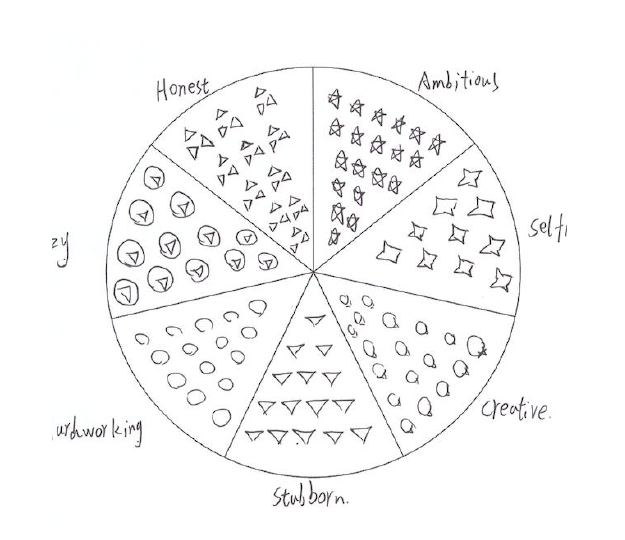}
    \caption{A semantically-resonant pattern design by the non-expert participants (P4) for personality concept set collected in our evaluation workshop.}
    \label{fig:p4-ab}
\end{figure}

\begin{figure}[t]
    \centering
    \includegraphics[width=\appendixfigurewidth]{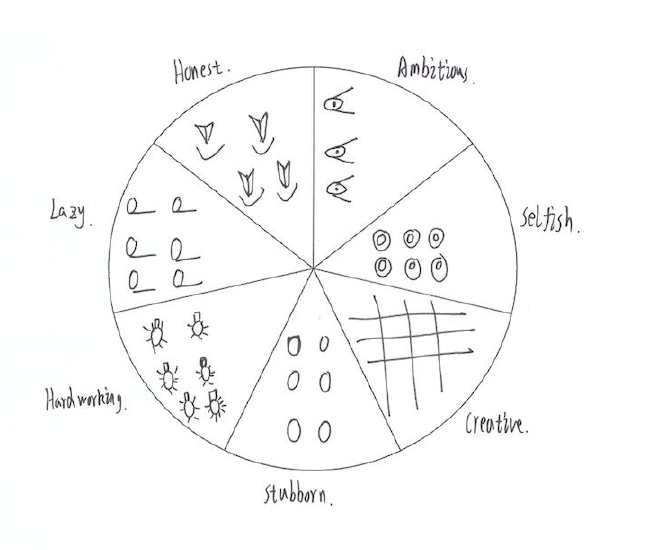}
    \caption{A semantically-resonant pattern design by the non-expert participants (P5) for personality concept set collected in our evaluation workshop.}
    \label{fig:p5-ab}
\end{figure}

\begin{figure}[t]
    \centering
    \includegraphics[width=\appendixfigurewidth]{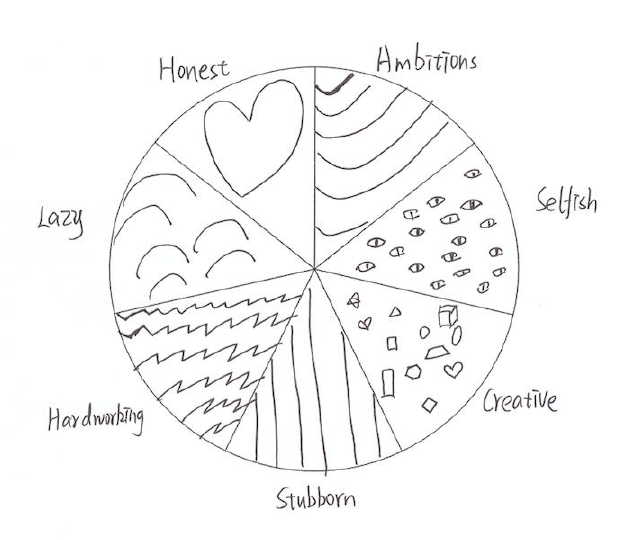}
    \caption{A semantically-resonant pattern design by the non-expert participants (P6) for personality concept set collected in our evaluation workshop.}
    \label{fig:p6-ab}
\end{figure}

\begin{figure}[t]
    \centering
    \includegraphics[width=\appendixfigurewidth]{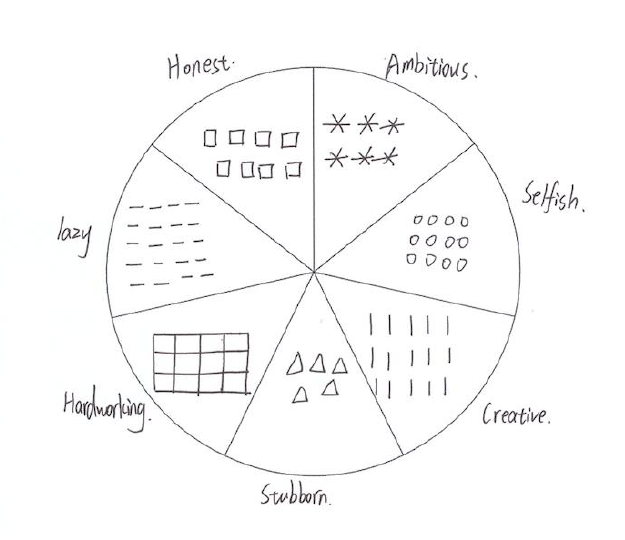}
    \caption{A semantically-resonant pattern design by the non-expert participants (P7) for personality concept set collected in our evaluation workshop.}
    \label{fig:p7-ab}
\end{figure}

\begin{figure}[t]
    \centering
    \includegraphics[width=\appendixfigurewidth]{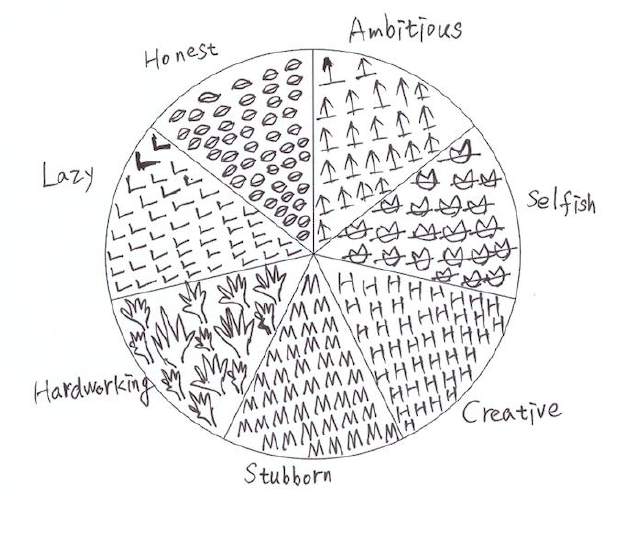}
    \caption{A semantically-resonant pattern design by the non-expert participants (P8) for personality concept set collected in our evaluation workshop.}
    \label{fig:p8-ab}
\end{figure}

\begin{figure}[t]
    \centering
    \includegraphics[width=\appendixfigurewidth]{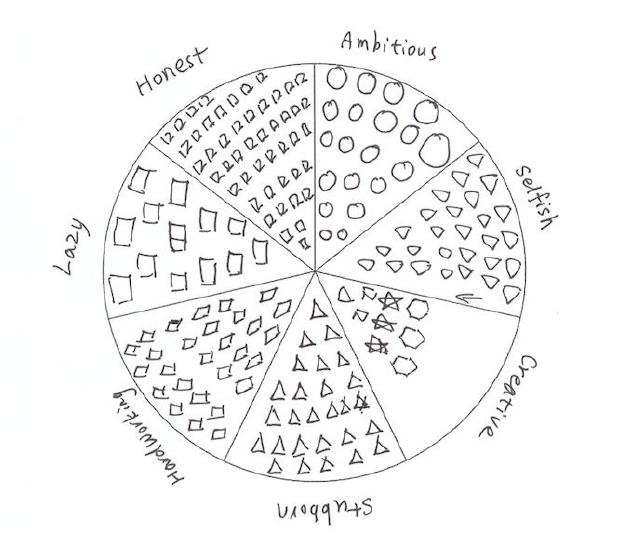}
    \caption{A semantically-resonant pattern design by the non-expert participants (P9) for personality concept set collected in our evaluation workshop.}
    \label{fig:p9-ab}
\end{figure}

\begin{figure}[t]
    \centering
    \includegraphics[width=\appendixfigurewidth]{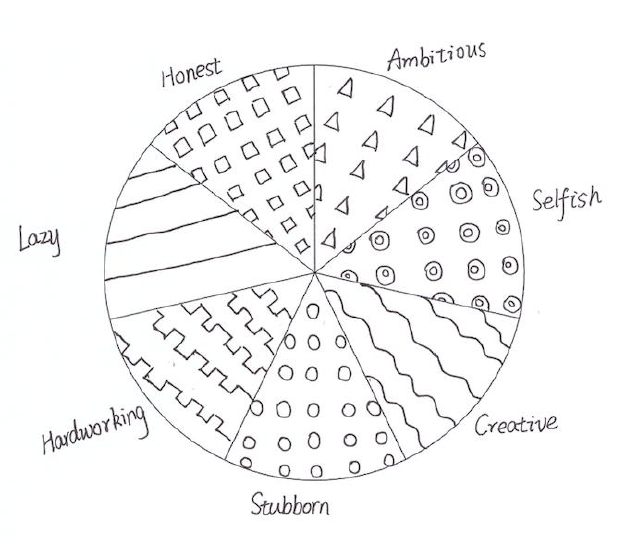}
    \caption{A semantically-resonant pattern design by the non-expert participants (P10) for personality concept set collected in our evaluation workshop.}
    \label{fig:p10-ab}
\end{figure}

\begin{figure}[t]
    \centering
    \includegraphics[width=\appendixfigurewidth]{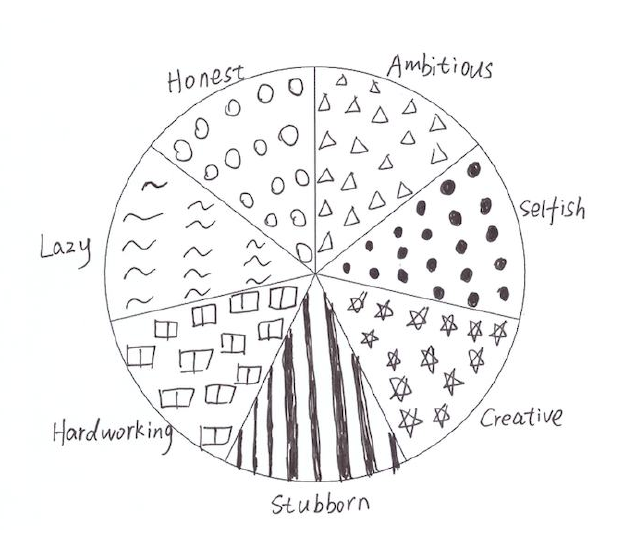}
    \caption{A semantically-resonant pattern design by the non-expert participants (P11) for personality concept set collected in our evaluation workshop.}
    \label{fig:p11-ab}
\end{figure}

\begin{figure}[t]
    \centering
    \includegraphics[width=\appendixfigurewidth]{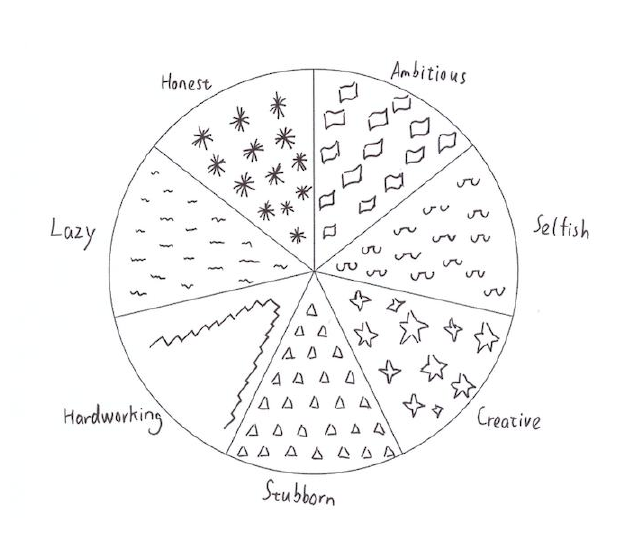}
    \caption{A semantically-resonant pattern design by the non-expert participants (P12) for personality concept set collected in our evaluation workshop.}
    \label{fig:p12-ab}
\end{figure}

\end{document}